\RequirePackage{fix-cm}
\documentclass[smallextended]{svjour3}       
\smartqed  

\usepackage{algorithmic}
\usepackage[numbers,sort&compress]{natbib}
\usepackage{textcomp}
\usepackage{amsmath,amssymb,amsfonts}
\usepackage{algorithmic}
\usepackage{graphicx}
\usepackage{textcomp}
\usepackage{xcolor}
\usepackage{soul}
\usepackage{subfigure}
\usepackage{tikz}
\usepackage{tabu}
\usepackage{threeparttable}
\usepackage{balance}
\usepackage{colortbl}
\usepackage{multirow}
\usepackage{adjustbox}
\usepackage{xcolor}

\definecolor{codegreen}{rgb}{0,0.6,0}
\definecolor{codegray}{rgb}{0.5,0.5,0.5}
\definecolor{codepurple}{rgb}{0.58,0,0.82}
\definecolor{backcolour}{rgb}{0.95,0.95,0.92}

\usepackage{listings}
\lstdefinestyle{mystyle}{
    float=tp,
    abovecaptionskip=-5pt,
    backgroundcolor=\color{backcolour},   
    commentstyle=\color{codegreen},
    keywordstyle=\color{magenta},
    numberstyle=\tiny\color{codegray},
    stringstyle=\color{codepurple},
    basicstyle=\ttfamily\footnotesize,
    breakatwhitespace=false,         
    breaklines=true,                 
    captionpos=b,                    
    keepspaces=true,                 
    numbers=left,                    
    numbersep=5pt,                  
    showspaces=false,                
    showstringspaces=false,
    showtabs=false,                  
    tabsize=2
}

\usepackage[utf8]{inputenc}
\usepackage{float}
\usepackage{amsmath}
\usepackage[english]{babel}
\usepackage[edges]{forest}
\usepackage{float}
\usepackage{chngpage}
\usepackage{tcolorbox}
\usepackage[ruled,vlined]{algorithm2e}
\usepackage[hyphens]{url}
\usepackage{hyperref}
\usepackage{graphics}
\newcommand*\circled[1]{\tikz[baseline=(char.base)]{
            \node[shape=circle,draw=red!60,inner sep=2pt] (char) {#1};}}
\newtheorem{defn}[theorem]{Definition}
\usepackage[figuresleft]{rotating}
\journalname{Empirical Software Engineering}
\begin{document}

\title{A Comparison of Reinforcement Learning Frameworks for Software Testing Tasks}


\author{Paulina Stevia Nouwou Mindom \and Amin Nikanjam \and Foutse Khomh} 

\authorrunning{Nouwou Mindom et al.} 

\institute{Paulina Stevia Nouwou Mindom \and Amin Nikanjam \and Foutse Khomh \at 
               Polytechnique Montréal, Québec, Canada \\
              \email{\{paulina-stevia.nouwou-mindom, amin.nikanjam, foutse.khomh\}@polymtl.ca} 
}

\date{Received: August 2022 / Accepted: June 2023}

\maketitle
\begin{abstract}
Software testing activities scrutinize the artifacts and the behavior of a software product to find possible defects and ensure that the product meets its expected requirements. Although various approaches of software testing have shown to be very promising in revealing defects in software, some of them lack automation or are partly automated which increases the testing time, the manpower needed, and overall software testing costs. 
Recently, Deep Reinforcement Learning (DRL) has been successfully employed in complex testing tasks such as game testing, regression testing, and test case prioritization to automate the process and provide continuous adaptation. Practitioners can employ DRL by implementing from scratch a DRL algorithm or using a DRL framework.
DRL frameworks offer well-maintained implemented state-of-the-art DRL algorithms to facilitate and speed up the development of DRL applications. Developers have widely used these frameworks to solve problems in various domains including software testing. However, to the best of our knowledge, there is no study that empirically evaluates the effectiveness and performance of implemented algorithms in DRL frameworks. 
Moreover, some guidelines are lacking from the literature that would help practitioners choose one DRL framework over another.
In this paper, therefore, we empirically investigate the applications of carefully selected DRL algorithms (based on the characteristics of algorithms and environments) on two important software testing tasks: test case prioritization in the context of Continuous Integration (CI) and game testing. 
For the game testing task, we conduct experiments on a simple game and use DRL algorithms to explore the game to detect bugs. Results show that some of the selected DRL frameworks such as Tensorforce outperform recent approaches in the literature. To prioritize test cases, we run extensive experiments on a CI environment where DRL algorithms from different frameworks are used to rank the test cases. We find some cases where our DRL configurations outperform the implementation of the baseline. Our results show that the performance difference between implemented algorithms in some cases is considerable, motivating further investigation. Moreover, empirical evaluations on some benchmark problems are recommended for researchers looking to select DRL frameworks, to make sure that DRL algorithms perform as intended.

\end{abstract}

\section{Introduction}
Software bugs and failures are costing trillions of dollars every year to the global economy according to a recent report by a software testing company Tricentis\footnote{\url{https://www.techrepublic.com/article/report-software-failure-caused-1-7-trillion-in-financial-losses-in-2017/}}. In 2017 alone, 606 software bugs costed the global economy about \$1.7 trillion dollars, affecting 3.7 billion people. To alleviate this issue, researchers and practitioners have been striving to develop efficient testing techniques and tools, to help improve the reliability of software systems before they are released to the public. 
Several strategies, such as random testing by Hamlet et al. \cite{hamlet1994random}, coverage-based testing by Zhu et al.\cite{zhu1997software} and search-based testing by Harman et al. \cite{harman2015achievements} have been proposed to evaluate that a software product does what it is supposed to do. More recently, 
Deep Reinforcement Learning (DRL) is being increasingly leveraged for software testing purposes as studied by Zheng et al.\cite{zheng2019wuji}, Bagherzadeh et al. \cite{bagherzadeh2021reinforcement}, Moghadam et al.\cite{moghadam2021autonomous}, Malialis et al.\cite{malialis2015distributed} thanks to the availability of multiple DRL frameworks providing implemented DRL algorithms, e.g., Advantage Actor Critic (A2C), Deep Q-Networks (DQN), Proximal Policy Optimization (PPO). 
For example, 
Kim et al. \cite{kim2018generating} leveraged the Keras-rl framework to apply DRL to test data generation. 
Similarly, Drozd et al. \cite{drozd2018fuzzergym} used the Tensorforce framework to apply DRL to Fuzzing testing  and Romdhana et al. \cite{romdhana2022deep} used the Stable-baselines framework for black box testing of android applications. 

However, given that these implemented DRL algorithms often make assumptions that could hold only for certain types of problems and not for others, it could be challenging for developers and researchers to select the most adequate DRL implementation for their problem. The choice of a DRL algorithm depends on the nature of the problem to solve, the available computation budget, and the desired generalizability of the trained models. Moreover, given the fact that DRL algorithms are often implemented differently in different DRL frameworks, it is unclear if the same results can be obtained using different frameworks.
To clear up these interrogations and help researchers and practitioners make informed decisions when choosing a DRL framework for their problem, in this paper, we examine and compare the applicability of different DRL frameworks for software engineering testing tasks. 
Specifically, we apply DRL algorithms from different frameworks to game testing and test case prioritization. The automation of game testing is critical because of the frequent requirements changes that occur during a game development process as studied by Santos et al. \cite{Santos18}. 
Recently, Yang et al. \cite{yang2018towards,yang2019bayes}, Koroglu et al. \cite{koroglu2018qbe}, Adamo et al. \cite{adamo2018reinforcement} applied different DRL algorithms to automate game testing and improve the fault identification process. Test case prioritization improves the testing process by finding optimal ordering of the test cases and detecting faults as early as possible. Bertolino et al. \cite{bertolino2020learning}, Spieker et al.\cite{spieker2017reinforcement} successfully applied DRL in prioritizing test cases for various configurations. Moreover, as these tasks have gained a lot of attention recently, by studying them we can provide meaningful results that can be used by the software engineering community.

In this paper, we perform a comprehensive comparison of different DRL algorithms implemented in three frameworks, i.e., Stable-baselines3 \cite{Stable-Baselines3}, Keras-rl \cite{framework:plappert2016kerasrl}, and Tensorforce \cite{schaarschmidt2018lift}. We investigate which DRL algorithms/frameworks may be more suitable for detecting bugs in games and solving the test case prioritization problem. 
Results show that the diversity of hyperparameters that each framework provide impacts its suitability for each of the studied software testing tasks. Given some algorithms, the Tensorforce framework tends to be more suitable for detecting bugs as it provides hyperparameters that allow a deeper exploration of the states of the environment while the Stable-baselines framework tends to be more suitable for the test case prioritization problem.

To summarize, our work makes the following contributions:
\begin{itemize}
    \item To evaluate the usefulness of DRL on game testing, we utilized three state-of-the-art DRL frameworks: Stable-baselines, Keras-rl, and Tensorforce. Specifically, we applied them to the Block Maze game for bug detection and collected the number of bugs, the state coverage, the code coverage, the cumulative reward, the average training and prediction times. We have compared a total of seven DRL configurations and some of them outperform the existing work.
    \item Based on eight publicly available datasets, we applied state-of-the-art DRL frameworks on two ranking models and collected results to evaluate their usefulness in prioritizing test cases. As metrics of comparison, we consider the Normalized Rank Percentile Average (NRPA), the Average Percentage of Faults Detected (APFD), the average training and prediction times for each DRL configuration. The results collected are compared with the baselines and we derive conclusions regarding the most accurate DRL frameworks for test case prioritization. We found out that in most datasets, the Stable-baselines framework originally used by Bagherzadeh et al. \cite{bagherzadeh2021reinforcement} performs better than Tensorforce and Keras-rl.
    \item We provide some recommendations for researchers looking to select a DRL framework  as we noticed differences in performance when considering the same algorithm among different frameworks. For example, the same DQN  algorithm from different frameworks show different results. 
\end{itemize}
 
The rest of this paper is organized as follows. In Section \ref{sec:Background}, we review the necessary background knowledge on the game testing problem, the test case prioritization problem, and DRL. The methodology followed in our study is described in Section \ref{studydesign}. We discuss the obtained empirical results in Section \ref{Validation}. Some recommendations for future work are mentioned in Section \ref{futurework}. We review related work in Section \ref{Related work}. Threats to validity of our study are discussed in Section \ref{threats}. Finally, we conclude the paper and discuss some future works in Section \ref{Conclusion}.

\section{Background} \label{sec:Background}
In this section, first we introduce DRL and present some state-of-art DRL frameworks. Secondly, we describe the terms and notations used to define the test case prioritization and game testing problems. 

\subsection{Deep Reinforcement Learning} \label{sec:Reinforcement Learning}
A DRL agent interacts with the environment that can be modelled as a Markov decision process $(\mathcal{S,A,P,\gamma})$ with the following components: 

\textbf{State of the environment:} A state $ s \in \mathcal{S} = \mathbb{R}^n $ represents the agent perception of the environment. 

\textbf{Action:} Based on the observation (i.e., state of the environment), the agent chooses among available actions in $\mathcal{A}$.

\textbf{State transition distribution:} 
$\mathcal{P} = \mathcal{P}(s_{t+1},r_{t}|s_{t},a_{t}) ~~ a_{t} \in \mathcal{A}$, defines the probability of the agent to move to the next state $s_{t+1}$, performing action $a_{t}$ receives $r_{t}$ as reward  given that it is in state $s_{t}$ .
The goal of the agent is to maximize the expected rewards discounted by $\gamma$. To make the decision to move to a state given its observation, the DRL agent follows a policy $\pi: \mathcal{S} \rightarrow \mathcal{A}$ which is a mapping from $\mathcal{S}$ to  $\mathcal{A}$.

\textbf{Episode:} An episode is a sequence of states of the environment, actions performed by an agent and rewards (an incentive mechanism that tells the agent about the effectiveness of the action)  which ends when the agent has reached a terminal state or has reached a maximum number of steps.

\textbf{Policy.} Given an agent, a policy $\pi$ is defined as a function $\pi:S \rightarrow A$ mapping each state $s \in S$ to an action $a \in A$. The policy indicates the agent's decision in each state of the underlined task. It can be a strategy from a human expert or learned from experiences accordingly.

DRL algorithms can be classified based on the following properties similar to the work by  Bagherzadeh et al.\cite{bagherzadeh2021reinforcement} :

\textbf{Model-based and Model-free DRL.} In model-based DRL, the agent knows the environment. It knows in advance the reaction of the environment to possible actions and the potential rewards it will get from taking each action. During training, the agent learns the optimal behavior by taking actions and observing the outcomes which include the next state and the immediate reward. On the contrary, in model-free DRL, the agent has to learn the dynamics of the environment by interacting with it. From the interaction with the environment, the agent learns an optimal policy for selecting an action. In this work, we are only interested in model-free DRL algorithms as some of the test case features (execution time) are unknown beforehand as well as the location of faults in 
a game.

\textbf{Value-based, policy-based, and actor-critic learning.}
At every state, value-based methods estimate the Q-value and select the action with the best Q-value. A Q-value shows how good an action might be given a state. Regarding policy-based methods, an initial policy is parameterized, then during training, the parameters are updated using gradient-based or gradient-free optimization techniques. Regarding actor-critic methods, the agent learns simultaneously from value-based and policy-based techniques. The policy function (actor) selects the action and the value function (critic) estimates the Q-values based on the action selected by the actor.

\textbf{Action and observation space.} The action space indicates the possible moves of the agent inside the environment. The observation space indicates what the agent can know about the environment. The action and observation space can be discrete or continuous. Specifically, the observation space can be a real number or high dimensional. While a discrete action space means that the agent chooses its action among distinct values, a continuous action space implies that the agent chooses actions among real values vectors. Not all DRL algorithms support discrete and continuous configurations for both the action and observation space, which limits the choice of algorithms to implement.

\textbf{On-policy vs Off-policy.} On-policy methods will collect data that is used to evaluate and improve a target policy and take actions. On the contrary, Off-policy methods will evaluate and improve a target policy that is different from the policy used to generate the data. Off-policy learners generally use a replay buffer to update the policy.

DRL methods use Deep Neural Networks (DNNs) to approximate the value function, or the model (state transition function and reward function) and tend to be 
a more manageable solution space in large complex environments.

\subsection{State-of-the-art DRL Frameworks} \label{State-of-the-art RL Algorithms and Frameworks}

In recent years, Lillicrap et al.\cite{lillicrap2015continuous}, Mnih et al. \cite{mnih2016asynchronous} introduced multiple model-free DRL algorithms; advancing the research around DRL. Different DRL frameworks such as Stable-baselines \cite{Stable-Baselines3,Stable-Baselines} and Tensorforce by  Schaarschmidt et al. \cite{schaarschmidt2018lift} have also been introduced to ease the implementation of DRL-based applications. These frameworks usually contain implementations of different DRL algorithms. While the developers may implement their own algorithm, in this work, we focus on comparing the implemented algorithms of existing DRL frameworks on software testing tasks. Table \ref{usageimplementation} provides a list of popular DRL frameworks, which are described below. 
 \begin{table*}[t]
 \large 
 \caption{Popular DRL frameworks. }
 \resizebox{\linewidth}{!}{
\begin{tabular}{|c|p{3cm}|p{3cm}|p{3cm}|p{3cm}|p{2cm}|}
\hline
\textbf{Framework} & \textbf{OpenAI baselines} \cite{framework:baselines} &\textbf{Stable-baselines3 }\cite{Stable-Baselines3} & \textbf{Tensorforce} \cite{schaarschmidt2018lift}& \textbf{Keras-rl} \cite{framework:plappert2016kerasrl} & \textbf{Dopamine} \cite{castro2018dopamine}\\ \hline
 \textbf{Github stars} & 12.1k & 2.6k &3.1k &  5.2k & 9.7k\\ \hline
 \textbf{\# of contributors} &  112 & 66& 54 &  39 & 10\\ \hline
 \textbf{Issues} & 391 open, 430 closed & 34 open, 453 closed &4 open, 617 closed &  10 open, 219 closed & 60 open, 82 closed\\ \hline
\textbf{Algorithms} & A2C, ACER, ACKTR, DDPG,
DQN, GAIL,
HER,PPO2,TRPO & A2C, DDPG, DQN,
 HER, PPO, SAC, TD3 & DQN, Dueling DQN, Double DQN, DDPG,
CDQN, A2C, A3C, TRPO,PPO, TA, Reinforce
 &  DQN, Double DQN, DDPG , CDQN, 
 Dueling DQN, SARSA & DQN, C51, Rainbow, IQN, SAC
  \\ \hline
\textbf{Tensorboard support} & Yes &Yes & Yes &No & Yes\\ \hline
\textbf{Pull requests} & 83 open 284 closed & 4 open, 199 closed &0 open, 225 closed & 33 open, 120 closed & 20 open, 20 closed\\ \hline
\textbf{Fork} & 4.2k & 546 & 511  &  1.3k & 1.3k\\ \hline
\textbf{Last update} &  Jan 31, 2020 & Dec 9, 2021 & Nov 10, 2021 & Nov 11, 2019 & Dec 14, 2021\\ \hline
\end{tabular}
}
\label{usageimplementation}
\end{table*}
\begin{itemize}
    \item OpenAI baselines \cite{framework:baselines} is the most popular DRL framework given its high GitHub star rating. It provides many state-of-the-art DRL algorithms. After installing the package, training a model only requires specifying the name of the algorithm as a parameter.
    
    \item Stable-baselines \cite{Stable-Baselines3,Stable-Baselines} is an improved version of OpenAI baselines with a more comprehensive documentation. In this paper, we used the version 3 of this framework, which is reported to be more reliable 
    because of its Pytorch \cite{NEURIPS2019_9015} backend that captures DNN policies. To train an agent, Stable-baselines has built-in functions that create a model depending on the DRL algorithm chosen. 
    
        \item Keras-rl \cite{framework:plappert2016kerasrl} provides the dueling extension of the DQN algorithm and SARSA algorithm that are not offered by Stable-baselines version 3. However, Keras-rl offers less algorithms than the previous frameworks.
        The training of an agent requires a few steps: the definition of the DNN that will be used for the training, the instantiation of the agent, its compilation, and finally the call of the training function. 
        
    \item Tensorforce \cite{schaarschmidt2018lift} provides the same algorithms as the Stable-baselines framework with some additions: Trust-Region Policy Optimization (TRPO), Dueling DQN, Reinforce, and Tensorforce Agent (TA). Tensorforce offers built-in functions to create and train an agent. Also, it offers the flexibility to train the agent without using the built-in functions, which allows it to capture the performance metrics of the agent, such as the reward. Finally, the training starts in a loop function depending on the number of episodes. Tensorforce relies on TensorFlow \cite{Abadi_TensorFlow_Large-scale_machine_2015} as backend.

    \item Dopamine \cite{castro2018dopamine} is a more recent framework that proposes an improved variant of the Deep Q-Networks (DQN) and the Soft Actor-Critic (SAC) algorithm. In addition to a TensorFlow backend for creating DNNs, Dopamine is configured using the gin\footnote{\url{https://github.com/google/gin-config}} framework, to specify and configure hyperparameters. The training of an agent requires instantiating the model and then starting the training with built-in functions.
\end{itemize}
Based on their popularity and ease of implementation, we choose to rely on Stable-baselines, Tensorforce, and Keras-rl frameworks. Table \ref{tab:Comparison between RL frameworks} summarizes the implemented DRL algorithms available in theses frameworks. 
\begin{table}[t]
\resizebox{\linewidth}{!}{
\begin{threeparttable}
\caption{Comparison between DRL frameworks}
\centering
\begin{tabular}{|c|c|c|c|c|}
\hline
\multicolumn{1}{|l|}{\textbf{RL Frameworks}} & \textbf{Algorithms} & \multicolumn{1}{c|}{\textbf{Learn.}} & \multicolumn{1}{c|}{\textbf{On/Off}} & \multicolumn{1}{c|}{\textbf{Act.}} \\ \hline
\multirow{7}{*}{\textbf{Stable-baselines}} & DQN  \cite{mnih2013playing}       &             Value                &    Off-policy                &       Dis                    \\ \cline{2-5} 
                                  & DDPG  \cite{lillicrap2015continuous}      &       Policy                      &   Off-policy                          &      Cont                     \\ \cline{2-5} 
                                  & A2C  \cite{mnih2016asynchronous}       &    Actor-Critic                         &     On-policy                        &    Both                       \\ \cline{2-5} 
                                  & TD3   \cite{fujimoto2018addressing}      & Policy                            &      Off-policy                       &   Cont                        \\ \cline{2-5} 
                                 
                                  & SAC  \cite{haarnoja2018soft}       &       Actor-Critic                      &      Off-policy                       &     Cont                      \\ \cline{2-5} 
                                  & PPO    \cite{schulman2017proximal}     &      Actor-Critic                       &        On-policy                     &         Both                  \\ \hline
\multirow{6}{*}{\textbf{Keras-rl}}         & DQN  \cite{mnih2013playing}       &             Value                &    Off-policy     & Dis                       \\ \cline{2-5} 
                                  & Dueling DQN \cite{wang2016dueling}&     Value                          &  Off-policy                            &     Dis                      \\ \cline{2-5} 
                                  & Double DQN  \cite{mnih2013playing}       &             Value                &    Off-policy & Dis                           \\ \cline{2-5} 
                                  & SARSA  \cite{sutton1998introduction}     &     Value                        &               On-policy              &   Dis                        \\ \cline{2-5} 
                                  & CDQN    \cite{gu2016continuous}    &   Value                          &    On-policy                         &  Cont                         \\ \cline{2-5} 
                                  & DDPG  \cite{lillicrap2015continuous}      &       Policy                      &   Off-policy                          &      Cont                            \\ \hline
\multirow{8}{*}{\textbf{Tensorforce}}   &   DQN  \cite{mnih2013playing}       &             Value                &    Off-policy &Dis                            \\ \cline{2-5} 
                                  & Double DQN  \cite{mnih2013playing}       &             Value                &    Off-policy & Dis                           \\ \cline{2-5} 
                                  & CDQN  \cite{gu2016continuous}    &   Value                          &    On-policy                         &  Cont                            \\ \cline{2-5} 
                                  & PPO    \cite{schulman2017proximal}     &      Actor-Critic                       &        On-policy                     &         Both                           \\ \cline{2-5} 
                                  & DDPG   \cite{lillicrap2015continuous}      &       Policy                      &   Off-policy                          &      Cont                            \\ \cline{2-5} 
                                  & A2C  \cite{mnih2016asynchronous}       &    Actor-Critic                         &     On-policy                        &    Both                            \\ \cline{2-5} 
                                  & A3C   \cite{mnih2016asynchronous}       &    Actor-Critic                         &     On-policy                        &    Both                             \\ \cline{2-5} 
                                  & TRPO   \cite{schulman2015trust}     &  Actor-Critic                           &          On-policy                   &   Both                      
                                   \\ \cline{2-5} 
                                  & TA   \cite{schulman2015trust}     &  Actor-Critic                           &          On-policy                   &   Both                    
                                   \\ \cline{2-5} 
                                  & Reinforce   \cite{schulman2015trust}     &  Policy                           &          On-policy                   &   Both                         \\ \hline
\end{tabular}
\begin{tablenotes}
     \item Cont:continuous, Dis:discrete, Both: continuous and discrete
     
 \end{tablenotes}
 \label{tab:Comparison between RL frameworks}
\end{threeparttable}
}
\end{table}  
Stable-baselines, Keras-rl, and Tensorforce have respectively 6, 5, and 10 available implemented DRL algorithms. They all contain the DQN algorithms, which we apply to the test case prioritization and game testing problems. We also  apply the A2C algorithm from Stable-baselines and Tensorforce to both test case prioritization and game testing problems. In addition to A2C and DQN, we applied the DDPG algorithm to the test case prioritization problem. Similarly, we applied the PPO algorithm to both test case prioritization and game testing problems. Stable-baselines framework in its second version offers two versions of PPO algorithm: PPO1 that requires OPENMPI\footnote{\url{https://www.open-mpi.org}} for multiprocessing and PPO2 that uses vectorized environments for multiprocessing. In this work, we chose to leverage PPO2 for two reasons: First, the version of OPENMPI required by PPO1 is not compatible with our experimental environment. Second, PPO from the Tensorforce framework uses vectorized environments for parallelism. Thus, it is fair to compare PPO from Tensorforce with PPO2 from Stable-baselines. For the sake of reading, we refer to PPO2 as the PPO from Stable-baselines. Keras-rl does not have an implemented version of the A2C, nor PPO algorithms, which could be applied to the previously mentioned problems. Moreover, these selected DRL algorithms are suitable for this paper, as we are capable of comparing their results with the baselines by Zheng et al. \cite{zheng2019wuji} and Bagherzadeh et al. \cite{bagherzadeh2021reinforcement}. Zheng et al. \cite{zheng2019wuji} used their own implementation of the A2C algorithm to detect bugs on three games. Thus, among the selected DRL strategies, we consider the A2C algorithm from the DRL frameworks and evaluate and compare our results with the results reported by Zheng et al. Given that the applicability of DRL algorithms is limited by the type of their action space, Bagherzadeh et al. \cite{bagherzadeh2021reinforcement} chose DRL algorithms from Stable-baselines that are compatible with the type of action space of the prioritization techniques they considered (see Section \ref{Creation of the RL environment test prio}). We do the same, and evaluate and compare the obtained results. 

\subsection{Game Testing}
The process of testing a game is an essential activity before its official release. The complexity of game testing has led researchers to investigate ways to automate it \cite{alshahwan2018deploying,fraser2011evosuite}. In the following, we introduce few concepts that are important to understand automatic game testing. 
\begin{defn}
\label{gamedefinition} : \textbf{Game.} A game $G$ can be defined as a function $G: A^n \rightarrow (S \times R)^n$, where $A$ is the set of action that can be performed by the agent, $S$ is the set of states of the game and $R$ represents the set of rewards that comes from the game, and $n$ is the number of steps in the game. A player takes a sequence of actions ($n$ actions) based on the observations it received until the end of the game. If we consider the game as an environment that the agent interacts with, each state refers to observations of the environment perceived by the agent at every time stamp. The action is a set of decisions that can be made by the agent which can be rewarded positively or negatively by the environment. 
\begin{figure}
    \centering
    \includegraphics[width=0.8\textwidth]{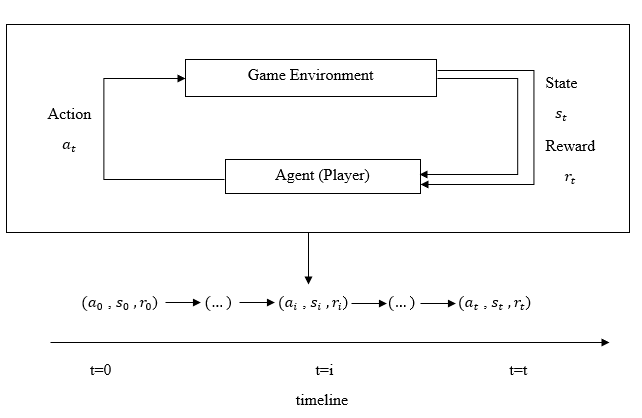}
    \caption{The interaction between a player and a game environment (inspired by Zheng et al. \cite{zheng2019wuji}).}
    \label{fig:GameEnvMDP}
\end{figure}
\end{defn}
Figure \ref{fig:GameEnvMDP} depicts the overall interaction between a player and a game. Given the state $s_t$ at time step $t$ the agent selects an action $a_t$ to interact with the game environment and receives a reward $s_t$  from the environment. The environment moves into a new state $s_{t+1}$, which affects the selection of the next action. 

\begin{defn}
\label{gamestatedefinition} : \textbf{Game state.} A state in the game refers to game's current status and can be represented as a fixed-length vector $(v_0, v_1, ..., v_n).$ Each element $v_i$ of the vector represents an aspect of the state of the game such as the position of a player, its speed or the location of the gold trophy in case of a Block Maze game. 
\end{defn}

\begin{defn}
\label{gametestingdefinition} : \textbf{Game tester.} Given a game $G$, a set of policies $\Pi$  to interact with $G$,  a set of states $S$ of $G$, and  a set of bugs  $B$  on $G$, a game tester $T$ is defined as a function $T_G: \Pi \rightarrow S \times B$. 
\end{defn}

A sequence of actions is a test case for a game. Since G is often a stochastic function, a test case may lead to multiple distinct states. A game tester aims to implement different strategies in order to explore the different states of the game to find bugs. In this paper, the game tester play the role of oracle to verify the presence or absence of a bug on an output state. Therefore, it implements different strategies in order to explore the different states of the game to find bugs. So, a test case generated by a game tester is a series of valid actions that can reach a state in which a bug might hide. 

\begin{defn}
\label{Testadequacydefinition} : \textbf{Test adequacy criteria.} 
We consider the state coverage and line coverage as  criteria to discover whether the
existing test cases have a good bug-revealing ability. The state coverage measures the number of visited state of the player during the play, and the code coverage measures the number of lines of code  related to the function of the game that have been covered during the play.
\end{defn}
Considering a $5 \times 5$ Block Maze game where bugs are injected and triggered when the player reached a location on the maze: 
\begin{itemize}
    \item A player has 4 possible actions (LEFT, RIGHT, UP, DOWN), a state is defined as the vector $(P,B)$ where $P$ is the player position at each step of the play and $B$ the position of a bug (the position that triggers a bug on the maze). 
    \item Initially the state of the Block Maze is $((0,0),(1,4))$.
    \item  A test case that leads to a bug can be \\
    \{RIGHT $\rightarrow$ RIGHT $\rightarrow$ RIGHT $\rightarrow$ RIGHT $\rightarrow$ DOWN\}, 
    \\ corresponding to the following states of the game  
    \\ $\{ ((0,0),\textbf{(1,4)}) \rightarrow ((0,1),\textbf{(1,4)}) \rightarrow ((0,2),\textbf{(1,4)}) \rightarrow \\
     ((0,3),\textbf{(1,4)}) \rightarrow ((0,4),\textbf{(1,4)}) \rightarrow (\textbf{(1,4),(1,4)})\}$.
\end{itemize}

As studied by Zheng et al. \cite{zheng2019wuji}, in this work, we consider the testing of large combat games with one agent.
\subsection{Test Case Prioritization} \label{testcaseprioritization}
Test Case Prioritization is the process of prioritizing test cases in a test suite. It allows to execute highly significant test cases first according to some measures, in order to detect faults as early as possible. In this paper, similar to \cite{bagherzadeh2021reinforcement}, we study test case prioritization in the context of Continuous Integration (CI). 

\begin{defn}
: \textbf{CI Cycles.} A CI cycle is composed of  a logical value and a set of test cases. The logical value indicates whether or not the cycle has failed. Failed cycles due to a test case failure are considered in this work, and we select a test case with at least one failed cycle.  
\end{defn}
\begin{defn}\label{Test case feature definition}
: \textbf{ Test case feature.} Each test case has an execution history and code-based features. The execution history shows a record of executions of test cases over the cycles. The execution history includes the execution verdict of a test case, the execution time, a sequence of verdicts from prior cycles, and the test age capturing the time the test case was introduced for the first time. The execution verdict indicates if the test case has failed or not. The execution time of a test case can be computed by averaging its previous execution times. The code-based features for a test case can indicate the changes that have been made, the impacted files with the number of lines of code which are relevant to predict the execution time 
and can be leveraged to prioritize test cases.
\end{defn}
\begin{defn}
: \textbf{Optimal ranking (Test Case prioritization)}. The test case prioritization process in this work is a ranking function that produces an ordered sequence based on the optimal ranking of test cases. The goal of prioritization is to get as close to this order as possible. The optimal ranking of a set of test cases is an order in which all test cases that fail are executed before test cases that pass. Furthermore, in this optimal ranking, test cases with a smaller time of execution should be executed sooner.
\end{defn}
\begin{defn}
:  \textbf{ DRL as a ranking process.} 
In this paper, we consider a prioritization approach that consists of continuously interacting with the CI environment while improving the ranking strategy. In the CI environment, a DRL agent is used to automatically and continuously learn a ranking strategy as closely as possible to the optimal one. Specifically, the agent is trained on the CI environment by replaying the execution logs of available test cases from previous cycles in order to rank test cases in subsequent cycles. The main idea, similar to other studies \cite{bagherzadeh2021reinforcement}, is to formulate the sequential interactions between CI and test case prioritization algorithm as a DRL problem. This way, state-of-the-art DRL techniques learn a strategy for test case prioritization, as close as possible to the optimal one, if we consider a predetermined optimal ranking as the ground truth. Using a CI environment simulator, the DRL agent is trained on the history of test execution and code-based features from previous cycles to prioritize test cases in next cycles. We can benefit from an adaptive training process with DRL, meaning that the agent receives feedback (i.e., reward) at the end of each cycle (or when the prediction accuracy is below a particular level). To adapt the learned policy, the execution logs of test cases can be replayed several times to ensure an efficient and continuous adaptation to changes in the system and regression test suite.
\end{defn}
Bagherzadeh et al. \cite{bagherzadeh2021reinforcement} also presented a detailed explanation of the terms \textbf{CI Cycles, Test case feature, Test Case prioritization, and Optimal ranking}.

\section{Study design}\label{studydesign}
In this section, we describe the methodology of our study which aims to compare different implemented DRL algorithms from existing frameworks. We also introduce the two problems that we selected for this comparison. 

\subsection{Research questions}
The goal of our work is to evaluate and compare implemented algorithms offered by different DRL frameworks. In order to achieve this goal, we focus on answering the following 
research questions. 
\begin{itemize}
   \item \textbf{RQ1:} How does the choice of DRL framework affect the performance of the software testing tasks?
   \item \textbf{RQ2:} Which combinations of DRL frameworks-algorithms perform better (i.e., get trained accurately and solve the problem effectively)? 
   \item \textbf{RQ3:} How stable are the results obtained from the DRL frameworks, over multiple runs? 
\end{itemize}
\subsection{Problem 1: Game testing using DRL} \label{sec:REINFORCEMENT LEARNING FOR GAME TESTING}
We aim to employ several DRL algorithms from different DRL frameworks in a game testing environment. More specifically, we use DRL to explore more states of a game where bugs might hide. Our work is based on wuji \cite{zheng2019wuji}, an automated game testing framework that combines Evolutionary Multi-Objective Optimization (EMOO) and DRL to detect bugs on a game. wuji randomly initializes a population of policies (represented by DNNs), adopts EMOO to diversify the state's exploration, then uses DRL to improve the capability of the agent to accomplish its mission. To train wuji on multiple DRL frameworks, we turn off EMOO 
and only consider the DRL part of wuji. In this way, we can focus on the effect of different DRL algorithms on detecting bugs.

\subsubsection{Creation of the DRL environment}\label{sec:createRLenvGame}

A game environment can be mapped into a DRL process by defining the state or observation, reward, action, end of an episode and the information related to the bug.
 
\textbf{Observation space:} As mentioned in Definition \ref{gamestatedefinition}, an observation is a set of features describing the state of the game. In our case, the observation of the agent is its position inside the maze.
 
\textbf{Action space:} The action space describes the available moves that can be made in the game. We consider a game with 4 discrete actions to choose: north, south, east, west.
 
\textbf{Reward function:} The reward function is the feedback from the environment regarding the agent's actions. It is designed so that the agent can accomplish its mission. The agent gets negatively rewarded when it reaches a non-valid position in the game or any other position that is not the goal position of the game, in all other cases it receives a positive reward.

The game testing task is representative of an SE testing task as its representation is similar to the baseline study work by Zheng et al. \cite{zheng2019wuji}  of detecting bugs on a Block Maze game.  In the game testing problem, the observation of the agent captures the state of the game where a bug might hide. The observation space has the size of a matrix $20 \times 20$ similar to the baseline study  by Zheng et al. \cite{zheng2019wuji}  and it is straightforward to look for bugs in a matrix. The action space describes the moves (north, south, east, west)  available to the agent to explore the game and find bugs. Finally, the reward function awards the agent based on its actions so that it can accomplish the game. The usage of a matrix as observation space has also been used in the literature. To promote the progress of DRL research, OpenAI integrates a collection of DRL tasks into the gym platform \cite{1606.01540}. Among these tasks, Atari environments have the observation space of a matrix. Our representation can  easily be extended to other games such as 3D games by extending the number of actions available to the agent or by  adding channels to the matrix, forming a 3D image. Further, Tufano et al. \cite{tufano2022using} study how to leverage DRL algorithms to detect performance bugs. Specifically, the authors injected artificially performance bugs on two 2D games, Cartpole \cite{Cartpole}, MsPacman \cite{MsPacman}, and investigated whether or not the DRL agents are able to detect the bugs. Similarly to our study, the moves available to the agents are left, right, up, and down for the MsPacman game. The observation space of MsPacman game has the size of a $84 \times 84$ matrix, which is also similar to our study. Bergdahl et al. \cite{bergdahl2020augmenting} employed DRL to increase test coverage, find game exploits, and discover bugs in a game. The authors studied sand-box environments where DRL agents receive positive reward for moving towards a goal and negative reward as a penalty for moving away from it, which is similar to our study.

\subsubsection{Experimental Setup } \label{Experimental Setup}
The Block Maze game has a discrete action space that limits the DRL configurations to which it can be applied. Therefore, we consider the following algorithms: DQN-SB, PPO-SB, A2C-SB, DQN-KR, DQN-TF, PPO-TF, and A2C-TF during our experiments. 

DRL algorithms from the studied DRL frameworks have their own hyperparameters settings. We employ the same values of the optimizer (the Adam optimizer \cite{kingma2014adam}), the DNN model (three fully-connected linear layers with 256, 128, 128 units as the hidden layers, connected to the output layer), the discount factor ($0.99$) and the learning rate ($0.25 \times 10^3$)  as the baseline work \cite{zheng2019wuji}, they are the ones we could exactly match with the different studied DRL algorithms. DQN-SB, DQN-TF, DQN-KR, PPO-TF, PPO-SB, A2C-SB, A2C-TF have respectively 19, 21, 7, 25, 19, 18, and 23 more hyperparameters whose values are provided in the replication package \cite{replication-package}.

We collected the results of each DRL algorithm for a total of 4 million steps and 10,000 steps. To counter the randomness effect during testing, we repeat each run 10 times to average the results. The experiments are run for approximately 30 days on the Niagara cluster servers provided by Digital Research Alliance of Canada (the Alliance)\footnote{\url{https://docs.scinet.utoronto.ca/index.php/Niagara_Quickstart}}. Each server has 40 cores at 2.4GHz with 202GB of main memory. Moreover, the testing experiments for 4 million steps are run on an ASUS desktop machine running Windows 10 on a 3.6 GHz Intel core i7 CPU with 16GB main memory. 
After each episode, the agent is reset before the next one. Zheng et al. \cite{zheng2019wuji} studied the detection of bugs by implementing a DRL approach. The game is tested by considering the winning score. We consider their work as a baseline and compare other DRL approaches with their results.

\subsubsection{Training of a DRL agent}
 
 Wuji randomly initializes DNN policies, then uses the A2C algorithm and an evolutionary multi-objective optimization algorithm to evolve the policies so that the agent can explore more states of the Block Maze game and accomplish its mission. In this paper, we are going to apply DQN, A2C, and PPO algorithms from Stable-baselines3 (SB), Keras-rl (KR) and Tensorforce (TF), respectively, to detect bugs in the Block Maze game. Stable-baselines3 is used here as opposed to Stable-baselines2 because the latter is in maintenance mode by its developers. Like Zheng et al. \cite{zheng2019wuji}, to train the agent, it interacts with the game (the environment). 
 The DRL agents use the gym interface during training to compute the best policy to play the game. During testing, the same OpenAI gym interface as the game environment is used.

 Regarding the reward distribution of the DRL agent, if it reaches the goal it receives 10 as reward. If its position is not valid, not within the environment space it receives $-1$ as reward otherwise it receives $-0.01$.

\subsubsection{Datasets} \label{sec:Datasets 3.2.4} 
\begin{figure}
    \centering
    \includegraphics[width=0.4\textwidth]{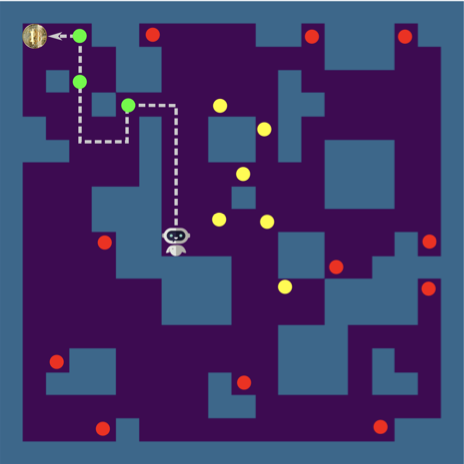}
    \caption{Block Maze with bugs (red, green and yellow dots).}
    \label{fig:blockmazewithbugs}
\end{figure}

A Block Maze game from Zheng et al. \cite{zheng2019wuji}, is selected in the evaluation Figure \ref{fig:blockmazewithbugs}. 
In the Block Maze game, the player's objective is to reach the goal coin. It has 4 possible actions to choose: {north, south, east, west}. Every action leads to moving into a neighbor cell in the grid in the corresponding direction, except that a collision on the block (dark green) results in no movement. 
To evaluate the effectiveness of our DRL approaches, 25 bugs are artificially injected to the Block Maze, and randomly distributed within the environment. A bug is a position in the Block Maze that is triggered if the robot (agent) reaches its location as in the map, as shown in Figure \ref{fig:blockmazewithbugs}. A bug has no direct impact on the game but can be located in invalid locations of the game environment such as the Block Maze obstacles or outside of the Block Maze observation space. Invalid locations on the other hand cause the end of the game. Therefore, in this study we consider 2 types of bugs. Type 1 refers to exploratory bugs that measure the exploration capabilities of the agent, and Type 2 refers to bugs at invalid locations of the Block Maze.

\subsubsection{Evaluation metrics} 
On top of metrics considered by Zheng et al. \cite{zheng2019wuji} including the number of bugs detected, the state and line coverage performed by the DRL configurations, we also measured average cumulative reward, training time, and testing time to assess the accuracy and effectiveness of the game testing process across different DRL approaches.

\begin{itemize}
    \item \textbf{Number of bugs detected}: the average number of bugs detected by our DRL agents after being trained.
    \item \textbf{The average cumulative reward} obtained by the DRL agents after being trained.
    \item \textbf{The line coverage}: the lines covered by each DRL approach during testing. 
    We use the Python library of coverage\footnote{\url{https://coverage.readthedocs.io}} to collect line coverage. This library gives you the result per Python file. As in our replication package \cite{replication-package}, both the gym environment and the actual game implementation are on the same file. Thus, the line coverage includes both the lines of code of the gym environment and the game implementation.
    \item \textbf{The state coverage}: the number of visited state during testing.
    \item \textbf{Training time:}  We collect the time consumed by the DRL agents to train their policy, which lasts for 10,000 steps.

\item \textbf{Prediction time:}  We collect the time consumed by the trained DRL agents to detect bugs for 10,000 steps, 4 million steps, or until reaching the goal coin of the game environment.

\end{itemize}

\subsubsection{Analysis Method}

We proceeded as follows to answer our research questions. In \textbf{RQ1}, we collected the number of bugs detected,  the average cumulative reward, the state and line coverage obtained by the player in the Block Maze game by using DRL algorithms from state-of-the-art frameworks (see Subsection \ref{Experimental Setup}). We also collected the training and testing times performed by these DRL configurations.  We relied on the implementation of algorithms provided by Stable-baselines3 \cite{Stable-Baselines3}, Keras-rl \cite{framework:plappert2016kerasrl} and Tensorforce \cite{schaarschmidt2018lift} frameworks. Moreover, we collected the training and  testing times of the DRL strategies as well as computed the state coverage and line coverage as adequacy criteria to assess their performance. To determine the best DRL strategy in \textbf{RQ2} we use the Welch’s ANOVA and Games-Howell post-hoc test \cite{welch1947generalization}, \cite{games1976pairwise}.We compare all DRL strategies across all runs in terms of bug's detected and average cumulative reward earned. Same as the study of bagherzadeh et al. \cite{bagherzadeh2021reinforcement}, the significance level is set to 0.05, difference with p-value $<= 0.05$ is considered significant. In \textbf{RQ3,} we investigate how the same algorithm performs, on average, across different DRL frameworks with multiple runs of testing. Specifically, the performance of trained agents with the same algorithm across different DRL frameworks resulting from multiple runs are evaluated based on metrics such as the number of bugs detected, the average cumulative reward, testing and prediction times collected in RQ1.

 Welch's ANOVA is a statistical test used to compare differences between groups by analyzing their means and their variances. Games-Howell post-hoc test completes Welch's ANOVA process by identifying groups that significantly differ from the others in respect to the mean. Games-Howell post-hoc test is used with Welch's ANOVA as the latter does not assume equal variances between groups \cite{games1976pairwise}. 

\subsection{Problem 2: Test case prioritization using DRL} \label{sec:Reinforcement Learning for Test case prioritization}
We aim to apply several DRL algorithms from different frameworks for test case prioritization in the context of CI. To do so, we follow a recent work on using DRL for test case prioritization by Bagherzadeh et al. \cite{bagherzadeh2021reinforcement}. The authors studied different approaches to prioritization techniques that can adapt and continuously improve while interacting with the CI environment. The interaction between the CI environment and test case prioritization is modeled as a DRL problem. They use state-of-the-art DRL techniques to learn prioritization strategies that are close to optimal ranking. The DRL agent is first trained offline on the test case execution history and code-based features of past cycles to prioritize the next cycles. At the end of each cycle, if the agent's accuracy in predicting the next cycles is less than a specified threshold, the test case execution history is replayed to improve the agent's policy. After offline training, the trained agent can be applied to rank the available test cases. Similarly, our approach for applying DRL techniques in the context of the CI environment is to train a DRL agent, based on the algorithms designed by Bagherzadeh et al.\cite{bagherzadeh2021reinforcement} that describe the ranking model in the context of CI and test case prioritization. We train a DRL agent using various DRL algorithms from popular frameworks, as described in subsection \ref{State-of-the-art RL Algorithms and Frameworks}.

\subsubsection{Creation of the DRL environment} \label{Creation of the RL environment test prio}
Test case prioritization can be mapped into a DRL problem by defining the details of the interaction of the agent with the environment, meaning observation, action, reward, and end condition of an episode. We map test case prioritization as a DRL problem by considering two ranking models: pointwise and pairwise, that have been employed by Bagherzadeh et al.\cite{bagherzadeh2021reinforcement}. 

 
 
\paragraph{Pointwise ranking function:}
Bagherzadeh et al.\cite{bagherzadeh2021reinforcement}  designed the pointwise ranking model as a class on which the observation space, action space, and reward function are defined.
This class consists of determining scores for each test case and then storing them in a temporary vector. At the end of the learning process, the test cases are sorted according to their scores stored in the temporary vector.

\textbf{Observation space:} The agent's observation is a record of the characteristics of a single test case with 4 numerical values. 
 
\textbf{Action space:} The action describes a score associated with each test case. The agent uses this score to order the test cases. Each action is a real number between 0 and 1.
 
\textbf{Reward function:} The reward function is computed here based on the normalized distance between the assigned ranking and the optimal ranking. The values range between 0 and 1.

\paragraph{Pairwise ranking function:}
Bagherzadeh et al.\cite{bagherzadeh2021reinforcement} designed the pairwise ranking model as a class on which the observation space, action space, and reward function are defined. 
This class uses the selection sorting algorithm \cite{knuth1997art} to rank the test cases. All test cases are divided into a pair: the sorted part on the left and the unsorted part on the right. At each time step, if a test case with high priority is found, its position is changed in the sorted part. The process continues until all test cases are sorted.
 
\textbf{Observation space:} An agent observation is a pair of test case records. 
 
\textbf{Action space:} The action space values are 0 or 1. The first value indicates that the first test case in the observation has a higher priority. 
 
\textbf{Reward function:} The reward function takes into account whether or not the test case with a higher priority fails. If it is the case, the agent receives the maximum reward of $1$ otherwise it receives $0$. In case the test cases in the pair have the same verdicts, the agent receives $0.5$ as a reward when the higher priority is given to the test case with less execution time otherwise it receives $0$.

The test case prioritization task is representative of  an SE testing task as its representation is similar to the baseline study work by Bagherzadeh et al. \cite{bagherzadeh2021reinforcement} of ranking test cases. The observation spaces of the ranking strategies capture the characteristics of the test cases which are used to rank them. Based on the score or priority of a test case a subsequent test case is selected. The reward function evaluates the capacity of the agent to rank test cases w.r.t the optimal ranking. Spieker et al. \cite{spieker2017reinforcement} applied DRL for the prioritization of test cases for various configurations. Similarly to our study, the observation of the environment captures the characteristics of a test case. The action space represents the priority of a test case for the current CI cycle which is also similar to this study. 
\subsubsection{Experimental Setup }
We implemented our ranking models using the DRL algorithms of the selected frameworks. We used the OpenAI gym library \cite{1606.01540:1606.01540} to mimic the CI environment using logs execution and relied on the implementation of the DRL algorithms provided by the Stable-baselines2 \cite{hill2019stable}, Keras-rl \cite{framework:plappert2016kerasrl} and Tensorforce \cite{schaarschmidt2018lift} frameworks. 
Stable-baselines2 is used here as it was originally used by Bagherzadeh et al. \cite{bagherzadeh2021reinforcement}. In any case, Stable-baselines2 and Stable-baselines3 provide for their implemented DRL algorithms the same hyperparameters. Moreover, to make sure Stable-baselines3 meets the performance of Stable-baselines2, its developers conducted experiments\footnote{\url{https://stable-baselines3.readthedocs.io/en/master/guide/migration.html}} to assess the performance of its implemented DRL algorithms and found them equivalent. So, a performance drop should not be expected by using either one of them. When applicable, we employ the default hyperparameters values of Stable-baselines2 for the experiments similarly to the original work \cite{bagherzadeh2021reinforcement}. Specifically, the architecture of the DNN model, the learning rate and the discount factor have the same values among all experiments. The details of all hyperparameters settings are documented in the replication package \cite{replication-package}. Regarding the APFD and NRPA metrics, for each dataset, we performed several experiments that correspond to the two pairwise and pointwise ranking models. It should be noted that the applicability of the DRL algorithms is restricted by the type of their action space. The pairwise  ranking model involves seven experiments for each data set, one for each DRL framework with DRL algorithms that can support a discrete action space (i.e., DQN-SB, DQN-KR, DQN-TF, A2C-SB, A2C-TF, PPO-TF, PPO-SB). Similarly, the pointwise-ranking model involves seven experiments for each dataset, one for each DRL framework with DRL algorithms that can support a continuous action space (i.e., DDPG-SB, DDPG-KR, DDPG-TF, A2C-SB, A2C-TF,PPO-TF, PPO-SB).  
The training process begins with training an agent by replaying the execution logs from the first cycle, followed by evaluating the trained agent on the second cycle. Then the logs from the second run are replayed to improve the agent, and so on.

Bagherzadeh et al.\cite{bagherzadeh2021reinforcement},  trained the agent for a minimum of $200 \times n \times \log_2 n$ episodes and one million steps for training each cycle, where $n$ refers to the number of test cases in the cycle. Training stops when the budget of steps per training instance is exhausted or when the sum of rewards in an episode cannot be improved for more than 100 consecutive episodes. 
After each episode, the agent is reset before the next one. To answer our questions, we recorded the rank of each test. Experiments are run 5 times for approximately 30 days allowing us to account for randomness,  on the Niagara cluster servers provided by Digital Research Alliance of Canada (the Alliance). 
Each server has 40 cores at 2.4GHz with 202GB of main memory. The total number of experiments is $320$.

\subsubsection{Comparison Baselines} \label{Comparison Baselines}
Bagherzadeh et al.\cite{bagherzadeh2021reinforcement} applied DRL using state-of-the-art DRL algorithms from the Stable-baselines framework to solve the test case prioritization problem. We use this work as a baseline and compare our suggested DRL strategies with their configurations. Bagherzadeh et al.\cite{bagherzadeh2021reinforcement} also presented the results of three benchmark works RL-BS1 \cite{spieker2017reinforcement}, RL-BS2 \cite{bertolino2020learning}, MART \cite{bertolino2020learning}. RL-BS1 applies DRL on simple history data sets. RL-BS2 applies DRL Shallow Network, Deep Neural Network, and Random Forest implementations on enriched datasets. MART is a supervised learning technique for ranking test cases. RL-BS1 and RL-BS2 show results with runs containing fewer than five test cases, which can inflate APFD and NRPA values when prioritization is not required. MART, as a deep learning technique, has no support for incremental learning \cite{zhang2019incremental} which is important for dealing with frequently changing CI environments. We will also compare our results with the mentioned baselines, i.e., RL-BS1, RL-BS2 and MART.

\subsubsection{Training of a DRL agent.}
The applicability of DRL algorithms depends on the action space of the ranking models. The pairwise  ranking model has discrete action space while the Pointwise ranking model has a continuous action space. For the sake of comparison of our selected DRL frameworks (Table \ref{tab:Comparison between RL frameworks}), DQN and A2C will be applied to the Pairwise  ranking model while DDPG will be applied to the Pointwise ranking model.
 
Regarding the test case prioritization problem, the agent is trained in a software-production environment, which is the case for many systems, especially safety-critical systems. During testing, the same OpenAI gym interface as the game environment is used.Nevertheless, after the training, the agent can be deployed into a real environment \cite{dulac2019challenges}. We follow the same procedure as Bagherzadeh et al.\cite{bagherzadeh2021reinforcement}. The agent is trained first on the available execution history. Then at the end of the cycle, the test cases are ranked and new execution logs are captured. The new logs are used to train the agent at the beginning of the next cycle.
 
\subsubsection{Integration of a DRL agent into CI Environments.}
To integrate the DRL agent into CI environments, the agent must first be trained on the execution history of available test cases and the history of test case-related code features \cite{bagherzadeh2021reinforcement}. Then, the trained agent is deployed to the production setting where the test case features can be used in each CI cycle to rank the test cases. During the testing process, if accuracy decreases, execution logs are captured and passed to the agent so that it can adapt to the changes.

\subsubsection{Datasets} \label{sec: subsection 336} \label{datasetTestCasePrio}
We ran our experiments on datasets used  by Bagherzadeh et al.\cite{bagherzadeh2021reinforcement}: Simple and enriched historical data sets. Simple historical datasets represent test situations where source code is not available and contain the age, average execution time, and verdicts of test cases. Enriched historical datasets represent test situations where source code is available but due to time constraints imposed by the CI, complete coverage analysis is not possible.  They are enriched with history data, execution history, and code characteristics from Apache Commons projects \cite{bertolino2020learning}. Table \ref{tab:Data sets} shows the list of datasets that we employ in this study and their characteristics.

\begin{table*}[t]
\caption{Datasets \cite{bagherzadeh2021reinforcement}}
\centering
\resizebox{\textwidth}{!}{
\begin{tabular}{ccccccc}
\multirow{2}{*}{Dataset}      & \multirow{2}{*}{Type}             & \multirow{2}{*}{Cycles} & \multirow{2}{*}{Logs}   & \multirow{2}{*}{Fail Rate(\%)} & \multirow{2}{*}{Failed Cycles} & Avg. Calc. Time (Avg)\\
&&&&&&Enriched Features \\ \hline
Paint-Control & Simple  & 332 & 25,568 & 19.36 & 252 & NA \\
IOFROL        & Simple   & 209    & 32,118 & 28.66          & 203           & NA                                      \\
Codec         & Enriched                     & 178    & 2,207  & 0              & 0             & 1.78                                    \\
Compress      & Enriched                     & 438    & 10,335 & 0.06           & 7             & 3.64                                    \\
Imaging       & Enriched                     & 147    & 4,482  & 0.04           & 2             & 5.60                                    \\
IO            & Enriched                     & 176    & 4,985  & 0.06           & 3             & 2.88                                    \\
Lang          & Enriched                     & 301    & 10,884 & 0.01           & 2             & 5.58                                    \\
Math          & \multicolumn{1}{l}{Enriched} & 57     & 3,822  & 0.01           & 7             & 9.46                                   
\end{tabular}
}
\label{tab:Data sets}
\end{table*}
The execution logs contain up to 438 CI cycles, and each CI cycle includes at least 6 test cases. Less than 6 test cases will not be relevant and can inflate the accuracy of the results \cite{bagherzadeh2021reinforcement}. The logs column indicates the number of test case execution logs which ranges from $2,207$ to $32,118$. Enriched datasets show a low rate of failed cycles and failure rate while the failure rates and number of failed cycles in simple datasets are high. The last column shows the average computation time of enriched features per cycle.

\subsubsection{Evaluation metrics}
We use two evaluation metrics to assess the accuracy of prioritization techniques across our DRL configurations. Bagherzadeh et al.\cite{bagherzadeh2021reinforcement} used both metrics. We present a description of them in the rest of this section.

\textbf{Normalized Rank Percentile Average (NRPA):} NRPA measures how close a predicted ranking of items is to the optimal ranking independently of the context of the problem or ranking criteria. The value can range from $0$ to $1$. The NRPA is defined as follows: $NRPA=\frac {RPA(s_e)}{RPA(s_o)}$. In this equation $s_e$ is the ordered sequence generated by a ranking algorithm $R$ that takes a set of $k$ items, and $s_o$ is the optimal ranking of the items. $RPA$ is defined as:

\begin{equation}
    RPA= \frac{\sum_{m \in s} \sum_{i=idx(s,m)}^{k} \lvert s \rvert - idx(s_o,m) + 1}{k^{2}(k+1)/2}
\end{equation}
where $idx(s,m)$ returns the position of $m$ in sequence $s$.

\textbf{Average Percentage of Faults Detected (APFD):} APFD measures the weighted average of the percentage of the fault detected by the execution of test cases in a certain order. It ranges from $0$ to $1$. Values close to $1$ imply fast fault detection. It is defined as follows:
\begin{equation}
    APFD(s_e)=1- \frac{\sum_{t \in s_e} idx(s_e,t)*t.v}{\lvert s_e \rvert *m} + \frac{1}{2*\lvert s_e \rvert}
\end{equation}
where $m$ is the total number of faults, $t$ is a test case among $s_e$ and $v$ its execution verdict, either 0 or 1.

However, NRPA can be misleading in the presence of failures as it treats all test cases the same regardless of their execution verdict. Bagherzadeh et al.\cite{bagherzadeh2021reinforcement} show that NRPA values contradict APFD values for some datasets, therefore recommending the use of APFD metric to measure how well a certain ranking can reveal faults early. Both APFD and NRPA metrics are suitable to measure the accuracy of the DRL ranking strategy, and are calculated during testing after the agent is trained. 

\textbf{Training time:} We collect the time consumed by the DRL agents to train their policy, which lasts for 200 episodes, for the pairwise and pointwise strategies.

\textbf{Prediction time:} For both pointwise and pairwise ranking models, we measured the time consumed by the DRL agents to rank a set of test cases. 
\subsubsection{Analysis Method}
To answer \textbf{RQ1}, we conducted experiments and collected the averages and standard deviations of APFD and NRPA for eight datasets (see Subsection \ref{datasetTestCasePrio}), as well as their training and prediction times using DRL algorithms from selected frameworks. We relied on the implementation of algorithms provided by Stable-baselines3 \cite{Stable-Baselines3}, Keras-rl \cite{framework:plappert2016kerasrl}, and Tensorforce \cite{schaarschmidt2018lift} frameworks. Furthermore, we collected from the study of Bagherzadeh et al.\cite{bagherzadeh2021reinforcement}, the averages and standard deviations of baseline configurations in terms of NRPA and APFD values. For each framework, we compare its best configuration with the baselines in terms of NRPA or APFD. 
We calculate Common Language Effect Size (CLES) \cite{mcgraw1992common}, \cite{arcuri2014hitchhiker}, between the best configuration of each framework and baselines to assess the effect size of differences. CLES estimates the probability that a randomly sampled value from one population is greater than a randomly sampled value from another population.
 In \textbf{RQ2}, we use the Welch’s ANOVA and Games-Howell post-hoc test \cite{welch1947generalization}, \cite{games1976pairwise} to indicate the best DRL algorithm.  All configurations across all cycles are compared using one NRPA or APFD value per cycle. Similar to the game testing problem, a difference with p-value $<= 0.05$ is considered significant in our assessments. In \textbf{RQ3,} we investigate how the same algorithm performs, on average, across different DRL frameworks with multiple runs of testing. Specifically, the performance of trained agents with the same algorithm across different DRL frameworks resulting from multiple runs are evaluated based on metrics such as NRPA, APFD, testing and prediction times collected in RQ1.

\subsection{Data Availability}
The source code of our implementation and the results of experiments are publicly available \cite{replication-package}.

\section{ Experimental results}\label{Validation}
We now report the results of our experiments. 

\subsection{Game Testing}

\begin{figure*}[t]
  \begin{minipage}{0.45\textwidth}
        \centering
        \includegraphics[width=\linewidth]{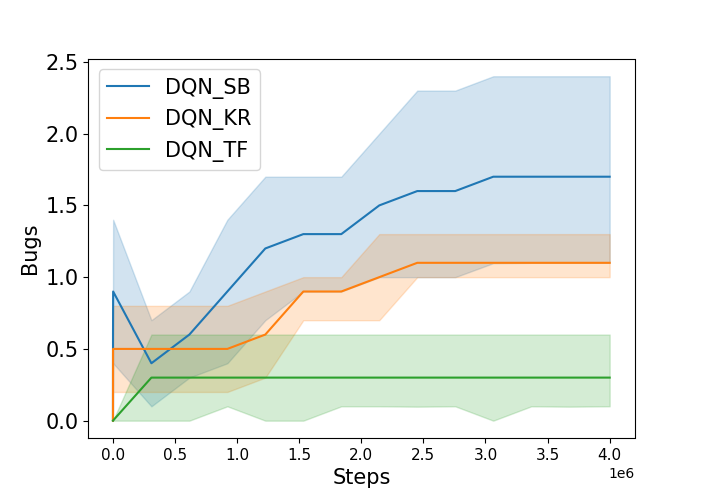}
        \caption{Number of bugs detected by \\ DQN agents from different frameworks.}
        \label{fig:Figure_DQNS-BUGS}
    \end{minipage}%
   \hfill
        \begin{minipage}{0.53\textwidth}
        \centering
        \includegraphics[width=\linewidth]{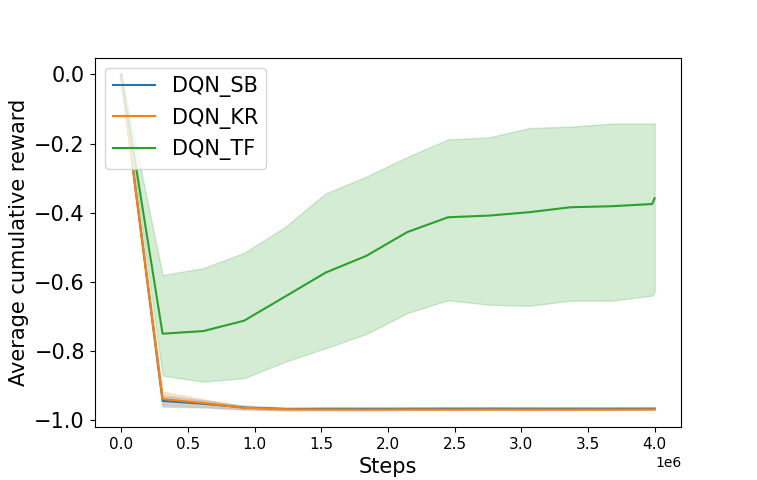}
        \caption{Average cumulative reward earned by \\ DQN agents from different frameworks.}
        \label{fig:Figure_DQNS-AVERAGE-REWARD}
            \end{minipage}%
    \end{figure*}

\hfill \break 
\textbf{RQ1:} Figures \ref{fig:Figure_DQNS-BUGS} and \ref{fig:Figure_DQNS-AVERAGE-REWARD}  show respectively the average number of detected bugs and average cumulative reward obtained by DQN algorithms from Stable-baselines3 (DQN\_SB), Keras-rl (DQN\_KR), and Tensorforce (DQN\_TF) frameworks. 
In Figures \ref{fig:Figure_DQNS-BUGS} and \ref{fig:Figure_DQNS-AVERAGE-REWARD} the x-axis represents the 4 million steps budget of training. In Figure \ref{fig:Figure_DQNS-BUGS} , the y-axis is the average number of bugs detected over 10 runs of the algorithm. In Figure \ref{fig:Figure_DQNS-AVERAGE-REWARD}, the y-axis is the average cumulative reward obtained by the DRL strategy over 10 runs of the algorithm. Among the DQN algorithms, the Stable-baselines performs better in terms of detecting bugs and Tensorforce performs better in terms of cumulative rewards. To explain these results, our intuition lies in the diversity of the hyperparameters provided by each DRL framework which affect the performance, as well as the difference between Tensorflow and Pytorch as the backend of the frameworks. 

\begin{figure*}[t]
  \begin{minipage}{0.45\textwidth}
        \centering
        \includegraphics[width=\linewidth]{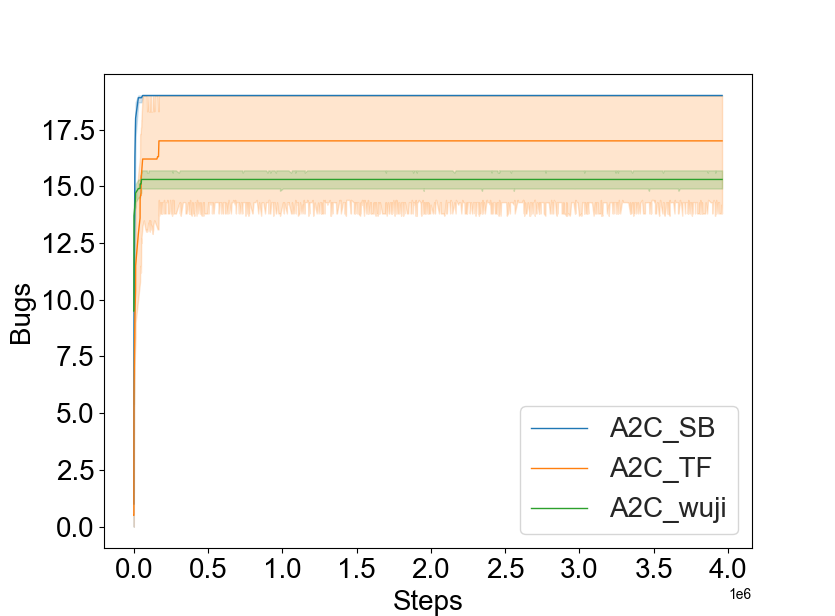}
        \caption{Number of bugs detected by \\ A2C agents from different frameworks.}
        \label{fig:Figure_A2CS-BUGS}
    \end{minipage}%
    \hfill
        \begin{minipage}{0.5\textwidth}
        \centering
        \includegraphics[width=\linewidth]{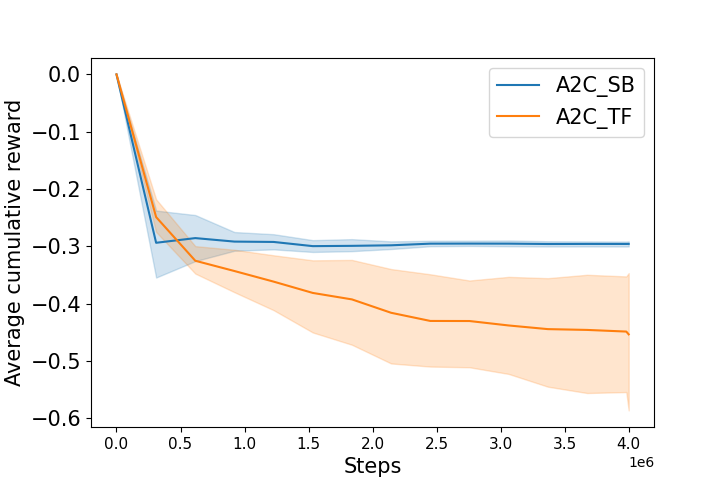}
        \caption{Average cumulative reward earned by \\ A2C agents from different frameworks.}
        \label{fig:Figure_A2CS-AVERAGE-REWARD}
            \end{minipage}%
    \end{figure*}

Figures \ref{fig:Figure_A2CS-BUGS} and \ref{fig:Figure_A2CS-AVERAGE-REWARD}  show respectively the average number of bugs and average cumulative reward obtained by the A2C algorithm from Stable-baselines3 (A2C\_SB), Tensorforce (A2C\_TF), and wuji \cite{zheng2019wuji} (A2C\_wuji) for a testing time of 4 million steps. 
  Recall that in this study, given that we compare DRL algorithms, we compare our results with the number of bugs detected by only the DRL part of wuji. Since the authors of wuji did not consider the average cumulative reward as a metric in the original work, we did not report it here. The reason is that we would not have any baselines to compare the results. A2C\_SB performs better than A2C\_wuji and A2C\_TF in terms of detecting bugs. In terms of rewards earned, the A2C algorithm from Stable-baselines3 performs better on average as it detects more bugs of Type 1 (see Table \ref{tab:Detection capabilities of the A2C and PPO}).   

\begin{figure*}
  \centering
  \begin{minipage}{0.45\textwidth}
        \centering
        \includegraphics[width=\linewidth]{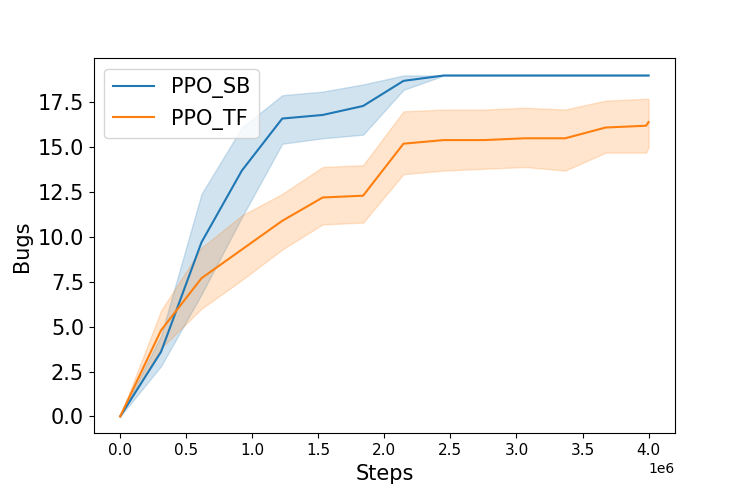}
        \caption{Number of bugs detected by \\ PPO agents from different frameworks.}
        \label{fig:Figure_PPOS-BUGS}
    \end{minipage}%
    \hfill
        \begin{minipage}{0.5\textwidth}
        \centering
        \includegraphics[width=\linewidth]{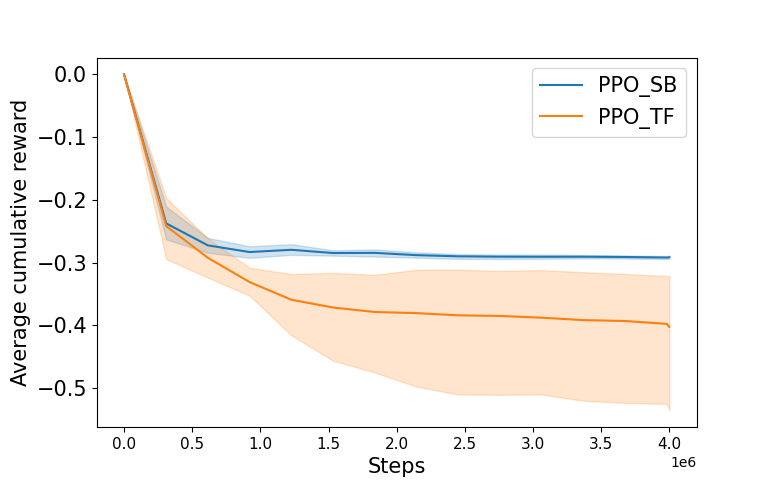}
        \caption{Average cumulative reward earned by \\ PPO agents from different frameworks.}
        \label{fig:Figure_PP0S-AVERAGE-REWARD}
            \end{minipage}%
    \end{figure*}

Figures \ref{fig:Figure_PPOS-BUGS} and \ref{fig:Figure_PP0S-AVERAGE-REWARD} show respectively the average number of detected bugs and average cumulative reward obtained by the PPO algorithms from Stable-baselines3 (PPO\_SB) and Tensorforce (PPO\_TF) frameworks. 
PPO\_SB has slightly (4.69\%) better performance in comparison to PPO\_TF in terms of bugs detected. Similarly, PPO\_SB performs better on average, in terms of rewards earned.

\begin{figure}[t]
        \centering
        \includegraphics[width=0.6\textwidth]{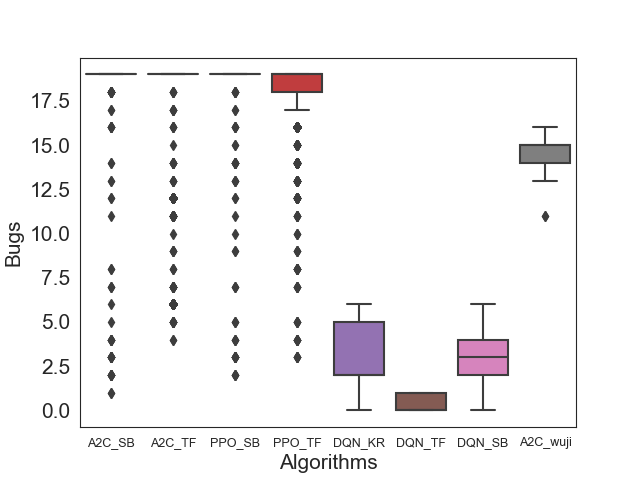}
        \caption{The number of bugs discovered using different strategies after 4 million steps for Block Maze.}
        \label{fig:bugs discovered using different strategies}
    \end{figure}

Figure \ref{fig:bugs discovered using different strategies} shows the statistical results of the number of bugs discovered by all the studied DRL configurations.
The A2C implementation of wuji \cite{zheng2019wuji} detects 19\% fewer bugs than the other A2C\_SB, A2C\_TF, PPO\_TF, PPO\_SB after 4 million steps during testing. Among the studied DQN strategies, DQN\_SB, DQN\_KR and DQN\_TF detect 88\%, 92\%, and 98\% fewer bugs respectively than the A2C implementation of wuji at the same step number. A2C algorithm takes advantage of all the benefits of value-based (like DQN) and policy-based DRL algorithms which explain why it detects more bugs than DQN algorithm.    

To assess the bug detection process, we calculate the rate of bug detection as the number of detected bugs to the number of total bugs. Table \ref{tab:Detection capabilities of DQN} and Table \ref{tab:Detection capabilities of the A2C and PPO} report the detection rate (in percentage) of the DRL strategies per type of bugs as defined in Subsection \ref{sec:Datasets 3.2.4}, for each testing budget (4 million steps and 10,000 steps).
\begin{table*}[t]
\caption{Detection rate (in percentage) of DQNs per type of bugs (values in bold indicate the best rate for each DRL strategy per each testing budget).}
\label{tab:Detection capabilities of DQN}
\centering
\resizebox{\textwidth}{!}{%
\begin{tabular}{|c|cccc|cccc|cccc|}
\hline
DRL strategies          & \multicolumn{4}{c|}{DQN\_SB}                                                                              & \multicolumn{4}{c|}{DQN\_TF}                                                                               & \multicolumn{4}{c|}{DQN\_KR}                                                                               \\ \hline
Testing budget in steps & \multicolumn{2}{c|}{4 million}                                 & \multicolumn{2}{c|}{10,000}              & \multicolumn{2}{c|}{4 million}                                 & \multicolumn{2}{c|}{10,000}               & \multicolumn{2}{c|}{4 million}                                & \multicolumn{2}{c|}{10,000}               \\ \hline
Type of bugs            & \multicolumn{1}{c|}{1}    & \multicolumn{1}{c|}{2}             & \multicolumn{1}{c|}{1}   & 2             & \multicolumn{1}{c|}{1}   & \multicolumn{1}{c|}{2}              & \multicolumn{1}{c|}{1}   & 2              & \multicolumn{1}{c|}{1}   & \multicolumn{1}{c|}{2}             & \multicolumn{1}{c|}{1}   & 2    \\ \hline
Detection percentage    & \multicolumn{1}{c|}{30.5} & \multicolumn{1}{c|}{\textbf{69.4}} & \multicolumn{1}{c|}{4.3} & \textbf{95.6} & \multicolumn{1}{c|}{0.0} & \multicolumn{1}{c|}{\textbf{100.0}} & \multicolumn{1}{c|}{0.0} & \textbf{100.0} & \multicolumn{1}{c|}{2.6} & \multicolumn{1}{c|}{\textbf{94.7}} & \multicolumn{1}{c|}{0.0} & \textbf{100.0} \\ \hline
\end{tabular}
}
\end{table*}
\begin{table*}[t]
\caption{Detection rate (in percentage) of A2C and PPO per type of bugs (values in bold indicate the best rate for each DRL strategy per each testing budget).}
\label{tab:Detection capabilities of the A2C and PPO}
\centering
\resizebox{\textwidth}{!}{%
\begin{tabular}{|c|cccc|cccc|cccc|cccc|}
\hline
DRL strategies & \multicolumn{4}{c|}{A2C\_TF} & \multicolumn{4}{c|}{A2C\_SB} & \multicolumn{4}{c|}{PPO\_TF} & \multicolumn{4}{c|}{PPO\_SB} \\ \hline

Testing budget in steps & \multicolumn{2}{c|}{4 million} & \multicolumn{2}{c|}{10,000} & \multicolumn{2}{c|}{4 million}                                 & \multicolumn{2}{c|}{10,000}      & \multicolumn{2}{c|}{4 million}                                 & \multicolumn{2}{c|}{10,000}               & \multicolumn{2}{c|}{4 million}                                 & 
\multicolumn{2}{c|}{10,000}               \\ \hline

Type of bugs            & \multicolumn{1}{c|}{1}             & \multicolumn{1}{c|}{2}    & \multicolumn{1}{c|}{1}    & 2             & \multicolumn{1}{c|}{1}             & \multicolumn{1}{c|}{2}    & \multicolumn{1}{c|}{1}    & 2    & \multicolumn{1}{c|}{1}             & \multicolumn{1}{c|}{2}    & \multicolumn{1}{c|}{1}             & 2    & \multicolumn{1}{c|}{1}             & \multicolumn{1}{c|}{2}    & \multicolumn{1}{c|}{1}             & 2 \\
\hline
Detection percentage    &\multicolumn{1}{c|}{\textbf{60.5}} & \multicolumn{1}{c|}{39.4} & \multicolumn{1}{c|}{47.6} & \textbf{52.3} & \multicolumn{1}{c|}{\textbf{64.2}} & \multicolumn{1}{c|}{35.7} & \multicolumn{1}{c|}{50.0} & 50.0 & \multicolumn{1}{c|}{\textbf{65.5}} & \multicolumn{1}{c|}{34.4} & \multicolumn{1}{c|}{\textbf{61.5}} & 38.4 & \multicolumn{1}{c|}{\textbf{56.8}} & \multicolumn{1}{c|}{43.1} & \multicolumn{1}{c|}{\textbf{57.5}} & 42.5\\
\hline
\end{tabular}
}
\end{table*}
The results show that the DQNs detect Type 2 more effectively, while the PPO strategies detect Type 1 more effectively.
%

We also analyze the line coverage obtained by each DRL strategy, as well as their state coverage on the Block Maze game. The line coverage is exactly the same for all strategies: 96\%. The other 4\% are mostly related to the code related to the player not reaching the goal of the Block Maze. Specifically, the lines of code on the Block Maze gym environment where we check if the player is at the goal location. In addition, the other 4\% are also related to the lines of codes instructing the termination of the game when a player reaches the goal. Finally, the line of code on the Block Maze gym environment that converts the maze to an RGB image is not reached either, as during testing we do not require it. 
The Block Maze has a total of 400 potential states to be visited by the DRL agent. Table \ref{tab:State coverage of DRL algorithms} shows the results of the state coverage obtained by the DRL algorithms from the frameworks we have evaluated. 
\begin{table}[t]
\caption{State coverage of DRL algorithms on the Block Maze game.}
\label{tab:State coverage of DRL algorithms}
\centering
\resizebox{\textwidth}{!}{
\begin{tabular}{|c|c|c|c|c|c|c|c|}
\hline
               & A2C\_SB & A2C\_TF & PPO\_SB & PPO\_TF & DQN\_SB & DQN\_TF & DQN\_KR \\ \hline
State coverage & \textbf{45.5 \%}     & \textbf{45.5 \%}     & \textbf{45.5 \%}     & 44.4 \%   & 9.3\%    &   8.8\%      & 9.9\%    \\ \hline
\end{tabular}
}
\end{table}
As expected, A2C\_SB, A2C\_TF, PPO\_TF and PPO\_SB have the largest state coverage as they are able to detect more bugs. The state coverage obtained by the DQNs are lower as they lead to fewer bugs detected, Stable-baselines framework still has better performance. Moreover, regarding the bugs that are not detected, the fact is that the DRL configurations are not able to cover all the observation state space. The code is relatively easy to cover, as opposed to the state coverage. Thus, detecting bugs by maximizing the state coverage could lead to better performance for game testing.

\begin{figure}[t]
        \centering
        \includegraphics[width=0.65\textwidth]{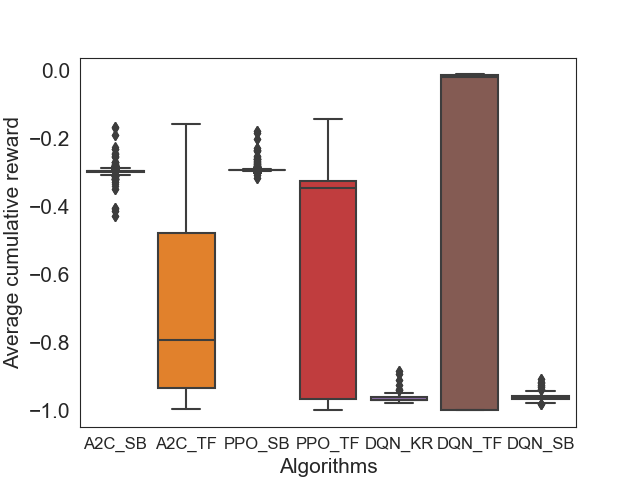}
        \caption{Average cumulative reward obtained by different DRL algorithms after 4 million steps for Block Maze.}
        \label{fig:Average cumulative reward obtained by different strategies after 4M steps for Block Maze}
    \end{figure}

In terms of winning the game (i.e., reaching the gold position of the Block Maze as illustrated in Figure \ref{fig:blockmazewithbugs}) none of our strategies are successful. Our results in Figure \ref{fig:Average cumulative reward obtained by different strategies after 4M steps for Block Maze} show that the DRL agents earn negative rewards for all steps during testing.

\begin{figure*}[t]
  \begin{minipage}{0.45\textwidth}
        \centering
        \includegraphics[width=\linewidth]{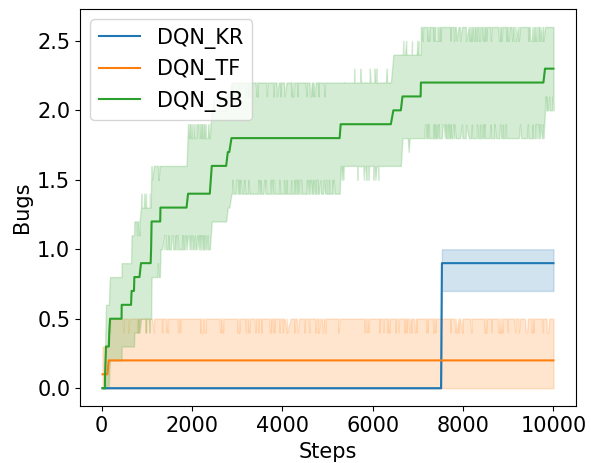}
        \caption{Number of bugs detected by \\ DQN agents from different frameworks \\ on a 10K budget steps.}
        \label{fig:dqns_bug10k}
    \end{minipage}%
   \hfill
        \begin{minipage}{0.48\textwidth}
        \centering
        \includegraphics[width=\linewidth]{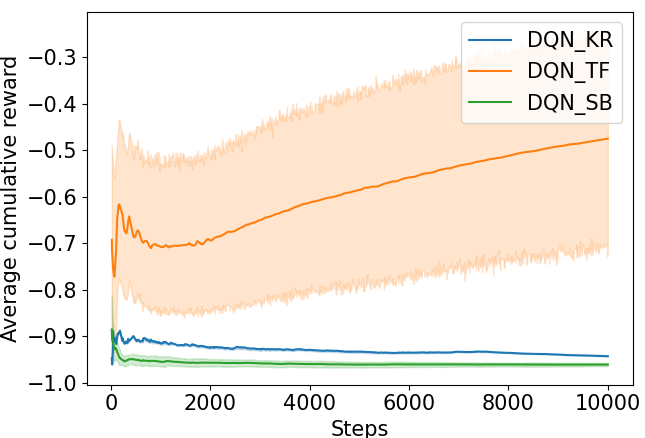}
        \caption{Average cumulative reward earned by DQN agents from different frameworks on a 10K budget steps.}
        \label{fig:dqns_rew10k}
            \end{minipage}%
    \end{figure*}

\begin{figure*}
  \begin{minipage}{0.45\textwidth}
        \centering
        \includegraphics[width=\linewidth]{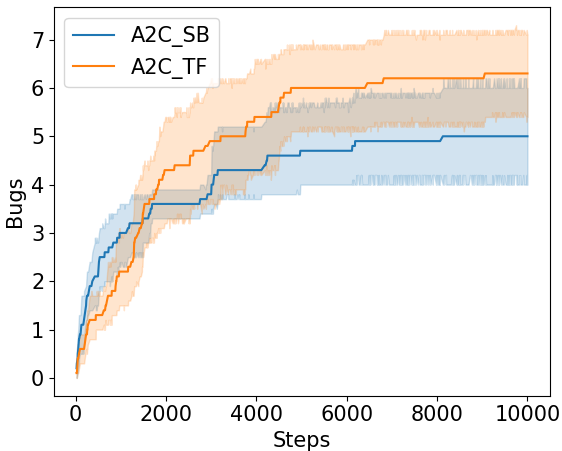}
        \caption{Number of bugs detected  by A2C agents from different frameworks on a 10k budget.}
        \label{fig:a2cs_bug10k}
    \end{minipage}%
    \hfill
        \begin{minipage}{0.5\textwidth}
        \centering
        \includegraphics[width=\linewidth]{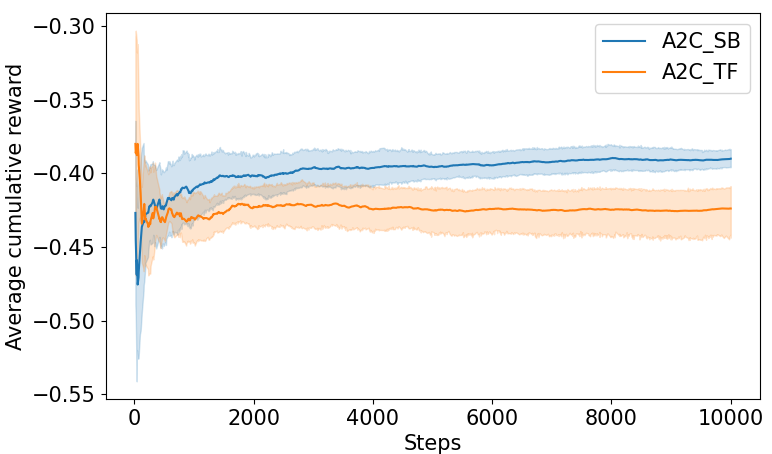}
        \caption{Average cumulative reward earned by A2C agents from different frameworks on a 10k budget.}
        \label{fig:a2cs_rew10k}
            \end{minipage}%
    \end{figure*}

\begin{figure*}
  \begin{minipage}{0.425\textwidth}
        \centering
        \includegraphics[width=\linewidth]{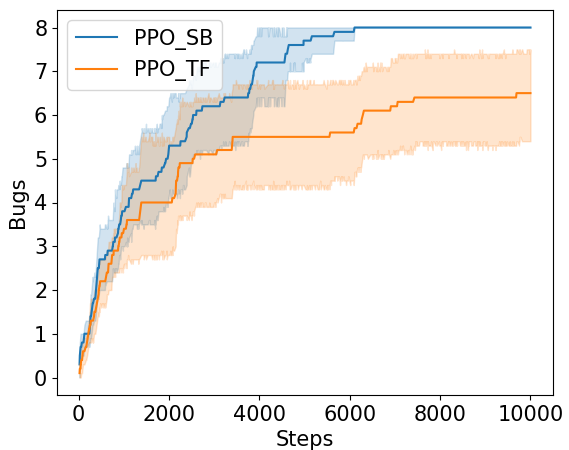}
        \caption{Number of bugs detected by PPO agents from different frameworks on a 10k budget.}
        \label{fig:ppos_bug10k}
    \end{minipage}%
    \hfill
        \begin{minipage}{0.5\textwidth}
        \centering
        \includegraphics[width=\linewidth]{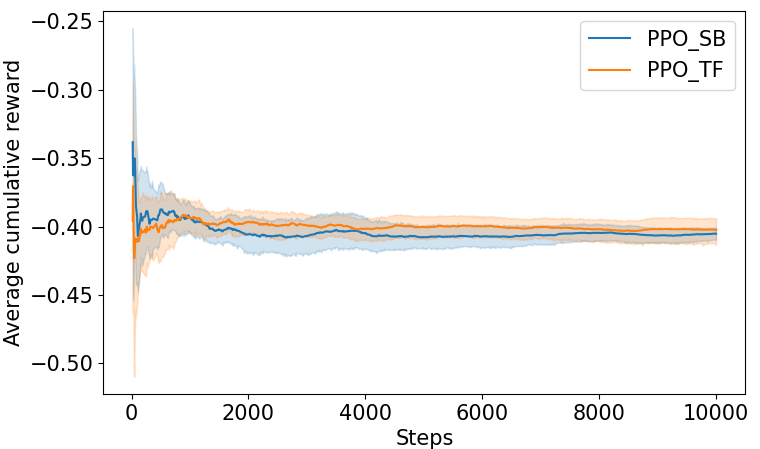}
        \caption{Average cumulative reward earned by PPO agents from different frameworks on a 10k budget.}
        \label{fig:ppos_rew10k}
            \end{minipage}%
    \end{figure*}

For a richer analysis, we collected our evaluation metrics on a reduced number of steps during test time (the evaluation metrics are collected on a 10,000 steps budget instead of 4 million) in order to answer RQ1. This analysis does not involve A2C-wuji as with this implementation the detection of bugs starts at 300000+ steps. 
Figures \ref{fig:dqns_bug10k} and \ref{fig:dqns_rew10k}  show respectively the average number of detected bugs and average cumulative reward over 10 runs of the algorithm obtained by DQN algorithms from Stable-baselines3 (DQN\_SB), Keras-rl (DQN\_KR), and Tensorforce (DQN\_TF) frameworks on a 10,000 budget steps.  
Similarly, Figures \ref{fig:a2cs_bug10k} and \ref{fig:a2cs_rew10k} show respectively the average number of detected bugs and average cumulative reward over 10 runs of the algorithm obtained by A2C algorithms from Stable-baselines3 (A2C\_SB), and Tensorforce (A2C\_TF) frameworks on a 10,000 budget steps.
Finally, Figures \ref{fig:ppos_bug10k} and \ref{fig:ppos_rew10k} show respectively the average number of detected bugs and average cumulative reward over 10 runs of the algorithm obtained by the PPOs algorithms from Stable-baselines3 (PPO\_SB), and Tensorforce (PPO\_TF) frameworks on a 10,000 budget steps.
Consistently, among the DQNs algorithms, Stable-baselines3 performs best in terms of detecting bugs and Tensorforce performs best in terms of rewards earned. Among A2C algorithms, on average, Tensorforce performs best in terms of detecting bugs, and Stable-baselines3 performs best in terms of rewards earned. Among the PPOs algorithms, Stable-baselines3 performs best in terms of detecting bugs, and Tensorforce performs best in terms of rewards earned. Similarly as with the 4 million steps budget, none of the DRL configurations is able to win the game.  
 
 Similarly as with the 4 million steps budget, none the DRL configurations is able to win the game.


In terms of training and prediction time, Tables \ref{tab:Results of Welch's anova test of  prediction (in milliseconds) of RL configurations on a 10k steps budget} and \ref{tab:Results of Welch's anova test of  training time (in milliseconds) of RL configurations on a 10k steps budget} show the results of Welch's ANOVA test and the CLES values for each of the DRL algorithms. 
\begin{table*}[t]
\caption{
Results of Welch's ANOVA test of  prediction time (in milliseconds) of DRL configurations  (in bold are DRL configurations where p-value is $<$ 0.05 and have greater performance w.r.t the effect size).}
\label{tab:Results of Welch's anova test of  prediction (in milliseconds) of RL configurations on a 10k steps budget}
\centering
\resizebox{\textwidth}{!}{
\begin{tabular}{|c|c|c|c|c|c|c|}
\hline
\begin{tabular}[c]{@{}c@{}}Testing budget\\  in steps\end{tabular} & A                & B                & mean(A)     & mean(B)     & pval     & CLES     \\ \hline
\multirow{21}{*}{10,000}                                           & \textbf{A2C\_SB} & A2C\_TF          & 1,31E+04    & 1,81E+05    & 0,00E+00 & 0,00E+00 \\ \cline{2-7} 
                                                                   & A2C\_SB          & \textbf{DQN\_KR} & 1,31E+04    & 1,08E+04    & 0,00E+00 & 1,00E+00 \\ \cline{2-7} 
                                                                   & A2C\_SB          & \textbf{DQN\_SB} & 1,31E+04    & 1,17E+04    & 0,00E+00 & 1,00E+00 \\ \cline{2-7} 
                                                                   & \textbf{A2C\_SB} & DQN\_TF          & 1,31E+04    & 1,86E+05    & 0,00E+00 & 0,00E+00 \\ \cline{2-7} 
                                                                   & \textbf{A2C\_SB} & PPO\_SB          & 1,31E+04    & 1,32E+04    & 9,70E-01 & 3,90E-01 \\ \cline{2-7} 
                                                                   & \textbf{A2C\_SB} & PPO\_TF          & 1,31E+04    & 1,83E+05    & 0,00E+00 & 0,00E+00 \\ \cline{2-7} 
                                                                   & A2C\_TF          & \textbf{DQN\_KR} & 1,81E+05    & 1,08E+04    & 0,00E+00 & 1,00E+00 \\ \cline{2-7} 
                                                                   & A2C\_TF          & \textbf{DQN\_SB} & 1,81E+05    & 1,17E+04    & 0,00E+00 & 1,00E+00 \\ \cline{2-7} 
                                                                   & \textbf{A2C\_TF} & DQN\_TF          & 1,81E+05    & 1,86E+05    & 1,00E-02 & 7,00E-02 \\ \cline{2-7} 
                                                                   & A2C\_TF          & \textbf{PPO\_SB} & 1,81E+05    & 1,32E+04    & 0,00E+00 & 1,00E+00 \\ \cline{2-7} 
                                                                   & \textbf{A2C\_TF} & PPO\_TF          & 1,81E+05    & 1,83E+05    & 9,30E-01 & 3,70E-01 \\ \cline{2-7} 
                                                                   & \textbf{DQN\_KR} & DQN\_SB          & 1,08E+04    & 1,17E+04    & 0,00E+00 & 0,00E+00 \\ \cline{2-7} 
                                                                   & \textbf{DQN\_KR} & DQN\_TF          & 1,08E+04    & 1,86E+05    & 0,00E+00 & 0,00E+00 \\ \cline{2-7} 
                                                                   & \textbf{DQN\_KR} & PPO\_SB          & 1,08E+04    & 1,32E+04    & 0,00E+00 & 0,00E+00 \\ \cline{2-7} 
                                                                   & \textbf{DQN\_KR} & PPO\_TF          & 1,08E+04    & 1,83E+05    & 0,00E+00 & 0,00E+00 \\ \cline{2-7} 
                                                                   & \textbf{DQN\_SB} & DQN\_TF          & 1,17E+04    & 1,86E+05    & 0,00E+00 & 0,00E+00 \\ \cline{2-7} 
                                                                   & \textbf{DQN\_SB} & PPO\_SB          & 1,17E+04    & 1,32E+04    & 0,00E+00 & 0,00E+00 \\ \cline{2-7} 
                                                                   & \textbf{DQN\_SB} & PPO\_TF          & 1,17E+04    & 1,83E+05    & 0,00E+00 & 0,00E+00 \\ \cline{2-7} 
                                                                   & DQN\_TF          & \textbf{PPO\_SB} & 1,86E+05    & 1,32E+04    & 0,00E+00 & 1,00E+00 \\ \cline{2-7} 
                                                                   & DQN\_TF          & \textbf{PPO\_TF} & 1,86E+05    & 1,83E+05    & 6,80E-01 & 6,90E-01 \\ \cline{2-7} 
                                                                   & \textbf{PPO\_SB} & PPO\_TF          & 1,32E+04    & 1,83E+05    & 0,00E+00 & 0,00E+00 \\ \hline
\multirow{21}{*}{4 million}                                        & A2C\_SB          & \textbf{A2C\_TF} & 1,63813E+12 & 1,64232E+12 & 0,00E+00 & 0,00E+00 \\ \cline{2-7} 
                                                                   & \textbf{A2C\_SB} & DQN\_KR          & 1,63813E+12 & 1,63861E+12 & 1,36E-11 & 0,00E+00 \\ \cline{2-7} 
                                                                   & A2C\_SB          & \textbf{DQN\_SB} & 1,63813E+12 & 1,63748E+12 & 5,17E-12 & 1,00E+00 \\ \cline{2-7} 
                                                                   & \textbf{A2C\_SB} & DQN\_TF          & 1,63813E+12 & 1,64331E+12 & 9,30E-12 & 0,00E+00 \\ \cline{2-7} 
                                                                   & \textbf{A2C\_SB} & PPO\_SB          & 1,63813E+12 & 1,63844E+12 & 0,00E+00 & 0,00E+00 \\ \cline{2-7} 
                                                                   & \textbf{A2C\_SB} & PPO\_TF          & 1,63813E+12 & 1,64089E+12 & 7,09E-13 & 0,00E+00 \\ \cline{2-7} 
                                                                   & A2C\_TF          & \textbf{DQN\_KR} & 1,64232E+12 & 1,63861E+12 & 4,89E-12 & 1,00E+00 \\ \cline{2-7} 
                                                                   & A2C\_TF          & \textbf{DQN\_SB} & 1,64232E+12 & 1,63748E+12 & 1,25E-11 & 1,00E+00 \\ \cline{2-7} 
                                                                   & \textbf{A2C\_TF} & DQN\_TF          & 1,64232E+12 & 1,64331E+12 & 0,00E+00 & 0,00E+00 \\ \cline{2-7} 
                                                                   & A2C\_TF          & \textbf{PPO\_SB} & 1,64232E+12 & 1,63844E+12 & 0,00E+00 & 1,00E+00 \\ \cline{2-7} 
                                                                   & A2C\_TF          & \textbf{PPO\_TF} & 1,64232E+12 & 1,64089E+12 & 0,00E+00 & 9,50E-01 \\ \cline{2-7} 
                                                                   & DQN\_KR          & \textbf{DQN\_SB} & 1,63861E+12 & 1,63748E+12 & 1,90E-11 & 1,00E+00 \\ \cline{2-7} 
                                                                   & \textbf{DQN\_KR} & DQN\_TF          & 1,63861E+12 & 1,64331E+12 & 2,60E-12 & 0,00E+00 \\ \cline{2-7} 
                                                                   & DQN\_KR          & \textbf{PPO\_SB} & 1,63861E+12 & 1,63844E+12 & 0,00E+00 & 1,00E+00 \\ \cline{2-7} 
                                                                   & \textbf{DQN\_KR} & PPO\_TF          & 1,63861E+12 & 1,64089E+12 & 4,63E-13 & 2,86E-02 \\ \cline{2-7} 
                                                                   & \textbf{DQN\_SB} & DQN\_TF          & 1,63748E+12 & 1,64331E+12 & 0,00E+00 & 0,00E+00 \\ \cline{2-7} 
                                                                   & \textbf{DQN\_SB} & PPO\_SB          & 1,63748E+12 & 1,63844E+12 & 0,00E+00 & 0,00E+00 \\ \cline{2-7} 
                                                                   & \textbf{DQN\_SB} & PPO\_TF          & 1,63748E+12 & 1,64089E+12 & 0,00E+00 & 0,00E+00 \\ \cline{2-7} 
                                                                   & DQN\_TF          & \textbf{PPO\_SB} & 1,64331E+12 & 1,63844E+12 & 0,00E+00 & 1,00E+00 \\ \cline{2-7} 
                                                                   & DQN\_TF          & \textbf{PPO\_TF} & 1,64331E+12 & 1,64089E+12 & 0,00E+00 & 1,00E+00 \\ \cline{2-7} 
                                                                   & \textbf{PPO\_SB} & PPO\_TF          & 1,63844E+12 & 1,64089E+12 & 0,00E+00 & 1,01E-02 \\ \hline
\end{tabular}
}
\begin{tablenotes}
     \item mean(A) and mean(B) refer to prediction time values.

 \end{tablenotes}
\end{table*}
\begin{table*}[t]
\caption{
Results of Welch's ANOVA test of  training time (in milliseconds) of DRL configurations on a 10k steps budget (in bold are DRL configurations where p-value is $<$ 0.05 and have greater performance w.r.t the effect size).}
\label{tab:Results of Welch's anova test of  training time (in milliseconds) of RL configurations on a 10k steps budget}
\centering
\begin{tabular}{|c|c|c|c|c|c|}
\hline
A                & B                & mean(A)  & mean(B)  & pval     & CLES     \\ \hline
A2C\_SB          & \textbf{A2C\_TF} & 1,63E+04 & 1,46E+04 & 3,43E-07 & 1,00E+00 \\ \hline
\textbf{A2C\_SB} & DQN\_KR          & 1,63E+04 & 6,37E+04 & 0,00E+00 & 0,00E+00 \\ \hline
A2C\_SB          & \textbf{DQN\_SB} & 1,63E+04 & 1,14E+03 & 0,00E+00 & 1,00E+00 \\ \hline
\textbf{A2C\_SB} & DQN\_TF          & 1,63E+04 & 9,28E+04 & 1,17E-13 & 0,00E+00 \\ \hline
A2C\_SB          & \textbf{PPO\_SB} & 1,63E+04 & 1,22E+04 & 3,50E-12 & 1,00E+00 \\ \hline
A2C\_SB          & \textbf{PPO\_TF} & 1,63E+04 & 1,43E+04 & 1,11E-05 & 1,00E+00 \\ \hline
\textbf{A2C\_TF} & DQN\_KR          & 1,46E+04 & 6,37E+04 & 0,00E+00 & 0,00E+00 \\ \hline
A2C\_TF          & \textbf{DQN\_SB} & 1,46E+04 & 1,14E+03 & 2,55E-15 & 1,00E+00 \\ \hline
\textbf{A2C\_TF} & DQN\_TF          & 1,46E+04 & 9,28E+04 & 4,84E-14 & 0,00E+00 \\ \hline
A2C\_TF          & \textbf{PPO\_SB} & 1,46E+04 & 1,22E+04 & 1,33E-08 & 1,00E+00 \\ \hline
A2C\_TF          & \textbf{PPO\_TF} & 1,46E+04 & 1,43E+04 & 7,69E-01 & 7,80E-01 \\ \hline
DQN\_KR          & \textbf{DQN\_SB} & 6,37E+04 & 1,14E+03 & 0,00E+00 & 1,00E+00 \\ \hline
\textbf{DQN\_KR} & DQN\_TF          & 6,37E+04 & 9,28E+04 & 6,83E-10 & 0,00E+00 \\ \hline
DQN\_KR          & \textbf{PPO\_SB} & 6,37E+04 & 1,22E+04 & 0,00E+00 & 1,00E+00 \\ \hline
DQN\_KR          & \textbf{PPO\_TF} & 6,37E+04 & 1,43E+04 & 0,00E+00 & 1,00E+00 \\ \hline
\textbf{DQN\_SB} & DQN\_TF          & 1,14E+03 & 9,28E+04 & 2,28E-14 & 0,00E+00 \\ \hline
\textbf{DQN\_SB} & PPO\_SB          & 1,14E+03 & 1,22E+04 & 0,00E+00 & 0,00E+00 \\ \hline
\textbf{DQN\_SB} & PPO\_TF          & 1,14E+03 & 1,43E+04 & 5,48E-13 & 0,00E+00 \\ \hline
DQN\_TF          & \textbf{PPO\_SB} & 9,28E+04 & 1,22E+04 & 7,26E-14 & 1,00E+00 \\ \hline
DQN\_TF          & \textbf{PPO\_TF} & 9,28E+04 & 1,43E+04 & 1,27E-14 & 1,00E+00 \\ \hline
\textbf{PPO\_SB} & PPO\_TF          & 1,22E+04 & 1,43E+04 & 4,85E-06 & 0,00E+00 \\ \hline
\end{tabular}
\begin{tablenotes}
     \item mean(A) and mean(B) refer to training time values.

 \end{tablenotes}
\end{table*}
In terms of  prediction time,  among the DQNs algorithms, Keras-rl and Stable-baselines3 frameworks have the best performance. Among the PPOs and A2Cs algorithms, Stable-baselines3 has the best results. In terms of training time, among the PPOs and DQNs algorithms, Stable-baselines3 has the best results. Among the A2Cs algorithms, Tensorforce has the best results.


In terms of state coverage, Table \ref{tab:State coverage of DRL algorithms on a 10K steps budget} shows the results of state coverage obtained by the DRL configurations on 10,000 budget steps.
\begin{table}[t]
\caption{State coverage of DRL algorithms on the Block Maze game on a 10K steps budget.}
\label{tab:State coverage of DRL algorithms on a 10K steps budget}
\centering
\resizebox{\textwidth}{!}{
\begin{tabular}{|c|c|c|c|c|c|c|c|}
\hline
               & PPO\_SB     & PPO\_TF & A2C\_TF & A2C\_SB & DQN\_SB & DQN\_KR & DQN\_TF \\ \hline
State coverage & \textbf{18.5\%} & 17.4\%    & 17\%      & 15.7\%    & 7.4\%    & 2.4\%     & 1.3\%     \\ \hline
\end{tabular}
}
\end{table}
Similarly, as with the 4 million steps budget, PPO\_TF, PPO\_SB, A2C\_SB and A2C\_TF have the largest state coverage. Nevertheless but expected with the fewer budget, all DRL configurations have lower state coverage. In terms of line coverage, all DRL configurations fare the same line coverage performance, similar to the 4 million steps budget: 96\%.

\begin{tcolorbox}
    Summary 1: From the results of RQ1, we can see that Stable-baselines3 and Tensorforce frameworks provide the best results in terms of bugs detected, reward earned and state coverage capability. Stable-baselines3 and Tensorforce frameworks provide a diverse set of hyperparameters for each of their DRL algorithms that contribute to their performance. For example, Stable-baselines3 with its DQN implementation provides a soft update coefficient to update the target network frequently and optimize the network efficiency.
\end{tcolorbox}
\hfill \break 
\textbf{RQ2:} We perform Welch’s ANOVA and Games-Howell post-hoc test to check for significant differences between our results. Tables \ref{tab:Welch’s ANOVA and Games-Howell post-hoc test regarding the average cumulative reward} and \ref{tab:Welch’s ANOVA and Games-Howell post-hoc test regarding the number bug} show respectively the results of Welch’s ANOVA and Games-Howell post-hoc test analysis in terms of average cumulative reward earned and number of bugs detected by the DRL algorithms. 
\begin{table*}[t]
\caption{Results of Welch’s ANOVA and Games-Howell post-hoc test regarding the average cumulative reward earned by DRL algorithms (in bold are DRL configurations where p-value is $<$ 0.05 and have greater performance w.r.t the effect size).}
\label{tab:Welch’s ANOVA and Games-Howell post-hoc test regarding the average cumulative reward}
\centering
\begin{tabular}{|c|c|c|c|c|c|}
\hline
A                & B                & mean(A)   & mean(B)   & pval     & CLES     \\ \hline
\textbf{A2C\_SB} & A2C\_TF          & -3,00E-01 & -7,20E-01 & 0,00E+00 & 9,60E-01 \\ \hline
\textbf{A2C\_SB} & DQN\_KR          & -3,00E-01 & -9,70E-01 & 0,00E+00 & 1,00E+00 \\ \hline
\textbf{A2C\_SB} & DQN\_SB          & -3,00E-01 & -9,60E-01 & 0,00E+00 & 1,00E+00 \\ \hline
\textbf{A2C\_SB} & DQN\_TF          & -3,00E-01 & -3,20E-01 & 0,00E+00 & 5,20E-01 \\ \hline
\textbf{A2C\_SB} & PPO\_SB          & -3,00E-01 & -2,90E-01 & 0,00E+00 & 3,30E-01 \\ \hline
\textbf{A2C\_SB} & PPO\_TF          & -3,00E-01 & -5,20E-01 & 0,00E+00 & 7,80E-01 \\ \hline
\textbf{A2C\_TF} & DQN\_KR          & -7,20E-01 & -9,70E-01 & 0,00E+00 & 8,40E-01 \\ \hline
\textbf{A2C\_TF} & DQN\_SB          & -7,20E-01 & -9,60E-01 & 0,00E+00 & 8,40E-01 \\ \hline
A2C\_TF          & \textbf{DQN\_TF} & -7,20E-01 & -3,20E-01 & 0,00E+00 & 2,20E-01 \\ \hline
A2C\_TF          & \textbf{PPO\_SB} & -7,20E-01 & -2,90E-01 & 0,00E+00 & 4,00E-02 \\ \hline
A2C\_TF          & \textbf{PPO\_TF} & -7,20E-01 & -5,20E-01 & 0,00E+00 & 3,10E-01 \\ \hline
DQN\_KR          & \textbf{DQN\_SB} & -9,70E-01 & -9,60E-01 & 0,00E+00 & 3,70E-01 \\ \hline
DQN\_KR          & \textbf{DQN\_TF} & -9,70E-01 & -3,20E-01 & 0,00E+00 & 7,00E-02 \\ \hline
DQN\_KR          & \textbf{PPO\_SB} & -9,70E-01 & -2,90E-01 & 0,00E+00 & 0,00E+00 \\ \hline
DQN\_KR          & \textbf{PPO\_TF} & -9,70E-01 & -5,20E-01 & 0,00E+00 & 7,00E-02 \\ \hline
DQN\_SB          & \textbf{DQN\_TF} & -9,60E-01 & -3,20E-01 & 0,00E+00 & 8,00E-02 \\ \hline
DQN\_SB          & \textbf{PPO\_SB} & -9,60E-01 & -2,90E-01 & 0,00E+00 & 0,00E+00 \\ \hline
DQN\_SB          & \textbf{PPO\_TF} & -9,60E-01 & -5,20E-01 & 0,00E+00 & 7,00E-02 \\ \hline
DQN\_TF          & \textbf{PPO\_SB} & -3,20E-01 & -2,90E-01 & 0,00E+00 & 4,80E-01 \\ \hline
DQN\_TF          & \textbf{PPO\_TF} & -3,20E-01 & -5,20E-01 & 0,00E+00 & 6,50E-01 \\ \hline
\textbf{PPO\_SB} & PPO\_TF          & -2,90E-01 & -5,20E-01 & 0,00E+00 & 7,80E-01 \\ \hline
\end{tabular}
\begin{tablenotes}
     \item mean(A) and mean(B) refer to average cumulative reward values.

 \end{tablenotes}
\end{table*}
\begin{table*}[t]
\caption{Results of Welch’s ANOVA and Games-Howell post-hoc test regarding the number bugs detected by DRL algorithms (in bold are DRL configurations where p-value is $<$ 0.05 and have greater performance w.r.t the effect size).}
\label{tab:Welch’s ANOVA and Games-Howell post-hoc test regarding the number bug}
\centering
\begin{tabular}{|c|c|c|c|c|c|}
\hline
A                  & B                & mean(A)  & mean(B)  & pval     & CLES     \\ \hline
\textbf{A2C\_SB}   & A2C\_TF          & 1,89E+01 & 1,69E+01 & 0,00E+00 & 6,80E-01 \\ \hline
\textbf{A2C\_SB}   & A2C\_wuji        & 1,89E+01 & 1,47E+01 & 0,00E+00 & 1,00E+00 \\ \hline
\textbf{A2C\_SB}   & DQN\_KR          & 1,89E+01 & 3,22E+00 & 0,00E+00 & 1,00E+00 \\ \hline
\textbf{A2C\_SB}   & DQN\_SB          & 1,89E+01 & 2,96E+00 & 0,00E+00 & 1,00E+00 \\ \hline
\textbf{A2C\_SB}   & DQN\_TF          & 1,89E+01 & 3,00E-01 & 0,00E+00 & 1,00E+00 \\ \hline
A2C\_SB            & PPO\_SB          & 1,89E+01 & 1,90E+01 & 9,80E-01 & 5,00E-01 \\ \hline
\textbf{A2C\_SB}   & PPO\_TF          & 1,89E+01 & 1,81E+01 & 0,00E+00 & 6,80E-01 \\ \hline
\textbf{A2C\_TF}   & A2C\_wuji        & 1,69E+01 & 1,47E+01 & 0,00E+00 & 9,30E-01 \\ \hline
\textbf{A2C\_TF}   & DQN\_KR          & 1,69E+01 & 3,22E+00 & 0,00E+00 & 1,00E+00 \\ \hline
\textbf{A2C\_TF}   & DQN\_SB          & 1,69E+01 & 2,96E+00 & 0,00E+00 & 1,00E+00 \\ \hline
\textbf{A2C\_TF}   & DQN\_TF          & 1,69E+01 & 3,00E-01 & 0,00E+00 & 1,00E+00 \\ \hline
A2C\_TF            & \textbf{PPO\_SB} & 1,69E+01 & 1,90E+01 & 0,00E+00 & 3,20E-01 \\ \hline
A2C\_TF            & \textbf{PPO\_TF} & 1,69E+01 & 1,81E+01 & 0,00E+00 & 4,00E-01 \\ \hline
\textbf{A2C\_wuji} & DQN\_KR          & 1,47E+01 & 3,22E+00 & 0,00E+00 & 1,00E+00 \\ \hline
\textbf{A2C\_wuji} & DQN\_SB          & 1,47E+01 & 2,96E+00 & 0,00E+00 & 1,00E+00 \\ \hline
\textbf{A2C\_wuji} & DQN\_TF          & 1,47E+01 & 3,00E-01 & 0,00E+00 & 1,00E+00 \\ \hline
A2C\_wuji          & \textbf{PPO\_SB} & 1,47E+01 & 1,90E+01 & 0,00E+00 & 0,00E+00 \\ \hline
A2C\_wuji          & \textbf{PPO\_TF} & 1,47E+01 & 1,81E+01 & 0,00E+00 & 1,00E-02 \\ \hline
\textbf{DQN\_KR}   & DQN\_SB          & 3,22E+00 & 2,96E+00 & 0,00E+00 & 5,50E-01 \\ \hline
\textbf{DQN\_KR}   & DQN\_TF          & 3,22E+00 & 3,00E-01 & 0,00E+00 & 9,50E-01 \\ \hline
DQN\_KR            & \textbf{PPO\_SB} & 3,22E+00 & 1,90E+01 & 0,00E+00 & 0,00E+00 \\ \hline
DQN\_KR            & \textbf{PPO\_TF} & 3,22E+00 & 1,81E+01 & 0,00E+00 & 0,00E+00 \\ \hline
DQN\_SB            & DQN\_TF          & 2,96E+00 & 3,00E-01 & 0,00E+00 & 9,60E-01 \\ \hline
DQN\_SB            & \textbf{PPO\_SB} & 2,96E+00 & 1,90E+01 & 0,00E+00 & 0,00E+00 \\ \hline
DQN\_SB            & \textbf{PPO\_TF} & 2,96E+00 & 1,81E+01 & 0,00E+00 & 0,00E+00 \\ \hline
DQN\_TF            & \textbf{PPO\_SB} & 3,00E-01 & 1,90E+01 & 0,00E+00 & 0,00E+00 \\ \hline
DQN\_TF            & \textbf{PPO\_TF} & 3,00E-01 & 1,81E+01 & 0,00E+00 & 0,00E+00 \\ \hline
\textbf{PPO\_SB}   & PPO\_TF          & 1,90E+01 & 1,81E+01 & 0,00E+00 & 6,90E-01 \\ \hline
\end{tabular}
\begin{tablenotes}
     \item mean(A) and mean(B) refer to the number of bugs detected values.

 \end{tablenotes}
\end{table*}
Tables \ref{tab:Welch’s ANOVA and Games-Howell post-hoc test regarding the average cumulative reward} and \ref{tab:Welch’s ANOVA and Games-Howell post-hoc test regarding the number bug} also report CLES between the DRL configurations. CLES values show the probability that one configuration detects more bugs than another or earns more rewards. Table \ref{tab:Welch’s ANOVA and Games-Howell post-hoc test regarding the average cumulative reward} shows that the A2Cs and PPOs earned significantly more rewards than the DQNs. Table \ref{tab:Welch’s ANOVA and Games-Howell post-hoc test regarding the number bug}  shows that on average the A2Cs and PPOs perform better than the DQNs with a high bug detection number between 12.23 and 15.28 and CLES values equal to 1. The A2C algorithms have similar performance, same as the PPO algorithms: while PPOs detect more bugs, they do not have statistically significant results in comparison to A2Cs.
Similarly Tables \ref{tab:Welch’s ANOVA and Games-Howell post-hoc test regarding the number bug on a 10k steps budget} and \ref{tab:Welch’s ANOVA and Games-Howell post-hoc test regarding the average cumulative reward on a 10k steps budget} show respectively the results of Welch's ANOVA post hoc tests regarding the bugs detected by the DRL algorithms and the rewards earned on 10,000 steps.
\begin{table*}[t]
\caption{Results of Welch’s ANOVA and Games-Howell post-hoc test regarding the number of bugs detected by DRL algorithms on a 10k steps budget (in bold are DRL configurations where p-value is $<$ 0.05 and have greater performance w.r.t the effect size).}
\label{tab:Welch’s ANOVA and Games-Howell post-hoc test regarding the number bug on a 10k steps budget}
\centering
\begin{tabular}{|c|c|c|c|c|c|}
\hline
A                & B                & mean(A)  & mean(B)  & pval     & CLES     \\ \hline
A2C\_SB          & \textbf{A2C\_TF} & 4,24E+00 & 5,05E+00 & 0,00E+00 & 3,80E-01 \\ \hline
\textbf{A2C\_SB} & DQN\_KR          & 4,24E+00 & 2,20E-01 & 0,00E+00 & 9,90E-01 \\ \hline
\textbf{A2C\_SB} & DQN\_SB          & 4,24E+00 & 1,74E+00 & 0,00E+00 & 9,20E-01 \\ \hline
\textbf{A2C\_SB} & DQN\_TF          & 4,24E+00 & 2,00E-01 & 0,00E+00 & 9,90E-01 \\ \hline
A2C\_SB          & \textbf{PPO\_SB} & 4,24E+00 & 6,61E+00 & 0,00E+00 & 1,80E-01 \\ \hline
A2C\_SB          & \textbf{PPO\_TF} & 4,24E+00 & 5,23E+00 & 0,00E+00 & 3,60E-01 \\ \hline
\textbf{A2C\_TF} & DQN\_KR          & 5,05E+00 & 2,20E-01 & 0,00E+00 & 9,90E-01 \\ \hline
\textbf{A2C\_TF} & DQN\_SB          & 5,05E+00 & 1,74E+00 & 0,00E+00 & 9,20E-01 \\ \hline
\textbf{A2C\_TF} & DQN\_TF          & 5,05E+00 & 2,00E-01 & 0,00E+00 & 9,90E-01 \\ \hline
A2C\_TF          & \textbf{PPO\_SB} & 5,05E+00 & 6,61E+00 & 0,00E+00 & 3,00E-01 \\ \hline
A2C\_TF          & \textbf{PPO\_TF} & 5,05E+00 & 5,23E+00 & 0,00E+00 & 4,80E-01 \\ \hline
DQN\_KR          & \textbf{DQN\_SB} & 2,20E-01 & 1,74E+00 & 0,00E+00 & 4,00E-02 \\ \hline
DQN\_KR          & DQN\_TF          & 2,20E-01 & 2,00E-01 & 7,00E-02 & 5,10E-01 \\ \hline
DQN\_KR          & \textbf{PPO\_SB} & 2,20E-01 & 6,61E+00 & 0,00E+00 & 0,00E+00 \\ \hline
DQN\_KR          & \textbf{PPO\_TF} & 2,20E-01 & 5,23E+00 & 0,00E+00 & 1,00E-02 \\ \hline
\textbf{DQN\_SB} & DQN\_TF          & 1,74E+00 & 2,00E-01 & 0,00E+00 & 9,60E-01 \\ \hline
DQN\_SB          & \textbf{PPO\_SB} & 1,74E+00 & 6,61E+00 & 0,00E+00 & 1,00E-02 \\ \hline
DQN\_SB          & \textbf{PPO\_TF} & 1,74E+00 & 5,23E+00 & 0,00E+00 & 7,00E-02 \\ \hline
DQN\_TF          & \textbf{PPO\_SB} & 2,00E-01 & 6,61E+00 & 0,00E+00 & 0,00E+00 \\ \hline
DQN\_TF          & \textbf{PPO\_TF} & 2,00E-01 & 5,23E+00 & 0,00E+00 & 1,00E-02 \\ \hline
\textbf{PPO\_SB} & PPO\_TF          & 6,61E+00 & 5,23E+00 & 0,00E+00 & 6,70E-01 \\ \hline
\end{tabular}
\begin{tablenotes}
     \item mean(A) and mean(B) refer to the number of bugs detected values.

 \end{tablenotes}
\end{table*}
\begin{table*}[t]
\caption{Results of Welch’s ANOVA and Games-Howell post-hoc test regarding the average cumulative reward on a 10k steps budget (in bold are DRL configurations where p-value is $<$ 0.05 and have greater performance w.r.t the effect size).}
\label{tab:Welch’s ANOVA and Games-Howell post-hoc test regarding the average cumulative reward on a 10k steps budget}
\centering
\begin{tabular}{|c|c|c|c|c|c|}
\hline
A                & B                & mean(A)  & mean(B)  & pval    & CLES    \\ \hline
\textbf{A2C\_SB} & A2C\_TF          & -4.00E-1 & -4.20E-1 & 0.00E+0 & 7.60E-1 \\ \hline
\textbf{A2C\_SB} & DQN\_KR          & -4.00E-1 & -9.30E-1 & 0.00E+0 & 1.00E+0 \\ \hline
\textbf{A2C\_SB} & DQN\_SB          & -4.00E-1 & -9.60E-1 & 0.00E+0 & 1.00E+0 \\ \hline
\textbf{A2C\_SB} & DQN\_TF          & -4.00E-1 & -5.90E-1 & 0.00E+0 & 7.20E-1 \\ \hline
\textbf{A2C\_SB} & PPO\_SB          & -4.00E-1 & -4.00E-1 & 0.00E+0 & 5.80E-1 \\ \hline
\textbf{A2C\_SB} & PPO\_TF          & -4.00E-1 & -4.00E-1 & 0.00E+0 & 5.30E-1 \\ \hline
\textbf{A2C\_TF} & DQN\_KR          & -4.20E-1 & -9.30E-1 & 0.00E+0 & 1.00E+0 \\ \hline
\textbf{A2C\_TF} & DQN\_SB          & -4.20E-1 & -9.60E-1 & 0.00E+0 & 1.00E+0 \\ \hline
\textbf{A2C\_TF} & DQN\_TF          & -4.20E-1 & -5.90E-1 & 0.00E+0 & 6.90E-1 \\ \hline
A2C\_TF          & \textbf{PPO\_SB} & -4.20E-1 & -4.00E-1 & 0.00E+0 & 2.80E-1 \\ \hline
A2C\_TF          & \textbf{PPO\_TF} & -4.20E-1 & -4.00E-1 & 0.00E+0 & 2.40E-1 \\ \hline
\textbf{DQN\_KR}          & DQN\_SB          & -9.30E-1 & -9.60E-1 & 0.00E+0 & 9.50E-1 \\ \hline
DQN\_KR          & \textbf{DQN\_TF}          & -9.30E-1 & -5.90E-1 & 0.00E+0 & 1.60E-1 \\ \hline
DQN\_KR          & \textbf{PPO\_SB} & -9.30E-1 & -4.00E-1 & 0.00E+0 & 0.00E+0 \\ \hline
DQN\_KR          & \textbf{PPO\_TF} & -9.30E-1 & -4.00E-1 & 0.00E+0 & 0.00E+0 \\ \hline
DQN\_SB          & \textbf{DQN\_TF} & -9.60E-1 & -5.90E-1 & 0.00E+0 & 1.40E-1 \\ \hline
DQN\_SB          & \textbf{PPO\_SB} & -9.60E-1 & -4.00E-1 & 0.00E+0 & 0.00E+0 \\ \hline
DQN\_SB          & \textbf{PPO\_TF} & -9.60E-1 & -4.00E-1 & 0.00E+0 & 0.00E+0 \\ \hline
DQN\_TF          & \textbf{PPO\_SB} & -5.90E-1 & -4.00E-1 & 0.00E+0 & 2.90E-1 \\ \hline
DQN\_TF          & \textbf{PPO\_TF} & -5.90E-1 & -4.00E-1 & 0.00E+0 & 2.80E-1 \\ \hline
\textbf{PPO\_SB} & \textbf{PPO\_TF} & -4.00E-1 & -4.00E-1 & 0.00E+0 & 4.50E-1 \\ \hline
\end{tabular}
\begin{tablenotes}
     \item mean(A) and mean(B) refer to average cumulative reward values.

 \end{tablenotes}
\end{table*}
Same as with the 4 million steps budget, the A2Cs and PPOs algorithms earned significantly more rewards than the DQNs algorithms (see Table \ref{tab:Welch’s ANOVA and Games-Howell post-hoc test regarding the average cumulative reward on a 10k steps budget}). In terms of number of bugs detected, Table \ref{tab:Welch’s ANOVA and Games-Howell post-hoc test regarding the number bug on a 10k steps budget} shows that the A2Cs detect fewer bugs than the PPOs algorithms with CLES values between [1.80E-01, 4.80E-01].
The following items summarize our results per DRL algorithm where $>$ denotes greater detected bugs and CLES values are greater than 60:

\textbf{A2C Algorithms:}
\begin{itemize}
    \item A2C\_SB $>$A2C\_TF $>$ A2C\_wuji
\end{itemize}

\textbf{PPO Algorithms:}
\begin{itemize}
    \item PPO\_SB $>$ PPO\_TF
\end{itemize}

\textbf{DQN Algorithms:}
\begin{itemize}
    \item DQN\_SB $>$ DQN\_TF
     \item DQN\_KR $>$ DQN\_TF
\end{itemize}

In terms of average cumulative reward, the following summarizes our results per DRL algorithm where CLES values are greater than 60.

\textbf{A2C Algorithms:}
\begin{itemize}
    \item A2C\_SB $>$ A2C\_TF
\end{itemize}

\textbf{PPO Algorithms:}
\begin{itemize}
    \item PPO\_SB $>$ PPO\_TF
\end{itemize}

Following are the results per DRL algorithm where $>$ denotes greater detected bugs based on 10,000 steps budget and where CLES values are greater than 60.

\textbf{PPO Algorithms:}
\begin{itemize}
    \item PPO\_SB $>$ PPO\_TF
\end{itemize}

\textbf{DQN Algorithms:}
\begin{itemize}
    \item DQN\_SB $>$ DQN\_TF
\end{itemize}

In terms of average cumulative reward, the following summarizes our results per DRL algorithm on the basis of a 10,000 steps budget where CLES values are greater than 60.

\textbf{A2C Algorithms:}
\begin{itemize}
    \item A2C\_SB $>$ A2C\_TF
\end{itemize}

\textbf{DQN Algorithms:}
\begin{itemize}
    \item DQN\_KR $>$ DQN\_SB
\end{itemize}

Moreover, on the basis of a 4 million steps budget, we observe with CLES values equal to $1$ that A2Cs and PPOs algorithms detect more bugs than DQN algorithms. Similarly, on the basis of a 10,000 steps budget, we observe with CLES values greater than 90 that A2Cs and PPOs algorithms detect more bugs than DQN algorithms. Practically it means for at least 90\% of episodes, PPOs algorithms detect more bugs.
\begin{tcolorbox}
\textbf{Finding 1: A2C and PPO algorithms show statistically significant performance compared to DQN algorithms in finding bugs in the examined game regardless of the DRL frameworks. }
\end{tcolorbox}

\begin{tcolorbox}
    Summary 2: From the results of RQ2, A2Cs and PPOs algorithms from the selected DRL frameworks provide the best results in terms of bugs detected, reward earned and state coverage capability. A2Cs and PPOs algorithms take advantage of actor-critic methods. The actor updates the policy distribution based on the critic’s estimate of the value function. Therefore, for small search spaces, like the Block Maze, the faster convergence rate of A2Cs and PPOs can lead to the detection of more bugs quickly as well as a wider state coverage capability.
\end{tcolorbox}
\hfill \break 
\textbf{RQ3:} Our findings show that we do not get similar results from the same DRL algorithm over the DRL frameworks. 
We explain this by the fact that each DRL framework that we used in this study do not provide the same hyperparameters regarding the DRL algorithm. Some of the hyperparameters are similar but not all of them. For example, the DQN algorithm from Stable-baselines has an additional hyperparameter called "gradient\_steps" to perform the gradient process as there are steps done during a rollout, instead of doing it after a complete rollout is done. These other hyperparameters, even with default values, can slightly improve efficiency as we observe in our results.
\hfill \break 
\subsection{Test case prioritization} \label{sec:Test case prioritization}
Tables \ref{tab:apfdandnrpa} and \ref{tab:apfdandnrpa2} show the averages and standard deviations of APFD and NRPA for the eight datasets, using different configurations (i.e., combinations of ranking model, DRL framework, and algorithm). The first column reports different DRL algorithms, the second column reports the ranking models followed by four datasets per table (a total of eight datasets). Each dataset column is subdivided into the DRL frameworks. In the rest of this section, we use \textit{[ranking model]-[RL algorithm]-[RL framework]} to refer to DRL configurations. For example, \textit{Pairwise-DQN-KR} corresponds to a configuration of the pairwise ranking model and the DQN algorithm from the Keras-rl framework. For each dataset (column), the relative performance rank of configurations in terms of APFD or NRPA are expressed with \circled{n}, where a lower rank indicates better performance. Again, we analyze the differences in the results by using Welch’s ANOVA and Games-Howell post-hoc test.

\begin{sidewaystable}
\centering
\caption{The average performance of different configurations in terms of APFD and NRPA, along with the results of the three baselines (RL-BS1, RL-BS2, and MART) for PAINT, IOFROL, CODEC, and IMAG datasets. The index in each cell shows the position of a configuration (row) with respect to others for each dataset (column) in terms of NRPA or APFD, based on statistical testing.}

\resizebox{\textwidth}{!}{%
\begin{tabular}{|c|c|ccc|ccc|ccc|ccc|}
\hline
                     &    & \multicolumn{3}{c|}{\begin{tabular}[c]{@{}c@{}}PAINT \\ (APFD)\end{tabular}}          & \multicolumn{3}{c|}{\begin{tabular}[c]{@{}c@{}}IOFROL\\  (APFD)\end{tabular}}          & \multicolumn{3}{c|}{\begin{tabular}[c]{@{}c@{}}CODEC  \\ (NRPA)\end{tabular}}               & \multicolumn{3}{c|}{\begin{tabular}[c]{@{}c@{}}IMAG\\  (NRPA)\end{tabular}}                 \\ \hline
                     & RM & \multicolumn{1}{c|}{SB}           & \multicolumn{1}{c|}{KR}           & TF            & \multicolumn{1}{c|}{SB}           & \multicolumn{1}{c|}{KR}            & TF            & \multicolumn{1}{c|}{SB}              & \multicolumn{1}{c|}{KR}             & TF             & \multicolumn{1}{c|}{SB}              & \multicolumn{1}{c|}{KR}             & TF             \\ \hline
DQN                  & PA & \multicolumn{1}{c|}{0.66±.2 \circled{3}} & \multicolumn{1}{c|}{0.49±.1 \circled{9}} & 0.56±.2 \circled{6}  & \multicolumn{1}{c|}{0.53±.1 \circled{4}} & \multicolumn{1}{c|}{0.49±.09 \circled{8}} & 0.50±.1 \circled{6}  & \multicolumn{1}{c|}{0.94±.06 \circled{3}}   & \multicolumn{1}{c|}{0.79±.07 \circled{9}}  & 0.86±.07 \circled{7}  & \multicolumn{1}{c|}{0.95±.06 \circled{1}}   & \multicolumn{1}{c|}{0.77±.06 \circled{8}}  & 0.87±.06 \circled{5}  \\ \hline
DDPG                 & PO & \multicolumn{1}{c|}{0.62±.2 \circled{4}} & \multicolumn{1}{c|}{0.57±.2 \circled{5}} & 0.52±.2 \circled{7}  & \multicolumn{1}{c|}{0.53±.1 \circled{4}} & \multicolumn{1}{c|}{0.52±.1 \circled{5}}  & 0.49±.1 \circled{7}  & \multicolumn{1}{c|}{0.89±.07 \circled{5}}   & \multicolumn{1}{c|}{0.76±.08 \circled{13}} & 0.78±.07 \circled{11} & \multicolumn{1}{c|}{0.85±.07 \circled{6}}   & \multicolumn{1}{c|}{0.75±.07 \circled{11}} & 0.77±.05 \circled{9}  \\ \hline
\multirow{2}{*}{PPO} & PA & \multicolumn{1}{c|}{0.69±.2 \circled{2}} & \multicolumn{1}{c|}{NA}           & 0.50±.18 \circled{8} & \multicolumn{1}{c|}{0.55±.1 \circled{2}} & \multicolumn{1}{c|}{NA}            & 0.49±.1 \circled{7}  & \multicolumn{1}{c|}{0.96±.05 \circled{1}}   & \multicolumn{1}{c|}{NA}             & 0.79±.07 \circled{9}  & \multicolumn{1}{c|}{0.95±.05 \circled{2}}   & \multicolumn{1}{c|}{NA}             & 0.78±.06 \circled{7}  \\ \cline{2-14} 
                     & PO & \multicolumn{1}{c|}{0.57±.2 \circled{5}} & \multicolumn{1}{c|}{NA}           & 0.49±.1 \circled{9}  & \multicolumn{1}{c|}{0.52±.1 \circled{5}} & \multicolumn{1}{c|}{NA}            & 0.49±.08 \circled{9} & \multicolumn{1}{c|}{0.93±.06 \circled{4}}   & \multicolumn{1}{c|}{NA}             & 0.78±.08 \circled{10} & \multicolumn{1}{c|}{0.91±.05 \circled{4}}   & \multicolumn{1}{c|}{NA}             & 0.75±.06 \circled{13} \\ \hline
A2C                  & PA & \multicolumn{1}{c|}{0.70±.2 \circled{1}} & \multicolumn{1}{c|}{NA}           & 0.57±.2 \circled{5}  & \multicolumn{1}{c|}{0.54±.1 \circled{3}} & \multicolumn{1}{c|}{NA}            & 0.67±.2 \circled{1}  & \multicolumn{1}{c|}{0.96 ± .04 \circled{2}} & \multicolumn{1}{c|}{NA}             & 0.78±.07 \circled{11} & \multicolumn{1}{c|}{0.95±.05 \circled{2}}   & \multicolumn{1}{c|}{NA}             & 0.76±.10 \circled{10} \\ \cline{2-14} 
                     & PO & \multicolumn{1}{c|}{0.57±.2 \circled{5}} & \multicolumn{1}{c|}{NA}           & 0.44±.1 \circled{10} & \multicolumn{1}{c|}{0.52±.1 \circled{5}} & \multicolumn{1}{c|}{NA}            & 0.49±.1 \circled{7}  & \multicolumn{1}{c|}{0.89 ± .06 \circled{6}} & \multicolumn{1}{c|}{NA}             & 0.83±.08 \circled{8}  & \multicolumn{1}{c|}{0.92 ± .05 \circled{3}} & \multicolumn{1}{c|}{NA}             & 0.85±.07 \circled{6}  \\ \hline
Optimal              & NA & \multicolumn{3}{c|}{0.79±.14}                                                         & \multicolumn{3}{c|}{0.89±.14}                                                          & \multicolumn{3}{c|}{NA}                                                                     & \multicolumn{3}{c|}{NA}                                                                     \\ \hline
RL-BS1               & PO & \multicolumn{3}{c|}{0.63±.16}                                                         & \multicolumn{3}{c|}{0.74±.24}                                                          & \multicolumn{3}{c|}{NA}                                                                     & \multicolumn{3}{c|}{NA}                                                                     \\ \hline
RL-BS2               & PO & \multicolumn{3}{c|}{NA}                                                               & \multicolumn{3}{c|}{NA}                                                                & \multicolumn{3}{c|}{0.90±.05}                                                               & \multicolumn{3}{c|}{0.89±.09}                                                               \\ \hline
MART                 & PA & \multicolumn{3}{c|}{NA}                                                               & \multicolumn{3}{c|}{NA}                                                                & \multicolumn{3}{c|}{0.96±.03}                                                               & \multicolumn{3}{c|}{0.90±.05}                                                               \\ \hline
\end{tabular}
\label{tab:apfdandnrpa}%
}

\caption{The average performance of different configurations in terms of APFD and NRPA, along with the results of the three baselines (RL-BS1, RL-BS2, and MART) for IO, COMP, LANG, and MATH datasets. The index in each cell shows the position of a configuration (row) with respect to others for each dataset (column) in terms of NRPA or APFD, based on statistical testing.}
\centering
\resizebox{\textwidth}{!}{%
\begin{tabular}{|c|c|ccc|ccc|ccc|ccc|}
\hline
                     &    & \multicolumn{3}{c|}{\begin{tabular}[c]{@{}c@{}}IO\\  (NRPA)\end{tabular}}           & \multicolumn{3}{c|}{\begin{tabular}[c]{@{}c@{}}COMP \\ (NRPA)\end{tabular}}           & \multicolumn{3}{c|}{\begin{tabular}[c]{@{}c@{}}LANG\\   (NRPA)\end{tabular}}         & \multicolumn{3}{c|}{\begin{tabular}[c]{@{}c@{}}MATH\\   (NRPA)\end{tabular}}               \\ \hline
                     & RM & \multicolumn{1}{c|}{SB}            & \multicolumn{1}{c|}{KR}       & TF             & \multicolumn{1}{c|}{SB}            & \multicolumn{1}{c|}{KR}       & TF               & \multicolumn{1}{c|}{SB}             & \multicolumn{1}{c|}{KR}       & TF             & \multicolumn{1}{c|}{SB}             & \multicolumn{1}{c|}{KR}             & TF             \\ \hline
DQN                  & PA & \multicolumn{1}{c|}{0.97±.02 \circled{2}} & \multicolumn{1}{c|}{0.76±.05} & 0.85±.06 \circled{7}  & \multicolumn{1}{c|}{0.97±.03 \circled{1}} & \multicolumn{1}{c|}{0.77±.05} & 0.85±.06 \circled{7}    & \multicolumn{1}{c|}{0.94±.04  \circled{2}} & \multicolumn{1}{c|}{0.78±.05} & 0.86±.06 \circled{6}  & \multicolumn{1}{c|}{0.93±.05 \circled{3}}  & \multicolumn{1}{c|}{0.77±.05 \circled{9}}  & 0.88±.07 \circled{4}  \\ \hline
DDPG                 & PO & \multicolumn{1}{c|}{0.90±.07 \circled{5}} & \multicolumn{1}{c|}{0.72±.07} & 0.75±.05 \circled{11} & \multicolumn{1}{c|}{0.86±.07 \circled{5}} & \multicolumn{1}{c|}{0.74±.06} & 0.76±.06 \circled{11}   & \multicolumn{1}{c|}{0.84±.08 \circled{7}}  & \multicolumn{1}{c|}{0.76±.05} & 0.76±.06 \circled{11} & \multicolumn{1}{c|}{0.87±.06  \circled{6}} & \multicolumn{1}{c|}{0.74±.06 \circled{12}} & 0.76±.05 \circled{11} \\ \hline
\multirow{2}{*}{PPO} & PA & \multicolumn{1}{c|}{0.97±.02 \circled{2}} & \multicolumn{1}{c|}{NA}       & 0.77±.05 \circled{9}  & \multicolumn{1}{c|}{0.97±.02 \circled{2}} & \multicolumn{1}{c|}{NA}       & 0.77 ± .05 \circled{9}  & \multicolumn{1}{c|}{0.94±.03 \circled{3}}  & \multicolumn{1}{c|}{NA}       & 0.77±.05 \circled{10} & \multicolumn{1}{c|}{0.95±.04 \circled{1}}  & \multicolumn{1}{c|}{NA}             & 0.77±.05 \circled{9}  \\ \cline{2-14} 
                     & PO & \multicolumn{1}{c|}{0.93±.04 \circled{3}} & \multicolumn{1}{c|}{NA}       & 0.82±.07 \circled{8}  & \multicolumn{1}{c|}{0.93±.04 \circled{3}} & \multicolumn{1}{c|}{NA}       & 0.77 ± .06 \circled{8}  & \multicolumn{1}{c|}{0.88±.06 \circled{5}}  & \multicolumn{1}{c|}{NA}       & 0.80±.07 \circled{9}  & \multicolumn{1}{c|}{0.84±.06 \circled{7}}  & \multicolumn{1}{c|}{NA}             & 0.76±.06 \circled{10} \\ \hline
A2C                  & PA & \multicolumn{1}{c|}{0.98±.02 \circled{1}} & \multicolumn{1}{c|}{NA}       & 0.75±.09 \circled{10} & \multicolumn{1}{c|}{0.97±.02 \circled{2}} & \multicolumn{1}{c|}{NA}       & 0.76 ± .08 \circled{10} & \multicolumn{1}{c|}{0.95±.03 \circled{1}}  & \multicolumn{1}{c|}{NA}       & 0.73±.1 \circled{12}  & \multicolumn{1}{c|}{0.95±.04 \circled{2}}  & \multicolumn{1}{c|}{NA}             & 0.57±.3 \circled{13}  \\ \cline{2-14} 
                     & PO & \multicolumn{1}{c|}{0.92±.04 \circled{4}} & \multicolumn{1}{c|}{NA}       & 0.87±.06 \circled{6}  & \multicolumn{1}{c|}{0.89±.05 \circled{4}} & \multicolumn{1}{c|}{NA}       & 0.85±.07 \circled{6}    & \multicolumn{1}{c|}{0.89±.05 \circled{4}}  & \multicolumn{1}{c|}{NA}       & 0.83±.07 \circled{8}  & \multicolumn{1}{c|}{0.88±.05 \circled{5}}  & \multicolumn{1}{c|}{NA}             & 0.82±.06 \circled{8}  \\ \hline
Optimal              & NA & \multicolumn{3}{c|}{NA}                                                             & \multicolumn{3}{c|}{NA}                                                               & \multicolumn{3}{c|}{NA}                                                              & \multicolumn{3}{c|}{NA}                                                                    \\ \hline
RL-BS1               & PO & \multicolumn{3}{c|}{NA}                                                             & \multicolumn{3}{c|}{NA}                                                               & \multicolumn{3}{c|}{NA}                                                              & \multicolumn{3}{c|}{NA}                                                                    \\ \hline
RL-BS2               & PO & \multicolumn{3}{c|}{0.84±.13}                                                       & \multicolumn{3}{c|}{0.90±.05}                                                         & \multicolumn{3}{c|}{0.89±.07}                                                        & \multicolumn{3}{c|}{0.95±.02}                                                              \\ \hline
MART                 & PA & \multicolumn{3}{c|}{0.93±.02}                                                       & \multicolumn{3}{c|}{0.96±.02}                                                         & \multicolumn{3}{c|}{0.94±.02}                                                        & \multicolumn{3}{c|}{0.95±.02}                                                              \\ \hline
\end{tabular}
\label{tab:apfdandnrpa2}%
}
\end{sidewaystable}

Table \ref{tab:The sum of training time (Minutes) for all cycles across datasets and configurations 1} and \ref{tab:The sum of training time (Minutes) for all cycles across datasets and configurations 2} show the overall training times for the first 10 cycles across datasets. Similarly, Tables \ref{tab:The average of prediction (ranking) time (Seconds) for all cycles across datasets and configurations. 1} and \ref{tab:The average of prediction (ranking) time (Seconds) for all cycles across datasets and configurations. 2} show the averages and standard deviations of prediction time (ranking) for the first 10 cycles across datasets. Each cell value represents a configuration as mentioned before. For each dataset, the relative performance ranks of configurations in terms of training/prediction time are expressed with \circled{n}, where a lower rank \circled{n} indicates better performance.  
\begin{sidewaystable}
\caption{Average training time (in minutes) of DRL configurations for the first 10 cycles across PAINT, IOFROL, CODEC, and IMAG datasets.}
\centering
\resizebox{\textwidth}{!}{%
\begin{tabular}{|c|c|ccc|ccc|ccc|ccc|}
\hline
\multirow{2}{*}{}    & \multirow{2}{*}{RM}     & \multicolumn{3}{c|}{PAINT}                                                                  & \multicolumn{3}{c|}{IOFROL}                                                                   & \multicolumn{3}{c|}{CODEC}                                                               & \multicolumn{3}{c|}{IMAG}                                                                  \\ \cline{3-14} 
                     &                         & \multicolumn{1}{c|}{SB}            & \multicolumn{1}{c|}{KR}             & TF               & \multicolumn{1}{c|}{SB}             & \multicolumn{1}{c|}{KR}             & TF                & \multicolumn{1}{c|}{SB}            & \multicolumn{1}{c|}{KR}           & TF              & \multicolumn{1}{c|}{SB}            & \multicolumn{1}{c|}{KR}           & TF                \\ \hline
DQN                  & PA                      & \multicolumn{1}{c|}{6.0±3.9  \circled{7}} & \multicolumn{1}{c|}{9.2±5.8 \circled{10}}  & 44.3±27.5 \circled{13}  & \multicolumn{1}{c|}{45.6±52.4 \circled{6}} & \multicolumn{1}{c|}{68.3±76.8 \circled{7}} & 315.1±356.8 \circled{13} & \multicolumn{1}{c|}{2.1±1.4 \circled{6}}  & \multicolumn{1}{c|}{3.4±2.4 \circled{8}} & 16.6±11.1 \circled{13} & \multicolumn{1}{c|}{0.9±1.6 \circled{4}}  & \multicolumn{1}{c|}{2.8±2.9 \circled{7}} & 23.7±20.4 \circled{14}   \\ \hline
DDPG                 & PO                      & \multicolumn{1}{c|}{2.2±1.4 \circled{3}}  & \multicolumn{1}{c|}{10.5±6.4 \circled{11}} & 8.1±.3 \circled{9}      & \multicolumn{1}{c|}{19.4±22.3 \circled{5}} & \multicolumn{1}{c|}{76.6±86.4 \circled{8}} & 347.9±398.4 \circled{14} & \multicolumn{1}{c|}{0.8±.6 \circled{3}}   & \multicolumn{1}{c|}{3.7±2.3 \circled{9}} & 16.8±10.8 \circled{14} & \multicolumn{1}{c|}{0.8±.7 \circled{3}}   & \multicolumn{1}{c|}{3.3±3.2 \circled{8}} & 14.1±14.0 \circled{12}   \\ \hline
\multirow{2}{*}{PPO} & \multicolumn{1}{l|}{PA} & \multicolumn{1}{c|}{3.8±2.0 \circled{5}}  & \multicolumn{1}{c|}{NA}             & 6.8±2.1 \circled{8}     & \multicolumn{1}{c|}{6.1±7.1 \circled{2}}   & \multicolumn{1}{c|}{NA}             & 220.1±227.3 \circled{15} & \multicolumn{1}{c|}{1.3±.5 \circled{4}}   & \multicolumn{1}{c|}{NA}           & 3.4±.7 \circled{8}     & \multicolumn{1}{c|}{1.67±.9 \circled{5}}  & \multicolumn{1}{c|}{NA}           & 3.1 ± .9 \circled{9}     \\ \cline{2-14} 
                     & \multicolumn{1}{l|}{PO} & \multicolumn{1}{c|}{2.6±3.2 \circled{4}}  & \multicolumn{1}{c|}{NA}             & 5.7±1.7 \circled{6}     & \multicolumn{1}{c|}{7.5±8.4 \circled{3}}   & \multicolumn{1}{c|}{NA}             & 120.8±69.6 \circled{16}  & \multicolumn{1}{c|}{0.55±.2 \circled{1}}  & \multicolumn{1}{c|}{NA}           & 3.4 ± .5 \circled{7}   & \multicolumn{1}{c|}{0.43±.3 \circled{1}}  & \multicolumn{1}{c|}{NA}           & 4.1 ± 1.4 \circled{10}   \\ \hline
A2C                  & PA                      & \multicolumn{1}{c|}{1.6±.9 \circled{2}}   & \multicolumn{1}{c|}{NA}             & 49.1±22.7 \circled{14}  & \multicolumn{1}{c|}{11.1±12.1 \circled{4}} & \multicolumn{1}{c|}{NA}             & 221.3±299.2 \circled{17} & \multicolumn{1}{c|}{0.6 ± .3 \circled{2}} & \multicolumn{1}{c|}{NA}           & 8.5 ± .8 \circled{10}  & \multicolumn{1}{c|}{0.6±.4 \circled{2}}   & \multicolumn{1}{c|}{NA}           & 14.7 ± 14.8 \circled{13} \\ \cline{2-14} 
                     & PO                      & \multicolumn{1}{c|}{1.5±.3 \circled{1}}   & \multicolumn{1}{c|}{NA}             & 36.0±21.8  \circled{12} & \multicolumn{1}{c|}{1.3±.3 \circled{1}}    & \multicolumn{1}{c|}{NA}             & 282.4±332.1 \circled{18} & \multicolumn{1}{c|}{1.7±.2 \circled{5}}   & \multicolumn{1}{c|}{NA}           & 14.8±9.9 \circled{11}  & \multicolumn{1}{c|}{1.7 ± .5 \circled{6}} & \multicolumn{1}{c|}{NA}           & 14.0±13.5 \circled{11}   \\ \hline
\end{tabular}
\label{tab:The sum of training time (Minutes) for all cycles across datasets and configurations 1}
}
\caption{Average training time (in minutes) of DRL configurations for the first 10 cycles across IO, COMP, LANG, and MATH datasets.}
\centering
\resizebox{\textwidth}{!}{%
\begin{tabular}{|c|c|ccc|ccc|ccc|ccc|}
\hline
\multirow{2}{*}{}    & \multirow{2}{*}{RM}     & \multicolumn{3}{c|}{IO}                                                                       & \multicolumn{3}{c|}{COMP}                                                                     & \multicolumn{3}{c|}{LANG}                                                                        & \multicolumn{3}{c|}{MATH}                                                                           \\ \cline{3-14} 
                     &                         & \multicolumn{1}{c|}{SB}             & \multicolumn{1}{c|}{KR}             & TF                & \multicolumn{1}{c|}{SB}             & \multicolumn{1}{c|}{KR}             & TF                & \multicolumn{1}{c|}{SB}             & \multicolumn{1}{c|}{KR}                & TF                & \multicolumn{1}{c|}{SB}               & \multicolumn{1}{c|}{KR}               & TF                  \\ \hline
DQN                  & PA                      & \multicolumn{1}{c|}{2.8 ± 1.4 \circled{6}} & \multicolumn{1}{c|}{4.2 ± 2.1 \circled{7}} & 7.8±.7 \circled{9}       & \multicolumn{1}{c|}{2.8 ± 1.4 \circled{6}} & \multicolumn{1}{c|}{4.2 ± 2.2 \circled{7}} & 8.3±.9 \circled{9}       & \multicolumn{1}{c|}{4.0 ± 4.7 \circled{6}} & \multicolumn{1}{c|}{17.3 ± 27.7 \circled{11}} & 8.5±1.6 \circled{10}     & \multicolumn{1}{c|}{15.7 ± 17.7 \circled{9}} & \multicolumn{1}{c|}{23.0 ± 26.1 \circled{9}} & 11.7±5.9 \circled{6}       \\ \hline
DDPG                 & PO                      & \multicolumn{1}{c|}{1.1±.5 \circled{5}}    & \multicolumn{1}{c|}{4.9±.2.4 \circled{8}}  & 20.7±10.4 \circled{13}   & \multicolumn{1}{c|}{1.1±.5 \circled{5}}    & \multicolumn{1}{c|}{4.9±2.4 \circled{8}}   & 21.9±11.2 \circled{13}   & \multicolumn{1}{c|}{1.5±1.8 \circled{5}}   & \multicolumn{1}{c|}{6.7±8.1 \circled{7}}      & 31.4±37.5 \circled{13}   & \multicolumn{1}{c|}{6.4±7.4 \circled{5}}     & \multicolumn{1}{c|}{26.0±29.3 \circled{10}}  & 131.9±166.0 \circled{13}   \\ \hline
\multirow{2}{*}{PPO} & \multicolumn{1}{l|}{PA} & \multicolumn{1}{c|}{0.6±.2 \circled{2}}    & \multicolumn{1}{c|}{NA}             & 8.8 ± 2.0 \circled{11}   & \multicolumn{1}{c|}{0.6±.2 \circled{2}}    & \multicolumn{1}{c|}{NA}             & 9.3± 2.1 \circled{10}    & \multicolumn{1}{c|}{1.3±1.3 \circled{4}}   & \multicolumn{1}{c|}{NA}                & 6.6 ± 2.9 \circled{9}    & \multicolumn{1}{c|}{2.8±3.1 \circled{2}}     & \multicolumn{1}{c|}{NA}               & 13.7 ± 9.3 \circled{7}     \\ \cline{2-14} 
                     & \multicolumn{1}{l|}{PO} & \multicolumn{1}{c|}{0.5±.2 \circled{1}}    & \multicolumn{1}{c|}{NA}             & 8.3 ± 1.8 \circled{10}   & \multicolumn{1}{c|}{0.5±.2 \circled{1}}    & \multicolumn{1}{c|}{NA}             & 11.7± 3.1 \circled{11}   & \multicolumn{1}{c|}{0.7±0.8 \circled{1}}   & \multicolumn{1}{c|}{NA}                & 5.1 ± 2.9 \circled{8}    & \multicolumn{1}{c|}{2.7±2.9 \circled{1}}     & \multicolumn{1}{c|}{NA}               & 18.8 ± 9.6 \circled{8}     \\ \hline
A2C                  & PA                      & \multicolumn{1}{c|}{0.8±.3 \circled{3}}    & \multicolumn{1}{c|}{NA}             & 22.4 ± 11.7 \circled{14} & \multicolumn{1}{c|}{0.8± .3 \circled{4}}   & \multicolumn{1}{c|}{NA}             & 22.1 ± 11.4 \circled{14} & \multicolumn{1}{c|}{1.1±1.2 \circled{3}}   & \multicolumn{1}{c|}{NA}                & 43.4 ± 41.2 \circled{14} & \multicolumn{1}{c|}{4.0±.4.4 \circled{4}}    & \multicolumn{1}{c|}{NA}               & 116.2 ± 130.6 \circled{12} \\ \cline{2-14} 
                     & PO                      & \multicolumn{1}{c|}{0.7±.3 \circled{4}}    & \multicolumn{1}{c|}{NA}             & 18.5 ± 9.7 \circled{12}  & \multicolumn{1}{c|}{0.7±.3 \circled{3}}    & \multicolumn{1}{c|}{NA}             & 18.3±9.5 \circled{12}    & \multicolumn{1}{c|}{1.0±1.1 \circled{2}}   & \multicolumn{1}{c|}{NA}                & 21.1±32.5 \circled{12}   & \multicolumn{1}{c|}{3.6±3.9 \circled{3}}     & \multicolumn{1}{c|}{NA}               & 114.5 ± 131.2 \circled{11} \\ \hline
\end{tabular}
\label{tab:The sum of training time (Minutes) for all cycles across datasets and configurations 2}
}

\caption{
The average of prediction (ranking) time (in seconds) of DRL configurations for the first 10 cycles across PAINT, IOFROL, CODEC, and IMAG datasets.}
\centering
\resizebox{\textwidth}{!}{%
\begin{tabular}{|c|c|ccc|ccc|ccc|ccc|}
\hline
\multirow{2}{*}{}    & \multirow{2}{*}{RM}     & \multicolumn{3}{c|}{PAINT}                                                           & \multicolumn{3}{c|}{IOFROL}                                                             & \multicolumn{3}{c|}{CODEC}                                                               & \multicolumn{3}{c|}{IMAG}                                                                \\ \cline{3-14} 
                     &                         & \multicolumn{1}{c|}{SB}          & \multicolumn{1}{c|}{KR}           & TF            & \multicolumn{1}{c|}{SB}           & \multicolumn{1}{c|}{KR}           & TF              & \multicolumn{1}{c|}{SB}            & \multicolumn{1}{c|}{KR}            & TF             & \multicolumn{1}{c|}{SB}            & \multicolumn{1}{c|}{KR}            & TF             \\ \hline
DQN                  & PA                      & \multicolumn{1}{c|}{1.8±.2 \circled{7}} & \multicolumn{1}{c|}{0.1±.07 \circled{2}} & 8.1±.9 \circled{13}  & \multicolumn{1}{c|}{4.5±3.2 \circled{8}} & \multicolumn{1}{c|}{1.1±1.1 \circled{3}} & 16.6±11.5 \circled{12} & \multicolumn{1}{c|}{1.5±.09 \circled{7}}  & \multicolumn{1}{c|}{0.06±.03 \circled{2}} & 7.4±.8 \circled{9}    & \multicolumn{1}{c|}{1.5±.1 \circled{6}}   & \multicolumn{1}{c|}{0.05±.03 \circled{2}} & 7.7±.9 \circled{10}   \\ \hline
DDPG                 & PO                      & \multicolumn{1}{c|}{0.8±.3 \circled{3}} & \multicolumn{1}{c|}{0.07±.3 \circled{1}} & 8.1±.3 \circled{12}  & \multicolumn{1}{c|}{0.9±.1 \circled{2}}  & \multicolumn{1}{c|}{0.2±.2 \circled{1}}  & 10.3±2.0 \circled{11}  & \multicolumn{1}{c|}{0.7±.03 \circled{3}}  & \multicolumn{1}{c|}{0.05±.03 \circled{1}} & 9.4±.7 \circled{14}   & \multicolumn{1}{c|}{0.7±.1 \circled{3}}   & \multicolumn{1}{c|}{0.05±.02 \circled{1}} & 8.7±.8 \circled{14}   \\ \hline
\multirow{2}{*}{PPO} & \multicolumn{1}{l|}{PA} & \multicolumn{1}{c|}{1.1±.3 \circled{4}} & \multicolumn{1}{c|}{NA}           & 9.2±.8 \circled{15}  & \multicolumn{1}{c|}{2.2±1.3 \circled{6}} & \multicolumn{1}{c|}{NA}           & 67.4±45.2 \circled{14} & \multicolumn{1}{c|}{1.1±.0 \circled{4}}   & \multicolumn{1}{c|}{NA}            & 8.6±.7 \circled{13}   & \multicolumn{1}{c|}{1.1±.3 \circled{5}}   & \multicolumn{1}{c|}{NA}            & 8.3 ± .5 \circled{11} \\ \cline{2-14} 
                     & \multicolumn{1}{l|}{PO} & \multicolumn{1}{c|}{1.4±.5 \circled{5}} & \multicolumn{1}{c|}{NA}           & 8.9±1.3 \circled{14} & \multicolumn{1}{c|}{1.2±.2 \circled{4}}  & \multicolumn{1}{c|}{NA}           & 21.2±7.6 \circled{15}  & \multicolumn{1}{c|}{1.1±.1 \circled{5}}   & \multicolumn{1}{c|}{NA}            & 8.2±.4 \circled{10}   & \multicolumn{1}{c|}{1.1±.1 \circled{4}}   & \multicolumn{1}{c|}{NA}            & 8.3±1.0 \circled{13}  \\ \hline
A2C                  & PA                      & \multicolumn{1}{c|}{1.8±.5 \circled{8}} & \multicolumn{1}{c|}{NA}           & 7.9±1.1 \circled{10} & \multicolumn{1}{c|}{3.0±2.0 \circled{7}} & \multicolumn{1}{c|}{NA}           & 8.1±1.7 \circled{10}   & \multicolumn{1}{c|}{1.5 ± .3 \circled{6}} & \multicolumn{1}{c|}{NA}            & 8.4 ± .7 \circled{11} & \multicolumn{1}{c|}{1.6±.4 \circled{7}}   & \multicolumn{1}{c|}{NA}            & 8.3 ± .8 \circled{12} \\ \cline{2-14} 
                     & PO                      & \multicolumn{1}{c|}{1.5±.3 \circled{6}} & \multicolumn{1}{c|}{NA}           & 7.4±.6 \circled{9}   & \multicolumn{1}{c|}{1.3±.3 \circled{5}}  & \multicolumn{1}{c|}{NA}           & 7.8±1.3 \circled{9}    & \multicolumn{1}{c|}{1.7±.2 \circled{8}}   & \multicolumn{1}{c|}{NA}            & 8.4±.9 \circled{12}   & \multicolumn{1}{c|}{1.7 ± .5 \circled{8}} & \multicolumn{1}{c|}{NA}            & 7.7±.8 \circled{9}    \\ \hline
\end{tabular}
\label{tab:The average of prediction (ranking) time (Seconds) for all cycles across datasets and configurations. 1}
}
\caption{The average of prediction (ranking) time (in seconds) of DRL configurations for the first 10 cycles across IO, COMP, LANG, and MATH datasets.}
\centering
\resizebox{\textwidth}{!}{%
\begin{tabular}{|c|c|ccc|ccc|ccc|ccc|}
\hline
\multirow{2}{*}{}    & \multirow{2}{*}{RM}     & \multicolumn{3}{c|}{IO}                                                                     & \multicolumn{3}{c|}{COMP}                                                                   & \multicolumn{3}{c|}{LANG}                                                                  & \multicolumn{3}{c|}{MATH}                                                                  \\ \cline{3-14} 
                     &                         & \multicolumn{1}{c|}{SB}            & \multicolumn{1}{c|}{KR}              & TF              & \multicolumn{1}{c|}{SB}            & \multicolumn{1}{c|}{KR}              & TF              & \multicolumn{1}{c|}{SB}            & \multicolumn{1}{c|}{KR}             & TF              & \multicolumn{1}{c|}{SB}            & \multicolumn{1}{c|}{KR}            & TF               \\ \hline
DQN                  & PA                      & \multicolumn{1}{c|}{1.6 ± .1 \circled{6}} & \multicolumn{1}{c|}{0.07 ± .03 \circled{2}} & 7.8±.7 \circled{9}     & \multicolumn{1}{c|}{1.6 ± .1 \circled{7}} & \multicolumn{1}{c|}{0.07 ± .03 \circled{2}} & 8.3±.9 \circled{9}     & \multicolumn{1}{c|}{1.7 ± .2 \circled{8}} & \multicolumn{1}{c|}{1.4 ± 2.3 \circled{5}} & 8.5±1.6 \circled{12}   & \multicolumn{1}{c|}{2.2 ± .9 \circled{8}} & \multicolumn{1}{c|}{0.3 ± .3 \circled{2}} & 11.7±5.9 \circled{14}   \\ \hline
DDPG                 & PO                      & \multicolumn{1}{c|}{0.7±.1 \circled{3}}   & \multicolumn{1}{c|}{0.05±.02 \circled{1}}   & 8.9±.7 \circled{12}    & \multicolumn{1}{c|}{0.7±.1 \circled{3}}   & \multicolumn{1}{c|}{0.05±.03 \circled{1}}   & 9.8±.9 \circled{12}    & \multicolumn{1}{c|}{0.7±.1 \circled{2}}   & \multicolumn{1}{c|}{0.06±.03 \circled{1}}  & 9.6±.8 \circled{14}    & \multicolumn{1}{c|}{0.8±.3 \circled{3}}   & \multicolumn{1}{c|}{0.1±.07 \circled{1}}  & 9.6±1.1 \circled{12}    \\ \hline
\multirow{2}{*}{PPO} & \multicolumn{1}{l|}{PA} & \multicolumn{1}{c|}{1.1±.1 \circled{4}}   & \multicolumn{1}{c|}{NA}              & 10.2 ± .8 \circled{13} & \multicolumn{1}{c|}{1.1±.1 \circled{4}}   & \multicolumn{1}{c|}{NA}              & 10.6± .8 \circled{13}  & \multicolumn{1}{c|}{1.1±.1 \circled{3}}   & \multicolumn{1}{c|}{NA}             & 9.5 ± 1.1 \circled{13} & \multicolumn{1}{c|}{1.4±.4 \circled{5}}   & \multicolumn{1}{c|}{NA}            & 10.3 ± 2.0 \circled{13} \\ \cline{2-14} 
                     & \multicolumn{1}{l|}{PO} & \multicolumn{1}{c|}{1.1±.1 \circled{4}}   & \multicolumn{1}{c|}{NA}              & 8.7 ± .5 \circled{10}  & \multicolumn{1}{c|}{1.1±.1 \circled{4}}   & \multicolumn{1}{c|}{NA}              & 8.9 ± 1.0 \circled{11} & \multicolumn{1}{c|}{1.2±.1 \circled{4}}   & \multicolumn{1}{c|}{NA}             & 8.1 ± .6 \circled{10}  & \multicolumn{1}{c|}{1.1±.1 \circled{4}}   & \multicolumn{1}{c|}{NA}            & 8.9 ± 1.2 \circled{10}  \\ \hline
A2C                  & PA                      & \multicolumn{1}{c|}{1.4±.3 \circled{5}}   & \multicolumn{1}{c|}{NA}              & 8.8 ± .5 \circled{11}  & \multicolumn{1}{c|}{1.5± .3 \circled{6}}  & \multicolumn{1}{c|}{NA}              & 8.4 ± .4 \circled{10}  & \multicolumn{1}{c|}{1.6±.4 \circled{7}}   & \multicolumn{1}{c|}{NA}             & 8.4 ± 1.2 \circled{11} & \multicolumn{1}{c|}{2.0±.9 \circled{7}}   & \multicolumn{1}{c|}{NA}            & 9.0 ± 1.4 \circled{11}  \\ \cline{2-14} 
                     & PO                      & \multicolumn{1}{c|}{1.6±.3  \circled{7}}  & \multicolumn{1}{c|}{NA}              & 7.6 ± .8 \circled{8}   & \multicolumn{1}{c|}{1.4±.3 \circled{5}}   & \multicolumn{1}{c|}{NA}              & 7.3±.7 \circled{8}     & \multicolumn{1}{c|}{1.5±.3 \circled{6}}   & \multicolumn{1}{c|}{NA}             & 7.7±.6 \circled{9}     & \multicolumn{1}{c|}{1.6±.5 \circled{6}}   & \multicolumn{1}{c|}{NA}            & 8.3 ± .9 \circled{9}    \\ \hline
\end{tabular}
\label{tab:The average of prediction (ranking) time (Seconds) for all cycles across datasets and configurations. 2}
}
\end{sidewaystable}
\hfill \break
\textbf{RQ1:} As shown in Table \ref{tab:apfdandnrpa}, pairwise configurations perform best across Stable-baselines's algorithms. Pairwise-A2C-SB yields the best averages. Based on the post-hoc test, Pairwise-A2C-SB performs best across all datasets. Similarly, the Stable-baselines framework performs best regarding the pointwise ranking model. While Pairwise-A2C-SB has the best performance overall, Tensorforce has good performance on IOFROL dataset when implementing Pairwise-A2C configuration. IOFROL is a simple dataset with a high number of execution logs. When using this dataset, the training time of the DRL agent is long, which might explain why Tensorforce configurations perform well. Specifically, despite the high number of execution logs of the IOFROL dataset, Tensorforce still has good performance.

To show the importance of selecting the best DRL configuration, we measured the effect size of the differences between pairs of configurations based on CLES. As shown in Table \ref{tab:Common Language Effect Size}, the CLES values among one of the worst and best cases for the six enriched datasets are over 80\%, whereas they are 66\% and 71\% for the simple Paint-Control and IOFROL datasets, respectively. These results show that, for each dataset, we have, with high probability, a DRL configuration that has adequately learned a ranking strategy.

\begin{table}[t]
\caption{Common Language Effect Size between one of the worst and best configurations for each dataset based on accuracy.}
\label{tab:Common Language Effect Size}
\centering
\begin{tabular}{cccc}
Dataset      & Best Conf.                   & Worst Cons. & CLES    \\ \hline
Paint-Control & PAIRWISE-A2C-SB                      & POINTWISE-A2C-TF         & .66 \\
IOFROL        &  PAIRWISE-A2C-TF  &     PAIRWISE-DQN-KR      & .71 \\
Codec         &    PAIRWISE-A2C-SB                  &     POINTWISE-DDPG-KR     &  .97\\
Compress      &      PAIRWISE-A2C-SB               &   POINTWISE-DDPG-KR       &  \textbf{.99} \\
Imaging       &       PAIRWISE-DQN-SB         &    POINTWISE-DDPG-KR     &  .98  \\
IO            &  PAIRWISE-A2C-SB             &  POINTWISE-DDPG-KR       &  \textbf{.99}   \\
Lang          &         PAIRWISE-A2C-SB              &    PAIRWISE-A2C-TF     & .81 \\
Math          & PAIRWISE-A2C-SB  &    PAIRWISE-DQN-TF       & .81
\end{tabular}
\end{table}

\begin{figure*}[t]
  \begin{minipage}{0.5\textwidth}
        \includegraphics[width=\textwidth]{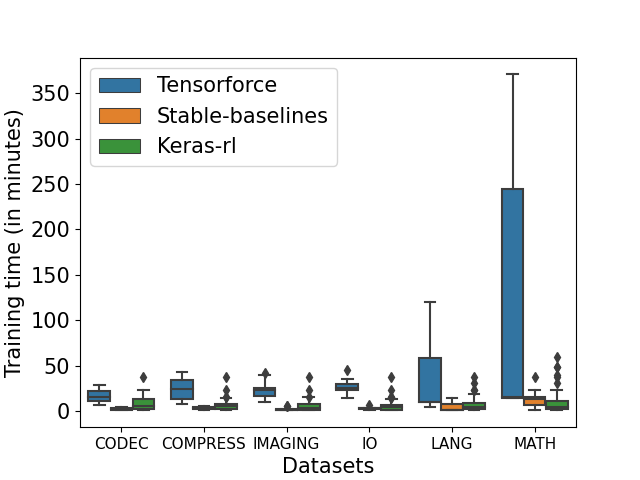}
        \caption{Training time of Pairwise-DQN \\ configuration accross DRL frameworks for \\ enriched datasets.}
        \label{fig:boxplot_DQN_train}
    \end{minipage}%
    \hfill
        \begin{minipage}{0.5\textwidth}
        \centering
        \includegraphics[width=\textwidth]{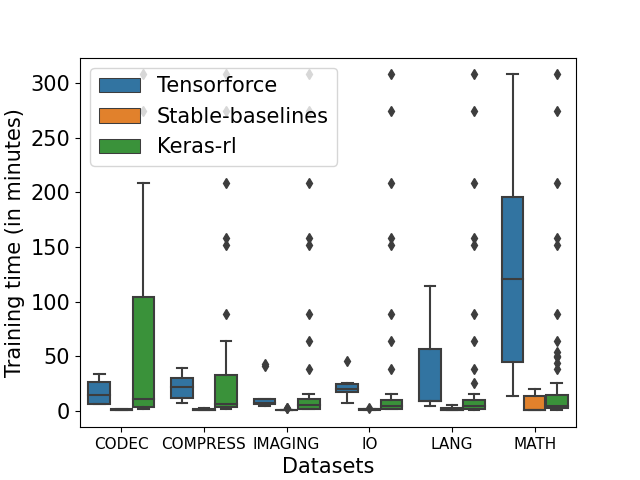}
        \caption{Training time  Pointwise-DDPG \\ configuration  accross DRL frameworks for \\ enriched datasets.}
        \label{fig:boxplot_DDPG_train}
            \end{minipage}
    \end{figure*}

In terms of training time, as shown in Tables \ref{tab:The sum of training time (Minutes) for all cycles across datasets and configurations 1} and \ref{tab:The sum of training time (Minutes) for all cycles across datasets and configurations 2}, both pairwise and pointwise configurations perform well for some datasets/frameworks. Figures \ref{fig:boxplot_DQN_train} and \ref{fig:boxplot_DDPG_train}  show the statistical analysis of the training time involving Pairwise-DQN and Pointwise-DDPG configurations, respectively.
The results show that Pointwise-DDPG-SB performs best followed by Pointwise-DDPG-KR. Regarding the DQN configurations, similarly, the Stable-baselines framework performs best.
It is worth mentioning that, since DRL agents are trained offline, the training time does not add any delay to the CI build process. 

In terms of prediction time, as shown in Table \ref{tab:The average of prediction (ranking) time (Seconds) for all cycles across datasets and configurations. 1} and \ref{tab:The average of prediction (ranking) time (Seconds) for all cycles across datasets and configurations. 2}, similar to the training time, both configurations (pairwise or pointwise) perform well for some of the datasets/frameworks. Based on the post-hoc test, Pairwise-DQN-SB performs best on average followed by Pairwise-DQN-KR. The prediction time among pointwise and pairwise configurations goes up to 11 seconds, notably for Pairwise-DQN-TF, which is non-negligible for CI builds. 

The last three rows of Table \ref{tab:apfdandnrpa} and \ref{tab:apfdandnrpa2} show the averages and standard deviations of baselines configurations in terms of NRPA and APFD values collected from \cite{bagherzadeh2021reinforcement}, for the datasets on which they were originally experimented. Tables \ref{tab:Pairwise-A2C-SB and Baselines}, \ref{tab:Pairwise-DQN-KR and Baselines}, and \ref{tab:Pairwise-A2C-TF and Baselines} show the results of CLES between the best configuration of each framework and selected baselines for all datasets, to assess the effect size of differences. 
\begin{table}[t]
\caption{Common Language Effect Size between Pairwise-A2C-SB and selected baselines.}
\label{tab:Pairwise-A2C-SB and Baselines}
\centering
\begin{tabular}{|c|c|c|c|}
\hline
\multirow{2}{*}{} & RL-BS1        & RL-BS2        & MART          \\ \cline{2-4} 
                  & CLES           & CLES           & CLES           \\ \hline
IO                & NA            & .792          & .762          \\ \hline
CODEC             & NA            & .743          & \textbf{.857} \\ \hline
IMAG              & NA            & .717          & .724          \\ \hline
COMP              & NA            & \textbf{.910} & .639          \\ \hline
LANG              & NA            & .766          & .699          \\ \hline
MATH              & NA            & .773          & .588          \\ \hline
PAINT.            & \textbf{.607} & NA            & NA            \\ \hline
IOFROL            & .344          & NA            & NA            \\ \hline
\end{tabular}
\end{table}
\begin{table}[t]
\caption{Common Language Effect Size between Pairwise-DQN-KR and selected baselines.}
\label{tab:Pairwise-DQN-KR and Baselines}
\centering
\begin{tabular}{|c|c|c|c|}
\hline
\multirow{2}{*}{} & RL-BS1 & RL-BS2        & MART          \\ \cline{2-4} 
                  & CLES    & CLES           & CLES           \\ \hline
IO                & NA     & .238          & .627          \\ \hline
CODEC             & NA     & .244          & \textbf{.761} \\ \hline
IMAG              & NA     & .242          & .599          \\ \hline
COMP              & NA     & \textbf{.348} & .458          \\ \hline
LANG              & NA     & .321          & .495          \\ \hline
MATH              & NA     & .272          & .568          \\ \hline
PAINT.            & \textbf{.282}   & NA            & NA            \\ \hline
IOFROL            &   .268     & NA            & NA            \\ \hline
\end{tabular}
\end{table}
\begin{table}[t]
\caption{Common Language Effect Size between Pairwise-A2C-TF and selected baselines.}
\label{tab:Pairwise-A2C-TF and Baselines}
\centering
\begin{tabular}{|c|c|c|c|}
\hline
\multirow{2}{*}{} & RL-BS1 & RL-BS2        & MART          \\ \cline{2-4} 
                  & CLES    & CLES           & CLES           \\ \hline
IO                & NA     & .238          & .619          \\ \hline
CODEC             & NA     & .237          & \textbf{.766} \\ \hline
IMAG              & NA     & .224          & .583          \\ \hline
COMP              & NA     & \textbf{.335} & .451          \\ \hline
LANG              & NA     &     .264          &  .451             \\ \hline
MATH              & NA     & .163          & .429          \\ \hline
PAINT.            &  .409      & NA            & NA            \\ \hline
IOFROL            &  \textbf{.559}      & NA            & NA            \\ \hline
\end{tabular}
\end{table}

The row RL-BS1 in Table \ref{tab:apfdandnrpa} and \ref{tab:apfdandnrpa2} shows the results of an RL-based solution reported by Bagherzadeh et al. \cite{bagherzadeh2021reinforcement}. For the Paint-Control dataset, Pairwise-A2C-SB fares slightly better than RL-BS1  with a CLES of $60.2$. Also, both solutions (RL-BS1, Pairwise-A2C-SB) are close to the optimal ranking (the row labeled as “Optimal” in Table \ref{tab:apfdandnrpa} and \ref{tab:apfdandnrpa2}). For dataset IOFROL, RL-BS1 performs better than Pairwise-A2C-SB: however, both solutions do not perform well as their values are lower than the optimal ranking. RL-BS1 performs better than  Pairwise-DQN-KR for both simple datasets. Moreover, RL-BS1 and Pairwise-A2C-TF perform equivalently on the IOFROL dataset. This is justified with CLES values reported in Tables \ref{tab:Pairwise-DQN-KR and Baselines} and \ref{tab:Pairwise-A2C-TF and Baselines}. These results are anyway lower than the optimal ranking. As pointed out by Bagherzadeh et al. \cite{bagherzadeh2021reinforcement}, the test execution history provided by simple datasets is not sufficient enough to learn an accurate test prioritization policy. 

The row of RL-BS2 in Tables \ref{tab:apfdandnrpa} and \ref{tab:apfdandnrpa2} shows the results of an RL-based solution reported by Bagherzadeh et al. \cite{bagherzadeh2021reinforcement}. For all datasets, Pairwise-A2C-SB fares significantly better than RL-BS2 with CLES values between $71.7$ and $91.0$ as shown in Table \ref{tab:Pairwise-A2C-SB and Baselines}.  In contrast, RL-BS2 performs better than Pairwise-A2C-TF and Pairwise-DQN-KR for all datasets: CLES values between Pairwise-A2C-TF and RL-BS2 range between $16.3$ and $33.5$, and between $23.8$ and $34.8$ for Pairwise-DQN-KR and RL-BS2. Thus, according to these results, Pairwise-A2C-SB improves the baselines in the use of DRL for test case prioritization.

The row labeled by MART (MART ranking model) in Table \ref{tab:apfdandnrpa} and \ref{tab:apfdandnrpa2} provides the results of the best ML-based solution reported by Bagherzadeh et al. \cite{bagherzadeh2021reinforcement}. For  MATH dataset, Pairwise-A2C-SB performs equivalently as MART. We observe $58.8$ as CLES value for  MATH dataset. For other datasets, Pairwise-A2C-SB fares better than MART. The CLES of Pairwise-A2C-SB vs. MART ranges between $58.8$ to $85.7$ with an average of $0.711$, i.e., in $71.1\%$ of the cycles, Pairwise-A2C-SB fares better than MART. Then, we can conclude that Pairwise-A2C-SB advances state-of-the-art compared to the best ML-based ranking technique (MART). Pairwise-A2C-TF and Pairwise-DQN-KR solutions perform similarly  to MART with $0.549$ and $0.584$ CLES averages respectively.

\begin{tcolorbox}
\textbf{Finding 2: The performance of DQN algorithms is close to the A2C algorithms in all evaluated DRL frameworks when applying to the Pairwise ranking model.}
\end{tcolorbox}

\begin{figure*}[t]
  \begin{minipage}{0.5\textwidth}
        \includegraphics[width=\textwidth]{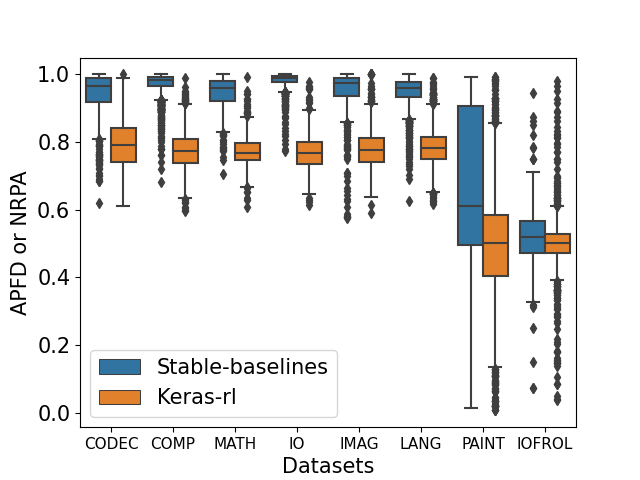}
    \end{minipage}%
    \hfill
        \begin{minipage}{0.5\textwidth}
        \centering
        \includegraphics[width=\textwidth]{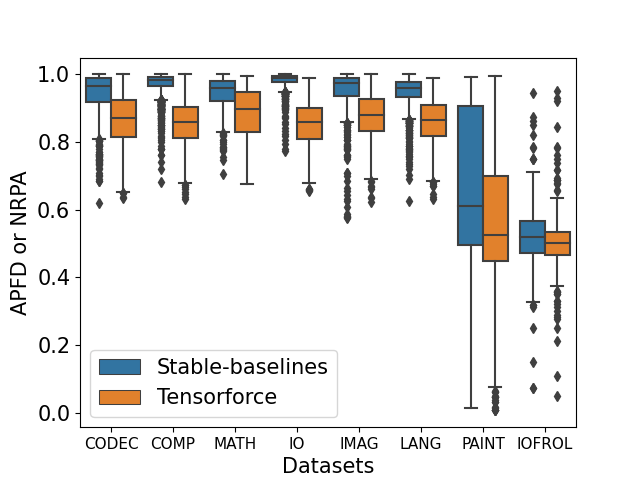}
            \end{minipage}%
            
               \caption{APFD (simple datasets) or NRPA (enriched datasets) of DQN-PAIRWISE configuration accross DRL frameworks for all datasets: Stable-baselines vs. Keras-rl (left) and Stable-baselines vs. Tensorforce (right).}
               \label{fig:boxplot_DQN_APFD_SB_KR_TF}
    \end{figure*}

\begin{tcolorbox}
    Summary 3: When it comes to the test case prioritization problem, where we have to take into account the ranking function and the size of the dataset, Stable-baselines leads in terms of accuracy followed by Tensorforce. In terms of inference time, Keras-rl provides the best results. Keras-rl with its DQN implementation is the lightest one in terms of implementation, hyperparameters it provides, and APIs it provides to evaluate the DRL agent.
\end{tcolorbox}
\hfill \break 
\textbf{RQ2:} Figure \ref{fig:boxplot_DQN_APFD_SB_KR_TF} shows the statistical results of APFD and NRPA metrics for the Pairwise-DQN configuration.

\begin{figure*}[t]
  \begin{minipage}{0.5\textwidth}
        \includegraphics[width=\textwidth]{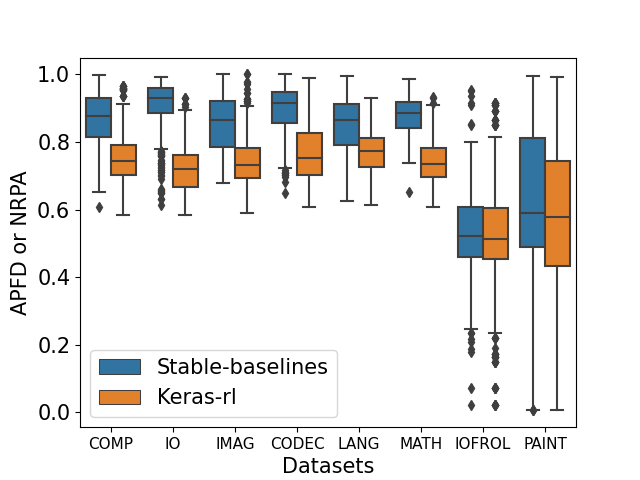}
    \end{minipage}%
    \hfill
        \begin{minipage}{0.5\textwidth}
        \centering
        \includegraphics[width=\textwidth]{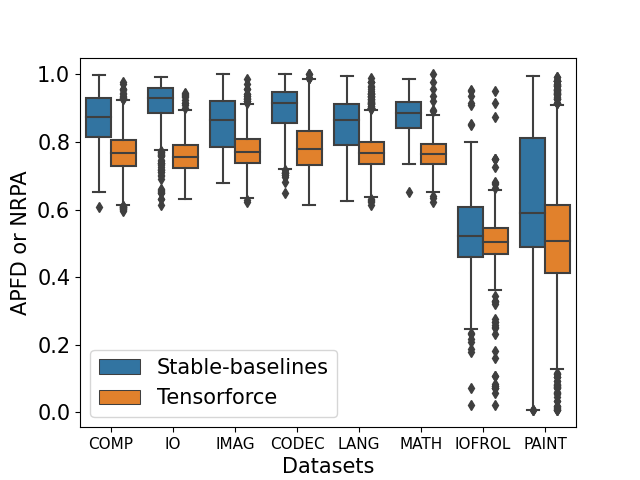}
            \end{minipage}
               \caption{APFD (simple datasets) or NRPA (enriched datasets) of DDPG-POINTWISE configuration accross DRL frameworks for all datasets: Stable-baselines vs. Keras-rl (left) and Stable-baselines vs. Tensorforce (right).}
               \label{fig:boxplot_DDPG_SB_TF_KR}
    \end{figure*}

The results show that the Stable-baselines framework performs better for all enriched datasets. Similarly, Figure \ref{fig:boxplot_DDPG_SB_TF_KR} shows the statistical results of APFD and NRPA metrics regarding the DDPG-Pointwise configuration.

According to the reported results, Stable-baselines perform best. Moreover, to analyze the accuracy of DRL algorithms w.r.t the relative performance, we performed two sets of Welch's ANOVA and Games-Howell post-hoc tests corresponding to the pairwise and pointwise ranking models, based on the result of all algorithms across datasets. Tables \ref{tab:Welch ANOVA and Games-Howell post-hoc tests regarding the ranking model} and \ref{tab:Welch ANOVA and Games-Howell post-hoc tests regarding the ranking model  for simple datasets}  show for each configuration the calculated mean, p-value and CLES.  
\begin{table}[t]
\caption{Results of Welch ANOVA and Games-Howell post-hoc tests on pairwise and pointwise ranking models for enriched datasets (in bold are DRL configurations where p-value is $<$ 0.05 and have greater performance w.r.t the effect size ).}
\label{tab:Welch ANOVA and Games-Howell post-hoc tests regarding the ranking model}
\centering
\resizebox{\textwidth}{!}{%
\begin{tabular}{c|c|c|c|c|c|c|}
\cline{2-7}
                                                                                                                     & A                & B                & mean(A)  & mean(B)  & pval     & CLES     \\ \hline
\multicolumn{1}{|c|}{\multirow{21}{*}{\begin{tabular}[c]{@{}c@{}}Pairwise and \\  enriched datasets\end{tabular}}}   & \textbf{A2C-SB}  & A2C-TF           & 9,68E-01 & 7,50E-01 & 4,28E-12 & 9,87E-01 \\ \cline{2-7} 
\multicolumn{1}{|c|}{}                                                                                               & \textbf{A2C-SB}  & DQN-KR           & 9,68E-01 & 7,80E-01 & 1,34E-12 & 9,86E-01 \\ \cline{2-7} 
\multicolumn{1}{|c|}{}                                                                                               & \textbf{A2C-SB}  & DQN-SB           & 9,68E-01 & 9,58E-01 & 0,00E+00 & 5,63E-01 \\ \cline{2-7} 
\multicolumn{1}{|c|}{}                                                                                               & \textbf{A2C-SB}  & DQN-TF           & 9,68E-01 & 8,42E-01 & 2,58E-12 & 9,31E-01 \\ \cline{2-7} 
\multicolumn{1}{|c|}{}                                                                                               & \textbf{A2C-SB}  & PPO-SB           & 9,68E-01 & 9,64E-01 & 1,25E-06 & 5,36E-01 \\ \cline{2-7} 
\multicolumn{1}{|c|}{}                                                                                               & \textbf{A2C-SB}  & PPO-TF           & 9,68E-01 & 7,79E-01 & 0,00E+00 & 9,84E-01 \\ \cline{2-7} 
\multicolumn{1}{|c|}{}                                                                                               & A2C-TF           & \textbf{DQN-KR}  & 7,50E-01 & 7,80E-01 & 0,00E+00 & 4,55E-01 \\ \cline{2-7} 
\multicolumn{1}{|c|}{}                                                                                               & A2C-TF           & \textbf{DQN-SB}  & 7,50E-01 & 9,58E-01 & 0,00E+00 & 2,31E-02 \\ \cline{2-7} 
\multicolumn{1}{|c|}{}                                                                                               & A2C-TF           & \textbf{DQN-TF}  & 7,50E-01 & 8,42E-01 & 0,00E+00 & 1,84E-01 \\ \cline{2-7} 
\multicolumn{1}{|c|}{}                                                                                               & A2C-TF           & \textbf{PPO-SB}  & 7,50E-01 & 9,64E-01 & 0,00E+00 & 1,46E-02 \\ \cline{2-7} 
\multicolumn{1}{|c|}{}                                                                                               & A2C-TF           & \textbf{PPO-TF}  & 7,50E-01 & 7,79E-01 & 0,00E+00 & 4,63E-01 \\ \cline{2-7} 
\multicolumn{1}{|c|}{}                                                                                               & DQN-KR           & \textbf{DQN-SB}  & 7,80E-01 & 9,58E-01 & 0,00E+00 & 2,49E-02 \\ \cline{2-7} 
\multicolumn{1}{|c|}{}                                                                                               & DQN-KR           & \textbf{DQN-TF}  & 7,80E-01 & 8,42E-01 & 0,00E+00 & 1,99E-01 \\ \cline{2-7} 
\multicolumn{1}{|c|}{}                                                                                               & DQN-KR           & \textbf{PPO-SB}  & 7,80E-01 & 9,64E-01 & 0,00E+00 & 1,57E-02 \\ \cline{2-7} 
\multicolumn{1}{|c|}{}                                                                                               & \textbf{DQN-KR}  & PPO-TF           & 7,80E-01 & 7,79E-01 & 1,00E+00 & 5,08E-01 \\ \cline{2-7} 
\multicolumn{1}{|c|}{}                                                                                               & \textbf{DQN-SB}  & DQN-TF           & 9,58E-01 & 8,42E-01 & 0,00E+00 & 8,98E-01 \\ \cline{2-7} 
\multicolumn{1}{|c|}{}                                                                                               & DQN-SB           & \textbf{PPO-SB}  & 9,58E-01 & 9,64E-01 & 0,00E+00 & 4,72E-01 \\ \cline{2-7} 
\multicolumn{1}{|c|}{}                                                                                               & \textbf{DQN-SB}  & PPO-TF           & 9,58E-01 & 7,79E-01 & 0,00E+00 & 9,72E-01 \\ \cline{2-7} 
\multicolumn{1}{|c|}{}                                                                                               & DQN-TF           & \textbf{PPO-SB}  & 8,42E-01 & 9,64E-01 & 0,00E+00 & 7,94E-02 \\ \cline{2-7} 
\multicolumn{1}{|c|}{}                                                                                               & \textbf{DQN-TF}  & PPO-TF           & 8,42E-01 & 7,79E-01 & 0,00E+00 & 7,96E-01 \\ \cline{2-7} 
\multicolumn{1}{|c|}{}                                                                                               & \textbf{PPO-SB}  & PPO-TF           & 9,64E-01 & 7,79E-01 & 3,30E-13 & 9,81E-01 \\ \hline
\multicolumn{1}{|c|}{\multirow{21}{*}{\begin{tabular}[c]{@{}c@{}}Pointwise and  \\  enriched datasets\end{tabular}}} & \textbf{A2C-SB}  & A2C-TF           & 9,68E-01 & 8,31E-01 & 3,68E-12 & 9,52E-01 \\ \cline{2-7} 
\multicolumn{1}{|c|}{}                                                                                               & \textbf{A2C-SB}  & DDPG-KR          & 9,68E-01 & 7,50E-01 & 3,85E-12 & 9,90E-01 \\ \cline{2-7} 
\multicolumn{1}{|c|}{}                                                                                               & \textbf{A2C-SB}  & DDPG-SB          & 9,68E-01 & 8,45E-01 & 0,00E+00 & 8,97E-01 \\ \cline{2-7} 
\multicolumn{1}{|c|}{}                                                                                               & \textbf{A2C-SB}  & DDPG-TF          & 9,68E-01 & 7,54E-01 & 0,00E+00 & 9,89E-01 \\ \cline{2-7} 
\multicolumn{1}{|c|}{}                                                                                               & \textbf{A2C-SB}  & PPO-SB           & 9,68E-01 & 9,14E-01 & 0,00E+00 & 8,22E-01 \\ \cline{2-7} 
\multicolumn{1}{|c|}{}                                                                                               & \textbf{A2C-SB}  & PPO-TF           & 9,68E-01 & 7,87E-01 & 0,00E+00 & 9,81E-01 \\ \cline{2-7} 
\multicolumn{1}{|c|}{}                                                                                               & \textbf{A2C-TF}  & DDPG-KR          & 8,31E-01 & 7,50E-01 & 2,19E-12 & 8,06E-01 \\ \cline{2-7} 
\multicolumn{1}{|c|}{}                                                                                               & A2C-TF           & \textbf{DDPG-SB} & 8,31E-01 & 8,45E-01 & 3,24E-03 & 4,09E-01 \\ \cline{2-7} 
\multicolumn{1}{|c|}{}                                                                                               & \textbf{A2C-TF}  & DDPG-TF          & 8,31E-01 & 7,54E-01 & 0,00E+00 & 7,75E-01 \\ \cline{2-7} 
\multicolumn{1}{|c|}{}                                                                                               & A2C-TF           & \textbf{PPO-SB}  & 8,31E-01 & 9,14E-01 & 0,00E+00 & 2,21E-01 \\ \cline{2-7} 
\multicolumn{1}{|c|}{}                                                                                               & \textbf{A2C-TF}  & PPO-TF           & 8,31E-01 & 7,87E-01 & 0,00E+00 & 7,06E-01 \\ \cline{2-7} 
\multicolumn{1}{|c|}{}                                                                                               & DDPG-KR          & \textbf{DDPG-SB} & 7,50E-01 & 8,45E-01 & 0,00E+00 & 1,64E-01 \\ \cline{2-7} 
\multicolumn{1}{|c|}{}                                                                                               & DDPG-KR          & \textbf{DDPG-TF} & 7,50E-01 & 7,54E-01 & 4,58E-01 & 4,24E-01 \\ \cline{2-7} 
\multicolumn{1}{|c|}{}                                                                                               & DDPG-KR          & \textbf{PPO-SB}  & 7,50E-01 & 9,14E-01 & 3,58E-12 & 5,01E-02 \\ \cline{2-7} 
\multicolumn{1}{|c|}{}                                                                                               & DDPG-KR          & \textbf{PPO-TF}  & 7,50E-01 & 7,87E-01 & 1,78E-12 & 3,63E-01 \\ \cline{2-7} 
\multicolumn{1}{|c|}{}                                                                                               & \textbf{DDPG-SB} & DDPG-TF          & 8,45E-01 & 7,54E-01 & 3,75E-14 & 8,12E-01 \\ \cline{2-7} 
\multicolumn{1}{|c|}{}                                                                                               & DDPG-SB          & \textbf{PPO-SB}  & 8,45E-01 & 9,14E-01 & 0,00E+00 & 3,29E-01 \\ \cline{2-7} 
\multicolumn{1}{|c|}{}                                                                                               & \textbf{DDPG-SB} & PPO-TF           & 8,45E-01 & 7,87E-01 & 4,16E-13 & 7,58E-01 \\ \cline{2-7} 
\multicolumn{1}{|c|}{}                                                                                               & DDPG-TF          & \textbf{PPO-SB}  & 7,54E-01 & 9,14E-01 & 0,00E+00 & 5,83E-02 \\ \cline{2-7} 
\multicolumn{1}{|c|}{}                                                                                               & DDPG-TF          & \textbf{PPO-TF}  & 7,54E-01 & 7,87E-01 & 0,00E+00 & 4,28E-01 \\ \cline{2-7} 
\multicolumn{1}{|c|}{}                                                                                               & \textbf{PPO-SB}  & PPO-TF           & 9,14E-01 & 7,87E-01 & 0,00E+00 & 9,05E-01 \\ \hline
\end{tabular}

}
\begin{tablenotes}
     \item mean(A) and mean(B) refer to NRPA values.

 \end{tablenotes}
\end{table}
\begin{table}[t]
\caption{Results of Welch ANOVA and Games-Howell post-hoc tests on pairwise and pointwise ranking models for simple datasets (in bold are DRL configurations where p-value is $<$ 0.05 and have greater performance w.r.t the effect size ).}
\label{tab:Welch ANOVA and Games-Howell post-hoc tests regarding the ranking model  for simple datasets}
\centering
\resizebox{\textwidth}{!}{%
\begin{tabular}{c|c|c|c|c|c|c|}
\cline{2-7}
                                                                                                                  & A                & B                & mean(A)  & mean(B)  & pval     & CLES     \\ \hline
\multicolumn{1}{|c|}{\multirow{21}{*}{\begin{tabular}[c]{@{}c@{}}Pairwise and \\   simple datasets\end{tabular}}} & A2C-SB           & \textbf{A2C-TF}  & 5.44E-01 & 6.00E-01 & 1.17E-12 & 4.26E-01 \\ \cline{2-7} 
\multicolumn{1}{|c|}{}                                                                                            & \textbf{A2C-SB}  & DQN-KR           & 5.44E-01 & 4.93E-01 & 0.00E+00 & 5.98E-01 \\ \cline{2-7} 
\multicolumn{1}{|c|}{}                                                                                            & A2C-SB           & \textbf{DQN-SB}  & 5.44E-01 & 6.06E-01 & 8.85E-13 & 4.16E-01 \\ \cline{2-7} 
\multicolumn{1}{|c|}{}                                                                                            & A2C-SB           & \textbf{DQN-TF}  & 5.44E-01 & 5.50E-01 & 9.79E-01 & 4.91E-01 \\ \cline{2-7} 
\multicolumn{1}{|c|}{}                                                                                            & A2C-SB           & \textbf{PPO-SB}  & 5.44E-01 & 6.35E-01 & 8.76E-13 & 3.56E-01 \\ \cline{2-7} 
\multicolumn{1}{|c|}{}                                                                                            & \textbf{A2C-SB}  & PPO-TF           & 5.44E-01 & 5.00E-01 & 1.51E-12 & 5.78E-01 \\ \cline{2-7} 
\multicolumn{1}{|c|}{}                                                                                            & \textbf{A2C-TF}  & DQN-KR           & 6.00E-01 & 4.93E-01 & 3.73E-13 & 6.33E-01 \\ \cline{2-7} 
\multicolumn{1}{|c|}{}                                                                                            & \textbf{A2C-TF}  & DQN-SB           & 6.00E-01 & 6.06E-01 & 9.96E-01 & 4.93E-01 \\ \cline{2-7} 
\multicolumn{1}{|c|}{}                                                                                            & \textbf{A2C-TF}  & DQN-TF           & 6.00E-01 & 5.50E-01 & 1.28E-07 & 5.61E-01 \\ \cline{2-7} 
\multicolumn{1}{|c|}{}                                                                                            & A2C-TF           & \textbf{PPO-SB}  & 6.00E-01 & 6.35E-01 & 1.19E-04 & 4.60E-01 \\ \cline{2-7} 
\multicolumn{1}{|c|}{}                                                                                            & \textbf{A2C-TF}  & PPO-TF           & 6.00E-01 & 5.00E-01 & 0.00E+00 & 6.21E-01 \\ \cline{2-7} 
\multicolumn{1}{|c|}{}                                                                                            & DQN-KR           & \textbf{DQN-SB}  & 4.93E-01 & 6.06E-01 & 0.00E+00 & 3.55E-01 \\ \cline{2-7} 
\multicolumn{1}{|c|}{}                                                                                            & DQN-KR           & \textbf{DQN-TF}  & 4.93E-01 & 5.50E-01 & 0.00E+00 & 4.18E-01 \\ \cline{2-7} 
\multicolumn{1}{|c|}{}                                                                                            & DQN-KR           & \textbf{PPO-SB}  & 4.93E-01 & 6.35E-01 & 2.30E-12 & 2.96E-01 \\ \cline{2-7} 
\multicolumn{1}{|c|}{}                                                                                            & DQN-KR           & \textbf{PPO-TF}  & 4.93E-01 & 5.00E-01 & 8.17E-01 & 4.88E-01 \\ \cline{2-7} 
\multicolumn{1}{|c|}{}                                                                                            & \textbf{DQN-SB}  & DQN-TF           & 6.06E-01 & 5.50E-01 & 1.85E-07 & 5.71E-01 \\ \cline{2-7} 
\multicolumn{1}{|c|}{}                                                                                            & DQN-SB           & \textbf{PPO-SB}  & 6.06E-01 & 6.35E-01 & 2.29E-02 & 4.65E-01 \\ \cline{2-7} 
\multicolumn{1}{|c|}{}                                                                                            & \textbf{DQN-SB}  & PPO-TF           & 6.06E-01 & 5.00E-01 & 4.31E-13 & 6.34E-01 \\ \cline{2-7} 
\multicolumn{1}{|c|}{}                                                                                            & DQN-TF           & \textbf{PPO-SB}  & 5.50E-01 & 6.35E-01 & 5.67E-13 & 3.88E-01 \\ \cline{2-7} 
\multicolumn{1}{|c|}{}                                                                                            & \textbf{DQN-TF}  & PPO-TF           & 5.50E-01 & 5.00E-01 & 1.22E-10 & 5.70E-01 \\ \cline{2-7} 
\multicolumn{1}{|c|}{}                                                                                            & \textbf{PPO-SB}  & PPO-TF           & 6.35E-01 & 5.00E-01 & 2.32E-12 & 6.87E-01 \\ \hline
\multicolumn{1}{|c|}{\multirow{21}{*}{\begin{tabular}[c]{@{}c@{}}Pointwise and  \\ simple datasets\end{tabular}}} & \textbf{A2C-SB}  & A2C-TF           & 5.51E-01 & 4.60E-01 & 1.70E-12 & 6.34E-01 \\ \cline{2-7} 
\multicolumn{1}{|c|}{}                                                                                            & \textbf{A2C-SB}  & DDPG-KR          & 5.51E-01 & 5.39E-01 & 4.32E-01 & 5.16E-01 \\ \cline{2-7} 
\multicolumn{1}{|c|}{}                                                                                            & A2C-SB           & \textbf{DDPG-SB} & 5.51E-01 & 5.57E-01 & 9.95E-01 & 4.92E-01 \\ \cline{2-7} 
\multicolumn{1}{|c|}{}                                                                                            & \textbf{A2C-SB}  & DDPG-TF          & 5.51E-01 & 5.14E-01 & 1.05E-06 & 5.56E-01 \\ \cline{2-7} 
\multicolumn{1}{|c|}{}                                                                                            & A2C-SB           & \textbf{PPO-SB}  & 5.51E-01 & 5.53E-01 & 1.00E+00 & 4.97E-01 \\ \cline{2-7} 
\multicolumn{1}{|c|}{}                                                                                            & \textbf{A2C-SB}  & PPO-TF           & 5.51E-01 & 4.92E-01 & 1.87E-12 & 5.88E-01 \\ \cline{2-7} 
\multicolumn{1}{|c|}{}                                                                                            & A2C-TF           & \textbf{DDPG-KR} & 4.60E-01 & 5.39E-01 & 2.17E-12 & 3.88E-01 \\ \cline{2-7} 
\multicolumn{1}{|c|}{}                                                                                            & A2C-TF           & \textbf{DDPG-SB} & 4.60E-01 & 5.57E-01 & 0.00E+00 & 3.83E-01 \\ \cline{2-7} 
\multicolumn{1}{|c|}{}                                                                                            & A2C-TF           & \textbf{DDPG-TF} & 4.60E-01 & 5.14E-01 & 1.38E-14 & 4.16E-01 \\ \cline{2-7} 
\multicolumn{1}{|c|}{}                                                                                            & A2C-TF           & \textbf{PPO-SB}  & 4.60E-01 & 5.53E-01 & 1.83E-12 & 3.64E-01 \\ \cline{2-7} 
\multicolumn{1}{|c|}{}                                                                                            & A2C-TF           & \textbf{PPO-TF}  & 4.60E-01 & 4.92E-01 & 2.12E-06 & 4.48E-01 \\ \cline{2-7} 
\multicolumn{1}{|c|}{}                                                                                            & DDPG-KR          & \textbf{DDPG-SB} & 5.39E-01 & 5.57E-01 & 4.93E-01 & 4.78E-01 \\ \cline{2-7} 
\multicolumn{1}{|c|}{}                                                                                            & \textbf{DDPG-KR} & DDPG-TF          & 5.39E-01 & 5.14E-01 & 6.49E-03 & 5.36E-01 \\ \cline{2-7} 
\multicolumn{1}{|c|}{}                                                                                            & DDPG-KR          & \textbf{PPO-SB}  & 5.39E-01 & 5.53E-01 & 2.45E-01 & 4.81E-01 \\ \cline{2-7} 
\multicolumn{1}{|c|}{}                                                                                            & \textbf{DDPG-KR} & PPO-TF           & 5.39E-01 & 4.92E-01 & 1.38E-11 & 5.67E-01 \\ \cline{2-7} 
\multicolumn{1}{|c|}{}                                                                                            & \textbf{DDPG-SB} & DDPG-TF          & 5.57E-01 & 5.14E-01 & 4.78E-04 & 5.55E-01 \\ \cline{2-7} 
\multicolumn{1}{|c|}{}                                                                                            & \textbf{DDPG-SB} & PPO-SB           & 5.57E-01 & 5.53E-01 & 9.99E-01 & 5.05E-01 \\ \cline{2-7} 
\multicolumn{1}{|c|}{}                                                                                            & \textbf{DDPG-SB} & PPO-TF           & 5.57E-01 & 4.92E-01 & 1.79E-09 & 5.80E-01 \\ \cline{2-7} 
\multicolumn{1}{|c|}{}                                                                                            & DDPG-TF          & \textbf{PPO-SB}  & 5.14E-01 & 5.53E-01 & 2.15E-07 & 4.42E-01 \\ \cline{2-7} 
\multicolumn{1}{|c|}{}                                                                                            & \textbf{DDPG-TF} & PPO-TF           & 5.14E-01 & 4.92E-01 & 4.05E-02 & 5.33E-01 \\ \cline{2-7} 
\multicolumn{1}{|c|}{}                                                                                            & \textbf{PPO-SB}  & PPO-TF           & 5.53E-01 & 4.92E-01 & 3.83E-12 & 5.90E-01 \\ \hline
\end{tabular}
}

\begin{tablenotes}
     \item mean(A) and mean(B) refer to APFD values.

 \end{tablenotes}

\end{table}
The results show that for the enriched datasets A2C-SB has better performance on both the pairwise and pointwise ranking models. Regarding the simple datasets, none of the DRL configurations has learned an adequate ranking strategy, as the highest CLES value is $0.63$. This is explained by the fact that it cannot always be possible to learn a proper policy from simple data.

To compare the DRL configurations based on their training time, we performed two sets of Welch's ANOVA and Games-Howell post-hoc tests corresponding to the pairwise and pointwise ranking models, based on the result of all algorithms across datasets.for the 10 first cycles. The results are reported on Tables \ref{tab:Results of Welch ANOVA and Games-Howell post-hoc tests of training time (in milliseconds) on pairwise and pointwise ranking models for enriched data} and \ref{tab:Results of Welch ANOVA and Games-Howell post-hoc tests of training time (in milliseconds) on pairwise and pointwise ranking models for simple data}.
\begin{table}[t]
\caption{Results of Welch ANOVA and Games-Howell post-hoc tests of training time (in milliseconds) on pairwise and pointwise ranking models for enriched datasets (in bold are DRL configurations where p-value is $<$ 0.05 and have greater performance w.r.t the effect size).}
\label{tab:Results of Welch ANOVA and Games-Howell post-hoc tests of training time (in milliseconds) on pairwise and pointwise ranking models for enriched data}
\centering
\resizebox{\textwidth}{!}{%
\begin{tabular}{|c|c|c|c|c|c|c|}
\hline
                                                                                                 & A                & B                & mean(A)  & mean(B)  & pval     & CLES     \\ \hline
\multirow{21}{*}{\begin{tabular}[c]{@{}c@{}}Pairwise and    \\ enriched datasets\end{tabular}}   & \textbf{A2C-SB}  & A2C-TF           & 8,23E+04 & 2,19E+06 & 1,25E-14 & 1,44E-02 \\ \cline{2-7} 
                                                                                                 & \textbf{A2C-SB}  & DQN-KR           & 8,23E+04 & 5,28E+05 & 6,68E-12 & 1,12E-01 \\ \cline{2-7} 
                                                                                                 & \textbf{A2C-SB}  & DQN-SB           & 8,23E+04 & 2,97E+05 & 1,99E-09 & 1,70E-01 \\ \cline{2-7} 
                                                                                                 & \textbf{A2C-SB}  & DQN-TF           & 8,23E+04 & 2,44E+06 & 0,00E+00 & 1,66E-02 \\ \cline{2-7} 
                                                                                                 & A2C-SB           & \textbf{PPO-SB}  & 8,23E+04 & 6,11E+04 & 3,08E-01 & 6,30E-01 \\ \cline{2-7} 
                                                                                                 & \textbf{A2C-SB}  & PPO-TF           & 8,23E+04 & 5,30E+05 & 1,59E-06 & 5,55E-02 \\ \cline{2-7} 
                                                                                                 & A2C-TF           & \textbf{DQN-KR}  & 2,19E+06 & 5,28E+05 & 0,00E+00 & 8,78E-01 \\ \cline{2-7} 
                                                                                                 & A2C-TF           & \textbf{DQN-SB}  & 2,19E+06 & 2,97E+05 & 0,00E+00 & 9,30E-01 \\ \cline{2-7} 
                                                                                                 & \textbf{A2C-TF}  & DQN-TF           & 2,19E+06 & 2,44E+06 & 9,77E-01 & 5,10E-01 \\ \cline{2-7} 
                                                                                                 & A2C-TF           & \textbf{PPO-SB}  & 2,19E+06 & 6,11E+04 & 0,00E+00 & 9,93E-01 \\ \cline{2-7} 
                                                                                                 & A2C-TF           & \textbf{PPO-TF}  & 2,19E+06 & 5,30E+05 & 0,00E+00 & 8,55E-01 \\ \cline{2-7} 
                                                                                                 & DQN-KR           & \textbf{DQN-SB}  & 5,28E+05 & 2,97E+05 & 8,94E-03 & 6,27E-01 \\ \cline{2-7} 
                                                                                                 & \textbf{DQN-KR}  & DQN-TF           & 5,28E+05 & 2,44E+06 & 9,80E-13 & 1,23E-01 \\ \cline{2-7} 
                                                                                                 & DQN-KR           & \textbf{PPO-SB}  & 5,28E+05 & 6,11E+04 & 3,64E-13 & 9,09E-01 \\ \cline{2-7} 
                                                                                                 & \textbf{DQN-KR}  & PPO-TF           & 5,28E+05 & 5,30E+05 & 1,00E+00 & 2,97E-01 \\ \cline{2-7} 
                                                                                                 & \textbf{DQN-SB}  & DQN-TF           & 2,97E+05 & 2,44E+06 & 1,84E-13 & 7,15E-02 \\ \cline{2-7} 
                                                                                                 & DQN-SB           & \textbf{PPO-SB}  & 2,97E+05 & 6,11E+04 & 1,87E-11 & 8,68E-01 \\ \cline{2-7} 
                                                                                                 & \textbf{DQN-SB}  & PPO-TF           & 2,97E+05 & 5,30E+05 & 1,02E-01 & 1,89E-01 \\ \cline{2-7} 
                                                                                                 & DQN-TF           & \textbf{PPO-SB}  & 2,44E+06 & 6,11E+04 & 0,00E+00 & 9,92E-01 \\ \cline{2-7} 
                                                                                                 & DQN-TF           & \textbf{PPO-TF}  & 2,44E+06 & 5,30E+05 & 3,41E-12 & 8,39E-01 \\ \cline{2-7} 
                                                                                                 & \textbf{PPO-SB}  & PPO-TF           & 6,11E+04 & 5,30E+05 & 3,89E-07 & 3,92E-02 \\ \hline
\multirow{21}{*}{\begin{tabular}[c]{@{}c@{}}Pointwise and     \\ enriched datasets\end{tabular}} & \textbf{A2C-SB}  & A2C-TF           & 7,45E+04 & 2,12E+06 & 0,00E+00 & 1,57E-02 \\ \cline{2-7} 
                                                                                                 & \textbf{A2C-SB}  & DDPG-KR          & 7,45E+04 & 5,02E+05 & 2,81E-13 & 8,31E-02 \\ \cline{2-7} 
                                                                                                 & \textbf{A2C-SB}  & DDPG-SB          & 7,45E+04 & 1,21E+05 & 6,98E-01 & 3,79E-01 \\ \cline{2-7} 
                                                                                                 & \textbf{A2C-SB}  & DDPG-TF          & 7,45E+04 & 2,39E+06 & 1,30E-13 & 1,31E-02 \\ \cline{2-7} 
                                                                                                 & A2C-SB           & \textbf{PPO-SB}  & 7,45E+04 & 5,98E+04 & 6,54E-01 & 5,81E-01 \\ \cline{2-7} 
                                                                                                 & A2C-SB           & \textbf{PPO-TF}  & 7,45E+04 & 5,16E+05 & 1,37E-13 & 4,42E-02 \\ \cline{2-7} 
                                                                                                 & A2C-TF           & \textbf{DDPG-KR} & 2,12E+06 & 5,02E+05 & 4,73E-10 & 8,65E-01 \\ \cline{2-7} 
                                                                                                 & A2C-TF           & \textbf{DDPG-SB} & 2,12E+06 & 1,21E+05 & 4,36E-14 & 9,70E-01 \\ \cline{2-7} 
                                                                                                 & \textbf{A2C-TF}  & DDPG-TF          & 2,12E+06 & 2,39E+06 & 9,89E-01 & 4,70E-01 \\ \cline{2-7} 
                                                                                                 & A2C-TF           & \textbf{PPO-SB}  & 2,12E+06 & 5,98E+04 & 0,00E+00 & 9,91E-01 \\ \cline{2-7} 
                                                                                                 & A2C-TF           & \textbf{PPO-TF}  & 2,12E+06 & 5,16E+05 & 3,59E-10 & 7,93E-01 \\ \cline{2-7} 
                                                                                                 & DDPG-KR          & \textbf{DDPG-SB} & 5,02E+05 & 1,21E+05 & 6,78E-09 & 8,65E-01 \\ \cline{2-7} 
                                                                                                 & \textbf{DDPG-KR} & DDPG-TF          & 5,02E+05 & 2,39E+06 & 2,77E-09 & 1,23E-01 \\ \cline{2-7} 
                                                                                                 & DDPG-KR          & \textbf{PPO-SB}  & 5,02E+05 & 5,98E+04 & 0,00E+00 & 9,33E-01 \\ \cline{2-7} 
                                                                                                 & \textbf{DDPG-KR} & PPO-TF           & 5,02E+05 & 5,16E+05 & 1,00E+00 & 3,14E-01 \\ \cline{2-7} 
                                                                                                 & \textbf{DDPG-SB} & DDPG-TF          & 1,21E+05 & 2,39E+06 & 3,08E-13 & 2,82E-02 \\ \cline{2-7} 
                                                                                                 & DDPG-SB          & \textbf{PPO-SB}  & 1,21E+05 & 5,98E+04 & 3,66E-01 & 6,89E-01 \\ \cline{2-7} 
                                                                                                 & \textbf{DDPG-SB} & PPO-TF           & 1,21E+05 & 5,16E+05 & 8,66E-15 & 7,61E-02 \\ \cline{2-7} 
                                                                                                 & DDPG-TF          & \textbf{PPO-SB}  & 2,39E+06 & 5,98E+04 & 1,35E-13 & 9,94E-01 \\ \cline{2-7} 
                                                                                                 & DDPG-TF          & \textbf{PPO-TF}  & 2,39E+06 & 5,16E+05 & 2,48E-09 & 8,19E-01 \\ \cline{2-7} 
                                                                                                 & \textbf{PPO-SB}  & PPO-TF           & 5,98E+04 & 5,16E+05 & 0,00E+00 & 3,42E-02 \\ \hline
\end{tabular}
}
\begin{tablenotes}
     \item mean(A) and mean(B) refer to training time values.

 \end{tablenotes}
\end{table}

\begin{table}[t]
\caption{Results of Welch ANOVA and Games-Howell post-hoc tests of training time (in milliseconds) on pairwise and pointwise ranking models for simple datasets (in bold are DRL configurations where p-value is $<$ 0.05 and have greater performance w.r.t the effect size).}
\label{tab:Results of Welch ANOVA and Games-Howell post-hoc tests of training time (in milliseconds) on pairwise and pointwise ranking models for simple data}
\centering
\resizebox{\textwidth}{!}{%
\begin{tabular}{|c|c|c|c|c|c|c|}
\hline
                                                                                              & A                & B                & mean(A)  & mean(B)  & pval     & CLES     \\ \hline
\multirow{21}{*}{\begin{tabular}[c]{@{}c@{}}Pairwise and\\   simple datasets\end{tabular}}    & \textbf{A2C-SB}  & A2C-TF           & 3,84E+05 & 1,04E+07 & 5,52E-07 & 5,43E-02 \\ \cline{2-7} 
                                                                                              & \textbf{A2C-SB}  & DQN-KR           & 3,84E+05 & 2,33E+06 & 2,23E-05 & 1,50E-01 \\ \cline{2-7} 
                                                                                              & \textbf{A2C-SB}  & DQN-SB           & 3,84E+05 & 1,56E+06 & 1,83E-04 & 2,05E-01 \\ \cline{2-7} 
                                                                                              & \textbf{A2C-SB}  & DQN-TF           & 3,84E+05 & 1,08E+07 & 5,24E-07 & 4,33E-02 \\ \cline{2-7} 
                                                                                              & A2C-SB           & \textbf{PPO-SB}  & 3,84E+05 & 2,72E+05 & 7,21E-01 & 6,52E-01 \\ \cline{2-7} 
                                                                                              & \textbf{A2C-SB}  & PPO-TF           & 3,84E+05 & 5,36E+06 & 5,40E-03 & 1,20E-01 \\ \cline{2-7} 
                                                                                              & A2C-TF           & \textbf{DQN-KR}  & 1,04E+07 & 2,33E+06 & 1,21E-04 & 8,19E-01 \\ \cline{2-7} 
                                                                                              & A2C-TF           & \textbf{DQN-SB}  & 1,04E+07 & 1,56E+06 & 1,45E-05 & 8,60E-01 \\ \cline{2-7} 
                                                                                              & \textbf{A2C-TF}  & DQN-TF           & 1,04E+07 & 1,08E+07 & 1,00E+00 & 4,66E-01 \\ \cline{2-7} 
                                                                                              & A2C-TF           & \textbf{PPO-SB}  & 1,04E+07 & 2,72E+05 & 4,05E-07 & 9,85E-01 \\ \cline{2-7} 
                                                                                              & A2C-TF           & \textbf{PPO-TF}  & 1,04E+07 & 5,36E+06 & 2,04E-01 & 7,28E-01 \\ \cline{2-7} 
                                                                                              & DQN-KR           & \textbf{DQN-SB}  & 2,33E+06 & 1,56E+06 & 5,91E-01 & 6,28E-01 \\ \cline{2-7} 
                                                                                              & \textbf{DQN-KR}  & DQN-TF           & 2,33E+06 & 1,08E+07 & 9,70E-05 & 1,66E-01 \\ \cline{2-7} 
                                                                                              & DQN-KR           & \textbf{PPO-SB}  & 2,33E+06 & 2,72E+05 & 5,64E-06 & 8,80E-01 \\ \cline{2-7} 
                                                                                              & \textbf{DQN-KR}  & PPO-TF           & 2,33E+06 & 5,36E+06 & 2,89E-01 & 4,63E-01 \\ \cline{2-7} 
                                                                                              & \textbf{DQN-SB}  & DQN-TF           & 1,56E+06 & 1,08E+07 & 1,24E-05 & 1,48E-01 \\ \cline{2-7} 
                                                                                              & DQN-SB           & \textbf{PPO-SB}  & 1,56E+06 & 2,72E+05 & 2,38E-05 & 8,50E-01 \\ \cline{2-7} 
                                                                                              & \textbf{DQN-SB}  & PPO-TF           & 1,56E+06 & 5,36E+06 & 7,64E-02 & 3,44E-01 \\ \cline{2-7} 
                                                                                              & DQN-TF           & \textbf{PPO-SB}  & 1,08E+07 & 2,72E+05 & 3,88E-07 & 9,90E-01 \\ \cline{2-7} 
                                                                                              & DQN-TF           & \textbf{PPO-TF}  & 1,08E+07 & 5,36E+06 & 1,60E-01 & 7,28E-01 \\ \cline{2-7} 
                                                                                              & \textbf{PPO-SB}  & PPO-TF           & 2,72E+05 & 5,36E+06 & 4,10E-03 & 1,12E-01 \\ \hline
\multirow{21}{*}{\begin{tabular}[c]{@{}c@{}}Pointwise and \\    simple datasets\end{tabular}} & \textbf{A2C-SB}  & A2C-TF           & 3,92E+05 & 1,08E+07 & 5,15E-04 & 5,49E-02 \\ \cline{2-7} 
                                                                                              & \textbf{A2C-SB}  & DDPG-KR          & 3,92E+05 & 2,62E+06 & 1,49E-05 & 1,50E-01 \\ \cline{2-7} 
                                                                                              & \textbf{A2C-SB}  & DDPG-SB          & 3,92E+05 & 6,51E+05 & 9,34E-01 & 3,86E-01 \\ \cline{2-7} 
                                                                                              & \textbf{A2C-SB}  & DDPG-TF          & 3,92E+05 & 1,19E+07 & 7,12E-07 & 4,44E-02 \\ \cline{2-7} 
                                                                                              & A2C-SB           & \textbf{PPO-SB}  & 3,92E+05 & 2,95E+05 & 8,49E-01 & 5,78E-01 \\ \cline{2-7} 
                                                                                              & \textbf{A2C-SB}  & PPO-TF           & 3,92E+05 & 3,80E+06 & 5,76E-08 & 1,05E-01 \\ \cline{2-7} 
                                                                                              & A2C-TF           & \textbf{DDPG-KR} & 1,08E+07 & 2,62E+06 & 1,23E-02 & 8,04E-01 \\ \cline{2-7} 
                                                                                              & A2C-TF           & \textbf{DDPG-SB} & 1,08E+07 & 6,51E+05 & 7,86E-04 & 9,21E-01 \\ \cline{2-7} 
                                                                                              & \textbf{A2C-TF}  & DDPG-TF          & 1,08E+07 & 1,19E+07 & 1,00E+00 & 4,55E-01 \\ \cline{2-7} 
                                                                                              & A2C-TF           & \textbf{PPO-SB}  & 1,08E+07 & 2,95E+05 & 4,47E-04 & 9,82E-01 \\ \cline{2-7} 
                                                                                              & A2C-TF           & \textbf{PPO-TF}  & 1,08E+07 & 3,80E+06 & 5,22E-02 & 6,76E-01 \\ \cline{2-7} 
                                                                                              & DDPG-KR          & \textbf{DDPG-SB} & 2,62E+06 & 6,51E+05 & 1,54E-03 & 8,09E-01 \\ \cline{2-7} 
                                                                                              & \textbf{DDPG-KR} & DDPG-TF          & 2,62E+06 & 1,19E+07 & 1,40E-04 & 1,80E-01 \\ \cline{2-7} 
                                                                                              & DDPG-KR          & \textbf{PPO-SB}  & 2,62E+06 & 2,95E+05 & 4,99E-06 & 8,53E-01 \\ \cline{2-7} 
                                                                                              & \textbf{DDPG-KR} & PPO-TF           & 2,62E+06 & 3,80E+06 & 5,44E-01 & 4,01E-01 \\ \cline{2-7} 
                                                                                              & \textbf{DDPG-SB} & DDPG-TF          & 6,51E+05 & 1,19E+07 & 1,48E-06 & 7,40E-02 \\ \cline{2-7} 
                                                                                              & DDPG-SB          & \textbf{PPO-SB}  & 6,51E+05 & 2,95E+05 & 7,56E-01 & 6,55E-01 \\ \cline{2-7} 
                                                                                              & \textbf{DDPG-SB} & PPO-TF           & 6,51E+05 & 3,80E+06 & 3,55E-06 & 1,29E-01 \\ \cline{2-7} 
                                                                                              & DDPG-TF          & \textbf{PPO-SB}  & 1,19E+07 & 2,95E+05 & 5,65E-07 & 9,93E-01 \\ \cline{2-7} 
                                                                                              & DDPG-TF          & \textbf{PPO-TF}  & 1,19E+07 & 3,80E+06 & 1,61E-03 & 6,69E-01 \\ \cline{2-7} 
                                                                                              & \textbf{PPO-SB}  & PPO-TF           & 2,95E+05 & 3,80E+06 & 2,36E-08 & 1,03E-01 \\ \hline
\end{tabular}
}
\begin{tablenotes}
     \item mean(A) and mean(B) refer to training time values.

 \end{tablenotes}
\end{table}
We summarize the results as follows:
\begin{itemize}
       \item Pairwise and simple datasets:  
      \begin{itemize}
          \item DQN-SB $>$ DQN-KR $>$ DQN-TF
          \item A2C-SB $>$ A2C-TF 
      \end{itemize}
    \item Pairwise and enriched datasets:  
      \begin{itemize}
          \item DQN-SB $>$ DQN-KR $>$ DQN-TF
          \item A2C-SB $>$ A2C-TF 
         \item PPO-SB $>$ PPO-TF 
      \end{itemize}
           \item Pointwise and simple datasets: 
     \begin{itemize}
         \item DDPG-SB $>$ DDPG-KR $>$ DDPG-TF
         \item A2C-SB $>$  A2C-TF
         \item PPO-SB $>$ PPO-TF
     \end{itemize}
     \item Pointwise and enriched datasets: 
     \begin{itemize}
         \item DDPG-KR $>$ DDPG-SB $>$ DDPG-TF
         \item A2C-SB $>$  A2C-TF
         \item PPO-SB $>$ PPO-TF
     \end{itemize}

\end{itemize}

To compare the DRL configurations based on their prediction time, we again performed two sets of Welch's ANOVA and Games-Howell post-hoc tests corresponding to the pairwise and pointwise ranking models. The results are reported on Tables \ref{tab:Results of Welch ANOVA and Games-Howell post-hoc tests of testing time (in milliseconds) on pairwise and pointwise ranking models for enriched data} and \ref{tab:Results of Welch ANOVA and Games-Howell post-hoc tests of testing time (in milliseconds) on pairwise and pointwise ranking models for simple data}. 
\begin{table}[t]
\caption{Results of Welch ANOVA and Games-Howell post-hoc tests of testing time (in milliseconds) on pairwise and pointwise ranking models for enriched datasets (in bold are DRL configurations where p-value is $<$ 0.05 and have greater performance w.r.t the effect size).}
\label{tab:Results of Welch ANOVA and Games-Howell post-hoc tests of testing time (in milliseconds) on pairwise and pointwise ranking models for enriched data}
\centering
\resizebox{\textwidth}{!}{%
\begin{tabular}{|c|c|c|c|c|c|c|}
\hline
                                                                                                 & A                & B                & mean(A)  & mean(B)  & pval     & CLES     \\ \hline
\multirow{21}{*}{\begin{tabular}[c]{@{}c@{}}Pairwise and    \\ enriched datasets\end{tabular}}   & \textbf{A2C-SB}  & A2C-TF           & 1,67E+03 & 8,21E+03 & 0,00E+00 & 0,00E+00 \\ \cline{2-7} 
                                                                                                 & A2C-SB           & \textbf{DQN-KR}  & 1,67E+03 & 2,90E+02 & 0,00E+00 & 9,67E-01 \\ \cline{2-7} 
                                                                                                 & \textbf{A2C-SB}  & DQN-SB           & 1,67E+03 & 1,72E+03 & 8,69E-01 & 4,13E-01 \\ \cline{2-7} 
                                                                                                 & \textbf{A2C-SB}  & DQN-TF           & 1,67E+03 & 8,32E+03 & 0,00E+00 & 0,00E+00 \\ \cline{2-7} 
                                                                                                 & A2C-SB           & \textbf{PPO-SB}  & 1,67E+03 & 1,19E+03 & 0,00E+00 & 8,62E-01 \\ \cline{2-7} 
                                                                                                 & \textbf{A2C-SB}  & PPO-TF           & 1,67E+03 & 9,95E+03 & 2,98E-14 & 0,00E+00 \\ \cline{2-7} 
                                                                                                 & A2C-TF           & \textbf{DQN-KR}  & 8,21E+03 & 2,90E+02 & 3,61E-13 & 9,99E-01 \\ \cline{2-7} 
                                                                                                 & A2C-TF           & \textbf{DQN-SB}  & 8,21E+03 & 1,72E+03 & 7,15E-14 & 1,00E+00 \\ \cline{2-7} 
                                                                                                 & \textbf{A2C-TF}  & DQN-TF           & 8,21E+03 & 8,32E+03 & 9,93E-01 & 5,91E-01 \\ \cline{2-7} 
                                                                                                 & A2C-TF           & \textbf{PPO-SB}  & 8,21E+03 & 1,19E+03 & 0,00E+00 & 1,00E+00 \\ \cline{2-7} 
                                                                                                 & \textbf{A2C-TF}  & PPO-TF           & 8,21E+03 & 9,95E+03 & 7,14E-07 & 2,01E-01 \\ \cline{2-7} 
                                                                                                 & \textbf{DQN-KR}  & DQN-SB           & 2,90E+02 & 1,72E+03 & 1,05E-13 & 3,33E-02 \\ \cline{2-7} 
                                                                                                 & \textbf{DQN-KR}  & DQN-TF           & 2,90E+02 & 8,32E+03 & 0,00E+00 & 1,69E-03 \\ \cline{2-7} 
                                                                                                 & \textbf{DQN-KR}  & PPO-SB           & 2,90E+02 & 1,19E+03 & 0,00E+00 & 3,34E-02 \\ \cline{2-7} 
                                                                                                 & \textbf{DQN-KR}  & PPO-TF           & 2,90E+02 & 9,95E+03 & 0,00E+00 & 0,00E+00 \\ \cline{2-7} 
                                                                                                 & \textbf{DQN-SB}  & DQN-TF           & 1,72E+03 & 8,32E+03 & 7,84E-14 & 0,00E+00 \\ \cline{2-7} 
                                                                                                 & DQN-SB           & \textbf{PPO-SB}  & 1,72E+03 & 1,19E+03 & 1,84E-14 & 9,26E-01 \\ \cline{2-7} 
                                                                                                 & \textbf{DQN-SB}  & PPO-TF           & 1,72E+03 & 9,95E+03 & 4,03E-14 & 0,00E+00 \\ \cline{2-7} 
                                                                                                 & DQN-TF           & \textbf{PPO-SB}  & 8,32E+03 & 1,19E+03 & 4,21E-14 & 1,00E+00 \\ \cline{2-7} 
                                                                                                 & \textbf{DQN-TF}  & PPO-TF           & 8,32E+03 & 9,95E+03 & 3,84E-05 & 1,76E-01 \\ \cline{2-7} 
                                                                                                 & \textbf{PPO-SB}  & PPO-TF           & 1,19E+03 & 9,95E+03 & 0,00E+00 & 0,00E+00 \\ \hline
\multirow{21}{*}{\begin{tabular}[c]{@{}c@{}}Pointwise and     \\ enriched datasets\end{tabular}} & \textbf{A2C-SB}  & A2C-TF           & 1,59E+03 & 7,86E+03 & 0,00E+00 & 0,00E+00 \\ \cline{2-7} 
                                                                                                 & A2C-SB           & \textbf{DDPG-KR} & 1,59E+03 & 6,74E+01 & 0,00E+00 & 1,00E+00 \\ \cline{2-7} 
                                                                                                 & A2C-SB           & \textbf{DDPG-SB} & 1,59E+03 & 7,83E+02 & 2,22E-15 & 9,80E-01 \\ \cline{2-7} 
                                                                                                 & \textbf{A2C-SB}  & DDPG-TF          & 1,59E+03 & 8,98E+03 & 0,00E+00 & 0,00E+00 \\ \cline{2-7} 
                                                                                                 & A2C-SB           & \textbf{PPO-SB}  & 1,59E+03 & 1,16E+03 & 0,00E+00 & 8,68E-01 \\ \cline{2-7} 
                                                                                                 & \textbf{A2C-SB}  & PPO-TF           & 1,59E+03 & 8,56E+03 & 0,00E+00 & 0,00E+00 \\ \cline{2-7} 
                                                                                                 & A2C-TF           & \textbf{DDPG-KR} & 7,86E+03 & 6,74E+01 & 7,35E-14 & 1,00E+00 \\ \cline{2-7} 
                                                                                                 & A2C-TF           & \textbf{DDPG-SB} & 7,86E+03 & 7,83E+02 & 0,00E+00 & 1,00E+00 \\ \cline{2-7} 
                                                                                                 & \textbf{A2C-TF}  & DDPG-TF          & 7,86E+03 & 8,98E+03 & 0,00E+00 & 2,49E-01 \\ \cline{2-7} 
                                                                                                 & A2C-TF           & \textbf{PPO-SB}  & 7,86E+03 & 1,16E+03 & 0,00E+00 & 1,00E+00 \\ \cline{2-7} 
                                                                                                 & \textbf{A2C-TF}  & PPO-TF           & 7,86E+03 & 8,56E+03 & 0,00E+00 & 2,90E-01 \\ \cline{2-7} 
                                                                                                 & \textbf{DDPG-KR} & DDPG-SB          & 6,74E+01 & 7,83E+02 & 9,66E-15 & 0,00E+00 \\ \cline{2-7} 
                                                                                                 & \textbf{DDPG-KR} & DDPG-TF          & 6,74E+01 & 8,98E+03 & 0,00E+00 & 0,00E+00 \\ \cline{2-7} 
                                                                                                 & \textbf{DDPG-KR} & PPO-SB           & 6,74E+01 & 1,16E+03 & 1,05E-13 & 0,00E+00 \\ \cline{2-7} 
                                                                                                 & \textbf{DDPG-KR} & PPO-TF           & 6,74E+01 & 8,56E+03 & 0,00E+00 & 0,00E+00 \\ \cline{2-7} 
                                                                                                 & \textbf{DDPG-SB} & DDPG-TF          & 7,83E+02 & 8,98E+03 & 0,00E+00 & 0,00E+00 \\ \cline{2-7} 
                                                                                                 & \textbf{DDPG-SB} & PPO-SB           & 7,83E+02 & 1,16E+03 & 0,00E+00 & 6,26E-02 \\ \cline{2-7} 
                                                                                                 & \textbf{DDPG-SB} & PPO-TF           & 7,83E+02 & 8,56E+03 & 2,22E-16 & 0,00E+00 \\ \cline{2-7} 
                                                                                                 & DDPG-TF          & \textbf{PPO-SB}  & 8,98E+03 & 1,16E+03 & 0,00E+00 & 1,00E+00 \\ \cline{2-7} 
                                                                                                 & DDPG-TF          & \textbf{PPO-TF}  & 8,98E+03 & 8,56E+03 & 2,51E-04 & 6,13E-01 \\ \cline{2-7} 
                                                                                                 & \textbf{PPO-SB}  & PPO-TF           & 1,16E+03 & 8,56E+03 & 0,00E+00 & 0,00E+00 \\ \hline
\end{tabular}
}
\begin{tablenotes}
     \item mean(A) and mean(B) refer to testing time values.
 \end{tablenotes}
\end{table}

\begin{table}[t]
\caption{Results of Welch ANOVA and Games-Howell post-hoc tests of testing time (in milliseconds) on pairwise and pointwise ranking models for simple datasets (in bold are DRL configurations where p-value is $<$ 0.05 and have greater performance w.r.t the effect size).}
\label{tab:Results of Welch ANOVA and Games-Howell post-hoc tests of testing time (in milliseconds) on pairwise and pointwise ranking models for simple data}
\centering
\resizebox{\textwidth}{!}{%
\begin{tabular}{|c|c|c|c|c|c|c|}
\hline
                                                                                               & A                & B                & mean(A)  & mean(B)  & pval     & CLES     \\ \hline
\multirow{21}{*}{\begin{tabular}[c]{@{}c@{}}Pairwise and \\    simple datasets\end{tabular}}   & \textbf{A2C-SB}  & A2C-TF           & 2,46E+03 & 8,79E+03 & 2,85E-14 & 3,00E-04 \\ \cline{2-7} 
                                                                                               & A2C-SB           & \textbf{DQN-KR}  & 2,46E+03 & 6,23E+02 & 0,00E+00 & 8,55E-01 \\ \cline{2-7} 
                                                                                               & \textbf{A2C-SB}  & DQN-SB           & 2,46E+03 & 3,07E+03 & 3,61E-01 & 3,77E-01 \\ \cline{2-7} 
                                                                                               & \textbf{A2C-SB}  & DQN-TF           & 2,46E+03 & 1,24E+04 & 4,33E-15 & 3,90E-03 \\ \cline{2-7} 
                                                                                               & A2C-SB           & \textbf{PPO-SB}  & 2,46E+03 & 1,73E+03 & 6,20E-03 & 7,52E-01 \\ \cline{2-7} 
                                                                                               & \textbf{A2C-SB}  & PPO-TF           & 2,46E+03 & 3,20E+04 & 2,40E-06 & 0,00E+00 \\ \cline{2-7} 
                                                                                               & A2C-TF           & \textbf{DQN-KR}  & 8,79E+03 & 6,23E+02 & 2,99E-14 & 1,00E+00 \\ \cline{2-7} 
                                                                                               & A2C-TF           & \textbf{DQN-SB}  & 8,79E+03 & 3,07E+03 & 3,55E-14 & 9,42E-01 \\ \cline{2-7} 
                                                                                               & \textbf{A2C-TF}  & DQN-TF           & 8,79E+03 & 1,24E+04 & 2,97E-03 & 5,80E-01 \\ \cline{2-7} 
                                                                                               & A2C-TF           & \textbf{PPO-SB}  & 8,79E+03 & 1,73E+03 & 3,62E-14 & 1,00E+00 \\ \cline{2-7} 
                                                                                               & \textbf{A2C-TF}  & PPO-TF           & 8,79E+03 & 3,20E+04 & 2,97E-04 & 2,15E-01 \\ \cline{2-7} 
                                                                                               & \textbf{DQN-KR}  & DQN-SB           & 6,23E+02 & 3,07E+03 & 2,73E-14 & 1,39E-01 \\ \cline{2-7} 
                                                                                               & \textbf{DQN-KR}  & DQN-TF           & 6,23E+02 & 1,24E+04 & 1,75E-14 & 0,00E+00 \\ \cline{2-7} 
                                                                                               & \textbf{DQN-KR}  & PPO-SB           & 6,23E+02 & 1,73E+03 & 5,10E-11 & 1,62E-01 \\ \cline{2-7} 
                                                                                               & \textbf{DQN-KR}  & PPO-TF           & 6,23E+02 & 3,20E+04 & 5,53E-07 & 0,00E+00 \\ \cline{2-7} 
                                                                                               & \textbf{DQN-SB}  & DQN-TF           & 3,07E+03 & 1,24E+04 & 1,42E-14 & 8,13E-02 \\ \cline{2-7} 
                                                                                               & DQN-SB           & \textbf{PPO-SB}  & 3,07E+03 & 1,73E+03 & 3,54E-05 & 8,41E-01 \\ \cline{2-7} 
                                                                                               & \textbf{DQN-SB}  & PPO-TF           & 3,07E+03 & 3,20E+04 & 3,94E-06 & 1,41E-02 \\ \cline{2-7} 
                                                                                               & DQN-TF           & \textbf{PPO-SB}  & 1,24E+04 & 1,73E+03 & 0,00E+00 & 1,00E+00 \\ \cline{2-7} 
                                                                                               & \textbf{DQN-TF}  & PPO-TF           & 1,24E+04 & 3,20E+04 & 4,22E-03 & 2,12E-01 \\ \cline{2-7} 
                                                                                               & \textbf{PPO-SB}  & PPO-TF           & 1,73E+03 & 3,20E+04 & 1,34E-06 & 0,00E+00 \\ \hline
\multirow{21}{*}{\begin{tabular}[c]{@{}c@{}}Pointwise and  \\    simple datasets\end{tabular}} & \textbf{A2C-SB}  & A2C-TF           & 1,40E+03 & 8,08E+03 & 1,31E-14 & 0,00E+00 \\ \cline{2-7} 
                                                                                               & A2C-SB           & \textbf{DDPG-KR} & 1,40E+03 & 1,73E+02 & 8,22E-15 & 1,00E+00 \\ \cline{2-7} 
                                                                                               & A2C-SB           & \textbf{DDPG-SB} & 1,40E+03 & 8,82E+02 & 3,80E-08 & 9,04E-01 \\ \cline{2-7} 
                                                                                               & \textbf{A2C-SB}  & DDPG-TF          & 1,40E+03 & 9,73E+03 & 5,33E-15 & 0,00E+00 \\ \cline{2-7} 
                                                                                               & A2C-SB           & \textbf{PPO-SB}  & 1,40E+03 & 1,37E+03 & 9,99E-01 & 5,39E-01 \\ \cline{2-7} 
                                                                                               & \textbf{A2C-SB}  & PPO-TF           & 1,40E+03 & 1,49E+04 & 0,00E+00 & 0,00E+00 \\ \cline{2-7} 
                                                                                               & A2C-TF           & \textbf{DDPG-KR} & 8,08E+03 & 1,73E+02 & 6,66E-16 & 1,00E+00 \\ \cline{2-7} 
                                                                                               & A2C-TF           & \textbf{DDPG-SB} & 8,08E+03 & 8,82E+02 & 1,07E-14 & 1,00E+00 \\ \cline{2-7} 
                                                                                               & \textbf{A2C-TF}  & DDPG-TF          & 8,08E+03 & 9,73E+03 & 4,41E-07 & 2,07E-01 \\ \cline{2-7} 
                                                                                               & A2C-TF           & \textbf{PPO-SB}  & 8,08E+03 & 1,37E+03 & 1,30E-14 & 1,00E+00 \\ \cline{2-7} 
                                                                                               & \textbf{A2C-TF}  & PPO-TF           & 8,08E+03 & 1,49E+04 & 5,06E-09 & 1,47E-01 \\ \cline{2-7} 
                                                                                               & \textbf{DDPG-KR} & DDPG-SB          & 1,73E+02 & 8,82E+02 & 2,10E-10 & 0,00E+00 \\ \cline{2-7} 
                                                                                               & \textbf{DDPG-KR} & DDPG-TF          & 1,73E+02 & 9,73E+03 & 9,77E-15 & 0,00E+00 \\ \cline{2-7} 
                                                                                               & \textbf{DDPG-KR} & PPO-SB           & 1,73E+02 & 1,37E+03 & 0,00E+00 & 0,00E+00 \\ \cline{2-7} 
                                                                                               & \textbf{DDPG-KR} & PPO-TF           & 1,73E+02 & 1,49E+04 & 9,55E-15 & 0,00E+00 \\ \cline{2-7} 
                                                                                               & \textbf{DDPG-SB} & DDPG-TF          & 8,82E+02 & 9,73E+03 & 6,15E-14 & 0,00E+00 \\ \cline{2-7} 
                                                                                               & \textbf{DDPG-SB} & PPO-SB           & 8,82E+02 & 1,37E+03 & 3,23E-07 & 1,08E-01 \\ \cline{2-7} 
                                                                                               & \textbf{DDPG-SB} & PPO-TF           & 8,82E+02 & 1,49E+04 & 0,00E+00 & 0,00E+00 \\ \cline{2-7} 
                                                                                               & DDPG-TF          & \textbf{PPO-SB}  & 9,73E+03 & 1,37E+03 & 0,00E+00 & 1,00E+00 \\ \cline{2-7} 
                                                                                               & \textbf{DDPG-TF} & PPO-TF           & 9,73E+03 & 1,49E+04 & 8,25E-06 & 3,36E-01 \\ \cline{2-7} 
                                                                                               & \textbf{PPO-SB}  & PPO-TF           & 1,37E+03 & 1,49E+04 & 0,00E+00 & 0,00E+00 \\ \hline
\end{tabular}
}
\begin{tablenotes}
     \item mean(A) and mean(B) refer to testing time values.

 \end{tablenotes}
\end{table}
Here is the result  for the 10 first cycles:
\begin{itemize}
        \item Pairwise and simple datasets:  
      \begin{itemize}
        \item DQN-KR $>$ DQN-SB $>$ DQN-TF
          \item A2C-SB $>$ A2C-TF 
          \item PPO-SB $>$ PPO-TF
      \end{itemize}
    \item Pairwise and enriched datasets:  
      \begin{itemize}
        \item DQN-KR $>$ DQN-SB $>$ DQN-TF
          \item A2C-SB $>$ A2C-TF 
          \item PPO-SB $>$ PPO-TF
      \end{itemize}
           \item Pointwise and simple datasets: 
     \begin{itemize}
            \item DDPG-KR $>$ DDPG-SB $>$ DDPG-TF
         \item A2C-SB $>$  A2C-TF
         \item PPO-SB $>$ PPO-TF
     \end{itemize}
     \item Pointwise and enriched datasets: 
     \begin{itemize}
            \item DDPG-KR $>$ DDPG-SB $>$ DDPG-TF
         \item A2C-SB $>$  A2C-TF
         \item PPO-SB $>$ PPO-TF
     \end{itemize}

\end{itemize}

Based on the results presented, we can conclude that both pairwise and pointwise configurations perform well with Stable-baselines and Keras-rl frameworks in terms of prediction times. Nevertheless, Tensorforce configurations need more time for training and prediction times.

\begin{tcolorbox}
\textbf{Finding 3: Overall Stable-baselines framework has better performance than Keras-rl and Tensorforce framework, therefore can be recommended when using DRL for test case prioritization.}
\end{tcolorbox}

\begin{tcolorbox}
    Summary 4: In terms of accuracy, both pairwise and pointwise ranking strategies have good performance when applied to the DRL algorithms for enriched datasets. Nevertheless, none of the DRL algorithms learn a qualified strategy when it comes to simple datasets. A possible reason is the nature of simple datasets (see Subsection \ref{sec: subsection 336}). It cannot be expected that learning an accurate policy for complex software systems is always possible on the basis of simple data.
\end{tcolorbox}

\begin{figure*}[t]
  \begin{minipage}{0.55\textwidth}
  
        \includegraphics[width=\textwidth]{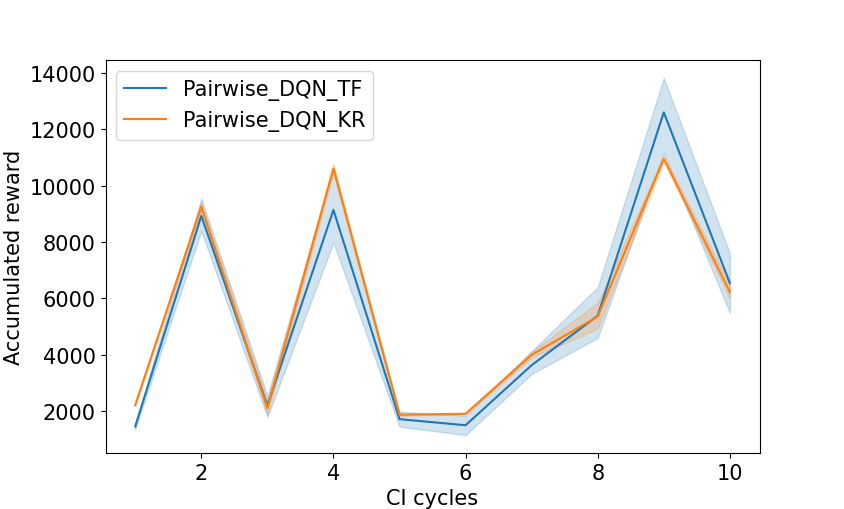}
        \caption{Accumulated reward during  training \\ of the Pairwise-DQN  configurations \\ for the first 10  CI cycles on CODEC  dataset.}
        \label{fig:dqn_pairwise_10_runs1}
    \end{minipage}
    \hfill
            \begin{minipage}{0.43\textwidth}
        
        \includegraphics[width=\textwidth]{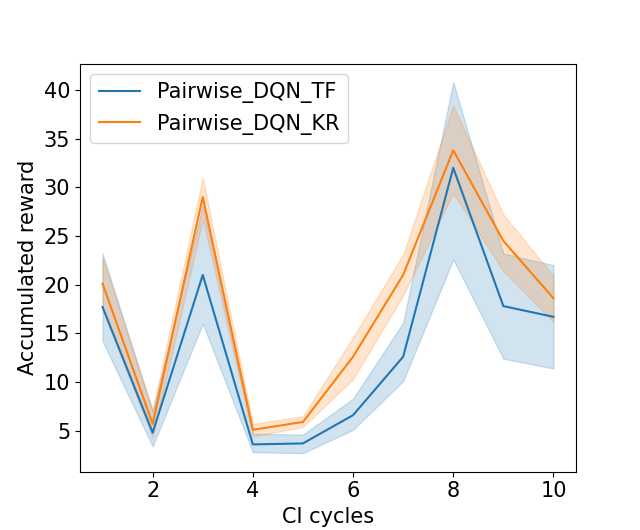}
        \caption{Accumulated reward during testing of the Pairwise-DQN configurations for the first 10 CI cycles  on the CODEC dataset.}
        \label{fig:dqn_pairwise_10_runs3}
            \end{minipage}
        
    \end{figure*}

\begin{figure}
    \centering

        \includegraphics[width=0.5\textwidth]{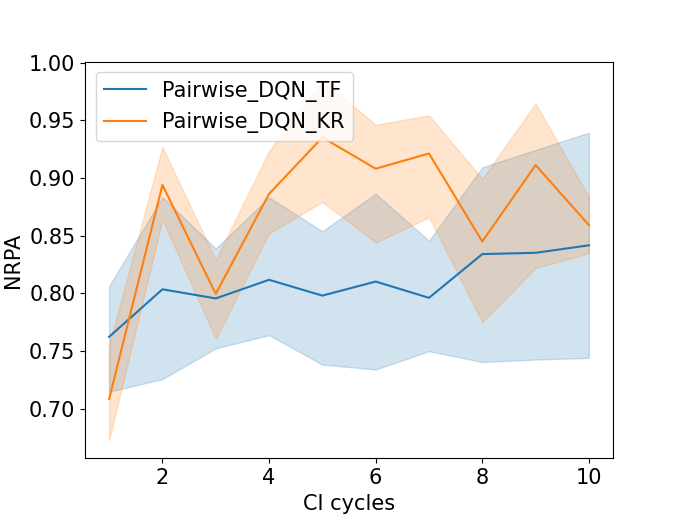}
        \caption{NRPA of the Pairwise-DQN  configurations for the first 10 CI cycles on the CODEC dataset.}
        \label{fig:dqn_pairwise_10_runs2}
\end{figure}

\hfill \break
\textbf{RQ3:} Figure ~\ref{fig:dqn_pairwise_10_runs1}, ~\ref{fig:dqn_pairwise_10_runs3},~\ref{fig:dqn_pairwise_10_runs2} show the results of the Pairwise-DQN configurations from Tensorforce and Keras-rl frameworks in terms of NRPA, accumulated reward obtained by agents during training and accumulated reward obtained by agents during testing on CODEC dataset.
The results are collected for the first 10 CI cycles over 5 different runs. Regarding the DQN algorithm, Keras-rl and Tensorforce have the same performance in terms of reward but perform differently in terms of NRPA. Similarly as with the others DRL algorithms, we do not observe stable results across the DRL frameworks.

\section{Recommendations about frameworks/algorithms selection} \label{futurework}
In this section, we discuss our recommendations regarding the selection of DRL frameworks/algorithms for researchers and practitioners. To derive some of the recommendations below and to investigate which hyperparameters are the most critical for the game testing problem, we conducted a manual hyperparameters tuning. Since the goal was not finding the best hyperparameters for each DRL algorithm in each DRL framework, we do not use automatic hyperparameter tuning.

The results of our analysis indicate that there are some differences in using the same algorithm from different DRL frameworks. This is due to the diversity of hyperparameters that are offered by different DRL frameworks. Among the studied DRL frameworks, the DQN algorithm from Keras-rl has the least number of hyperparameters (13 in total), leading to less flexibility in improving the agent's training process thus explaining its poor performance. 
Moreover, Table \ref{tab:Results of State coverage and number of bugs detected, performed by DQN_KR configuration for the game testing problem on a 10k steps budget over 5 runs} shows the results of the tuning of some hyperparameters provided by DQN-KR.
\begin{table*}[t]
\caption{
Results of State coverage and the number of bugs detected, performed by DQN-KR configuration for the game testing problem on a 10k steps budget over 5 runs.}
\label{tab:Results of State coverage and number of bugs detected, performed by DQN_KR configuration for the game testing problem on a 10k steps budget over 5 runs}
\centering
\resizebox{\textwidth}{!}{%
\begin{tabular}{|c|c|c|c|}
\hline
Hyperparameters                     & Value                    & Average number of bugs & Average state coverage \\ \hline
\multirow{3}{*}{Learning rate}      & \textbf{0.00025}         & 0.22                   & 10.2                   \\ \cline{2-4} 
                                    & 0.01                     & 1                      & 14.4                   \\ \cline{2-4} 
                                    & 0.001                    & 0.6                    & 6                      \\ \hline
\multirow{3}{*}{Batch size}         & \textbf{128}             & 0.22                   & 10.2                   \\ \cline{2-4} 
                                    & 64                       & 0.4                    & 12                     \\ \cline{2-4} 
                                    & 32                       & 0                      & 6.4                    \\ \hline
\multirow{2}{*}{Policy}             & \textbf{Epsilon greedy}  & 0.22                   & 10.2                   \\ \cline{2-4} 
                                    & Boltzman                 & 0.4                    & 4.4                    \\ \hline
\multirow{4}{*}{Enable dueling dqn} & \textbf{False}           & 0.22                   & 10.2                   \\ \cline{2-4} 
                                    & True, dueling type=avg   & 0                      & 6.8                    \\ \cline{2-4} 
                                    & True, dueling type=max   & 1                      & 9.6                    \\ \cline{2-4} 
                                    & True, dueling type=naive & 0.8                    & 11.4                   \\ \hline
\multirow{3}{*}{Test policy}        & \textbf{GreedyQPolicy()} & 0.22                   & 10.2                   \\ \cline{2-4} 
                                    & Epsilon greedy           & 0.2                    & 9.4                    \\ \cline{2-4} 
                                    & Boltzman                 & 0.4                    & 8.8                    \\ \hline
\multirow{3}{*}{Gamma}              & \textbf{0.99}            & 0.22                   & 10.2                   \\ \cline{2-4} 
                                    & 0.25                     & 0                      & 8                      \\ \cline{2-4} 
                                    & 0.025                    & 0.6                    & 6.4                    \\ \hline
\end{tabular}
}
\end{table*}
In bold are the values of the hyperparameters we initially use for our experiments. Then every time we vary each of them individually (see Column “Values” for their values) and collect the average number of bugs and state coverage. The results show that fine tuning the hyperparameters do not make DQN-KR significantly more performant. DQN-SB still has better performance. A DRL framework should offer a large number of hyperparameters to provide flexibility for tuning DRL agents and improve its efficiency.
\begin{tcolorbox}
\textbf{Recommendation 1: When applying a DRL algorithm from a DRL framework to an SE problem, we recommend choosing the DRL framework that offers the largest number of hyperparameters to have the flexibility to improve the agent efficiency.}
\end{tcolorbox}
In this paper, we studied two problems whose characteristics can be found in Table \ref{tab:General characteristics of the studied RL environments}. 
\begin{table}[t]
\caption{General characteristics of the studied DRL environments.}
\centering
\resizebox{\textwidth}{!}{%
\begin{tabular}{c|c|cc|}
\cline{2-4}
\multirow{2}{*}{}                         & \multirow{2}{*}{Game Testing} & \multicolumn{2}{c|}{Test case prioritization}               \\ \cline{3-4} 
                                          &                               & \multicolumn{1}{c|}{Pairwise strategy} & Pointwise strategy \\ \hline
\multicolumn{1}{|c|}{action space}        & Discrete                      & \multicolumn{1}{c|}{Discrete}          & Continuous-1D      \\ \hline
\multicolumn{1}{|c|}{observation space}   & Continuous-2D                 & \multicolumn{1}{c|}{Continuous-2D}     & Continuous-2D      \\ \hline
\multicolumn{1}{|c|}{Reward distribution} & Negative and positive values  & \multicolumn{1}{c|}{Positive values}   & Positive values    \\ \hline
\multicolumn{1}{|c|}{Environment type}    & Deterministic                 & \multicolumn{1}{c|}{Deterministic}     & Deterministic      \\ \hline
\end{tabular}
}

\label{tab:General characteristics of the studied RL environments}
\end{table}
Regardless of the studied frameworks, PPO’s and A2C’s algorithms have shown good performance when applied on the game testing problem. The PPO has shown slightly better performance as it has detected 1 to 2 more bugs than the A2C when used to detect bugs in the Block Maze game. Regarding the studied test case prioritization problem, Pairwise-A2C-SB yields the best performance. The implementations of PPO and A2C algorithms show good performance on discrete action space.

The studied problems have been implemented using both kinds of reward distribution (see Table \ref{tab:General characteristics of the studied RL environments}). In the game testing problem, the agent is positively rewarded only when it reaches the goal otherwise it is rewarded with small negative values. The results have shown that this kind of reward does not incentivize the agent to reach the goal regardless of the DRL implementation applied. In the test case prioritization problem, the agent is positively rewarded with small values even when it fails to rank test cases. Some DRL configurations have performed very well in terms of APFD or NRPA, close to the optimal value.  
\begin{tcolorbox}
\textbf{Recommendation 2: When designing a DRL problem, we recommend keeping the cumulative reward of the agent positive for better performance.}
\end{tcolorbox}


To showcase the difference between employing a simple DRL algorithm from the two frameworks, we performed some additional analysis of the hyperparameters offered by Stable-baselines and Tensorforce regarding the DQN algorithm and conducted some experiments. Here are our findings:
\begin{itemize}
    \item Stable-baselines3 provides a total of \textbf{25 hyperparameters} while Tensorforce provides \textbf{22 hyperparameters}.
    \item Table \ref{tab:DQN algorithm hyperparameters: differences between Stable-Baselines3 and Tensorforce} describes for Stable-baselines3 and Tensorforce, the hyperparameters that differ from each other. 
    \begin{table}[t]
\caption{DQN algorithm hyperparameters: differences between Stable-baselines3 and Tensorforce.}
\centering
\resizebox{\textwidth}{!}{%
\begin{tabular}{|c|c|c|}
\hline
Frameworks                          & Hyperparameters                          & Definitions                                                                                                                                \\ \hline
\multirow{9}{*}{Stable-baselines3} & Soft Update Coefficient                  & \begin{tabular}[c]{@{}c@{}}The target network of the DQN algorithm is \\ updated frequently by a little amount.\end{tabular}               \\ \cline{2-3} 
                                    & Gradient Step                            & \begin{tabular}[c]{@{}c@{}}Number of gradient steps to do \\ after each rollout.\end{tabular}                                              \\ \cline{2-3} 
                                    & Exploration Fraction                     & \begin{tabular}[c]{@{}c@{}}Fraction of the training period over \\ which the exploration rate is reduced\end{tabular}                     
                                                                                                                                      \\ \cline{2-3} 
                                    & Gradient Clipping     & \begin{tabular}[c]{@{}c@{}}The maximum value for \\ the gradient clipping\end{tabular}                                                     \\ \cline{2-3} 
                                    & Tensorboard Log                          & The log location for tensorboard                                                                                                           \\ \cline{2-3} 
                                    & Evaluate Environment & \begin{tabular}[c]{@{}c@{}}Whether to create a second environment that\\  will be used for evaluating the agent periodically.\end{tabular} \\ \cline{2-3} 
                                    & Seed                                     & Seed for the pseudo random generators                                                                                                      \\ \cline{2-3} 
                                    & Device                                   & Device on which the code should be run                                                                                                     \\ \cline{2-3} 
                                    & Model Setup                              & \begin{tabular}[c]{@{}c@{}}Whether or not to build the network at\\  the creation of the instance\end{tabular}                               \\ \hline
\multirow{7}{*}{Tensorforce}        & Variable Noise                           & \begin{tabular}[c]{@{}c@{}}Alternative exploration mechanism by \\ adding Gaussian noise to all trainable variables\end{tabular}                                                                                                                   \\ \cline{2-3} 
                                    & State Preprocessing                      & State preprocessing as layer or list of layers                                                                                             \\ \cline{2-3} 
                                    & Reward Preprocessing                     & Reward preprocessing as layer or list of layers                                                                                            \\ \cline{2-3} 
                                    & Return of processing                     & Return processing as layer or list of layers,                                                                                              \\ \cline{2-3} 
                                    & L2 Regularization                        & L2 regularization loss weight                                                                                                              \\ \cline{2-3} 
                                    & Entropy Regularization                   & \begin{tabular}[c]{@{}c@{}}To discourage the policy distribution \\ from being too certain\end{tabular}                                                                               \\ \cline{2-3} 
                                    & Huber Loss                               & Threshold of the Huber loss function                                                                                                       \\ \hline
\end{tabular}
}
\label{tab:DQN algorithm hyperparameters: differences between Stable-Baselines3 and Tensorforce}
\end{table}
An interesting hyperparameter is the \textit{variable noise} from Tensorforce, which adds Gaussian noise \cite{such2017deep} to all trainable variables as an exploration strategy. Adding noise to DRL agents during training has been shown to improve their exploration of the environment and their gains of reward throughout training \cite{fortunato2017noisy}.
\item We consider the \textit{variable noise=0.5} as an additional hyperparameter for the DQN algorithm from Tensorforce. Therefore, we collected the number of detected bugs and average reward of the DQN agent from Tensorforce for $50,000$ steps of training on the Block Maze game. Figures \ref{fig:dqn_var_noise} and \ref{fig:dqn_reward_var_noise} show that the DQN agent from Tensorforce is able to detect more bugs than initially (see Figure \ref{fig:Figure_DQNS-BUGS} and \ref{fig:Figure_DQNS-AVERAGE-REWARD}), with more gained reward. 
\begin{figure*}[t]
  \begin{minipage}{0.43\textwidth}
        \centering
        \includegraphics[width=\linewidth]{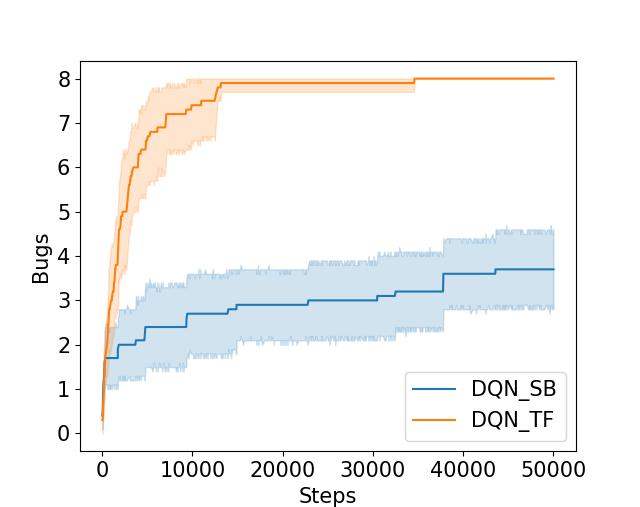}
        \caption{Number of bugs detected \\ by DQN Stable-baselines and DQN \\ (with Gaussian noise) Tensorforce.}
        \label{fig:dqn_var_noise}
    \end{minipage}%
    \hfill
        \begin{minipage}{0.57\textwidth}
        \centering
        \includegraphics[width=\linewidth]{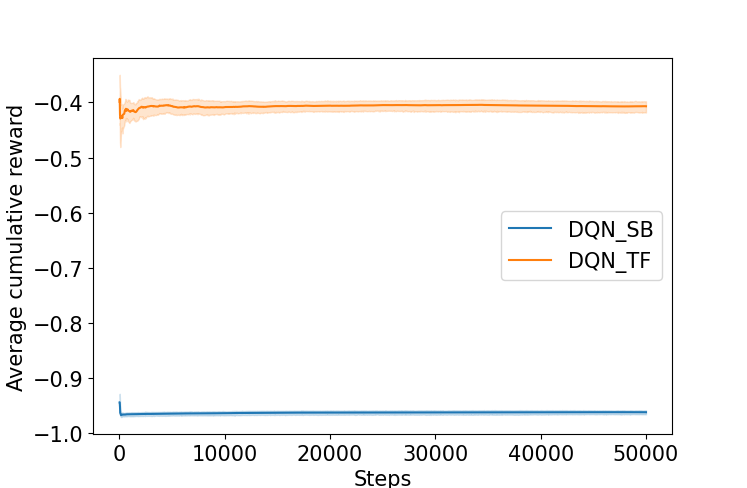}
        \caption{Average cumulative reward earned by DQN Stable-baselines and DQN (with \\Gaussian noise) Tensorforce.}
        \label{fig:dqn_reward_var_noise}
            \end{minipage}%
    \end{figure*}
    \item Furthermore we conducted more experiments to assess the effects of hyperparameters tuning on DQN-TF implementation regarding the game testing problem. Table \ref{tab:Results of State coverage and number of bugs detected, performed by DQN-TF configuration for the game testing problem on a 10k steps budget over 5 runs} shows the number of bugs and state coverage resulting from the hyperparameters tuning.
\begin{table*}[t]
\caption{
Results of State coverage and the number of bugs detected, performed by DQN-TF configuration for the game testing problem on a 10k steps budget over 5 runs.}
\label{tab:Results of State coverage and number of bugs detected, performed by DQN-TF configuration for the game testing problem on a 10k steps budget over 5 runs}
\centering
\begin{tabular}{|c|c|c|c|}
\hline
Hyperparameters                 & Value            & Average number of bugs & Average state coverage \\ \hline
\multirow{3}{*}{Learning rate}  & \textbf{0.00025} & 0.2                    & 4.4                    \\ \cline{2-4} 
                                & 0.01             & 1.6                    & 19.6                   \\ \cline{2-4} 
                                & 0.001            & 0.4                    & 12.6                   \\ \hline
\multirow{3}{*}{Batch size}     & \textbf{128}     & 0.2                    & 4.4                    \\ \cline{2-4} 
                                & 64               & 0.4                    & 7.4                    \\ \cline{2-4} 
                                & 32               & 0                      & 3.4                    \\ \hline
\multirow{3}{*}{Variable noise} & \textbf{0}       & 0.2                    & 4.4                    \\ \cline{2-4} 
                                & 0.5              & 8                      & 74                     \\ \cline{2-4} 
                                & 1                & 7.8                    & 73.8                   \\ \hline
\multirow{3}{*}{Gamma}          & \textbf{0.99}    & 0.2                    & 4.4                    \\ \cline{2-4} 
                                & 0.5              & 0                      & 3.8                    \\ \cline{2-4} 
                                & 0.1              & 0                      & 2.8                    \\ \hline
\end{tabular}
\end{table*}
As other results, in bold are the values of the hyperparameters we initially use for our experiments. Then every time we vary each of them individually (see Column “Value” for their values). As shown on Table \ref{tab:Results of State coverage and number of bugs detected, performed by DQN-TF configuration for the game testing problem on a 10k steps budget over 5 runs}, the only parameter that stands out is the variable noise which boosts DQN-TF performance. Such results indicate that a DRL framework with effective exploration strategies could improve the agent performance.

\item Similarly, Tables \ref{tab:Results of State coverage and number of bugs detected, performed by A2C-TF configuration for the game testing problem on a 10k steps budget over 5 runs} and \ref{tab:Results of State coverage and number of bugs detected, performed by PPO-TF configuration for the game testing problem on a 10k steps budget over 5 runs} show  the results of hyperparameters tuning regarding PPO-TF and A2C-TF implementations.
\begin{table*}[t]
\caption{
Results of State coverage and the number of bugs detected, performed by A2C-TF configuration for the game testing problem on a 10k steps budget over 5 runs.}
\label{tab:Results of State coverage and number of bugs detected, performed by A2C-TF configuration for the game testing problem on a 10k steps budget over 5 runs}
\centering
\resizebox{\textwidth}{!}{
\begin{tabular}{|c|c|c|c|}
\hline
Hyperparameters                         & Value            & Average number of bugs & Average state coverage \\ \hline
\multirow{3}{*}{Learning rate}          & \textbf{0.00025} & 6.6                    & 57.4                   \\ \cline{2-4} 
                                        & 0.01             & 1.8                    & 26.8                   \\ \cline{2-4} 
                                        & 0.001            & 1.6                    & 18                     \\ \hline
\multirow{3}{*}{Batch size}             & \textbf{128}     & 6.6                    & 57.4                   \\ \cline{2-4} 
                                        & 64               & 5                      & 55.4                   \\ \cline{2-4} 
                                        & 32               & 3.2                    & 44.4                   \\ \hline
\multirow{3}{*}{Variable noise}         & \textbf{0}       & 6.6                    & 57.4                   \\ \cline{2-4} 
                                        & 0.5              & 8                      & 74                     \\ \cline{2-4} 
                                        & 1                & 8                      & 74                     \\ \hline
\multirow{3}{*}{Gamma}                  & \textbf{0.99}    & 6.6                    & 57.4                   \\ \cline{2-4} 
                                        & 0.5              & 8                      & 74                     \\ \cline{2-4} 
                                        & 0                & 8                      & 74                     \\ \hline
\multirow{3}{*}{Entropy regularization} & \textbf{0}       & 6.6                    & 57.4                   \\ \cline{2-4} 
                                        & 0.5              & 8                      & 74                     \\ \cline{2-4} 
                                        & 1                & 8                      & 74                     \\ \hline
\multirow{3}{*}{L2 regularization}      & \textbf{0}       & 6.6                    & 57.4                   \\ \cline{2-4} 
                                        & 0.5              & 7.8                    & 73.5                   \\ \cline{2-4} 
                                        & 1                & 7.2                    & 72                     \\ \hline
\multirow{3}{*}{Exploration}            & \textbf{0}                & 6.6                    & 57.4                   \\ \cline{2-4} 
                                        & 0.5              & 7                      & 72                     \\ \cline{2-4} 
                                        & 1                & 8                      & 74                     \\ \hline
\end{tabular}
}
\end{table*}
\begin{table*}[t]
\caption{
Results of State coverage and number of bugs detected, performed by PPO-TF configuration for the game testing problem on a 10k steps budget over 5 runs.}
\label{tab:Results of State coverage and number of bugs detected, performed by PPO-TF configuration for the game testing problem on a 10k steps budget over 5 runs}
\centering
\resizebox{\textwidth}{!}{%
\begin{tabular}{|c|c|c|c|}
\hline
Hyperparameters                            & Value            & Average number of bugs & Average state coverage \\ \hline
\multirow{3}{*}{Learning rate}             & \textbf{0.00025} & 6.2                    & 68.4                   \\ \cline{2-4} 
                                           & 0.01             & 4                      & 52                     \\ \cline{2-4} 
                                           & 0.001            & 5.8                    & 63.4                   \\ \hline
\multirow{3}{*}{Batch size}                & \textbf{128}     & 6.2                    & 68.4                   \\ \cline{2-4} 
                                           & 64               & 5.6                    & 64.8                   \\ \cline{2-4} 
                                           & 32               & 8                      & 71.4                   \\ \hline
\multirow{3}{*}{Variable noise}            & \textbf{0}       & 6.2                    & 68.4                   \\ \cline{2-4} 
                                           & 0.5              & 8                      & 74                     \\ \cline{2-4} 
                                           & 1                & 8                      & 74                     \\ \hline
\multirow{3}{*}{Gamma}                     & \textbf{0.99}    & 6.2                    & 68.4                   \\ \cline{2-4} 
                                           & 0.5              & 7.2                    & 71.4                   \\ \cline{2-4} 
                                           & 0.1              & 7                      & 70.6                   \\ \hline
\multirow{3}{*}{Entropy regularization}    & \textbf{0}       & 6.2                    & 68.4                   \\ \cline{2-4} 
                                           & 0.5              & 8                      & 74                     \\ \cline{2-4} 
                                           & 1                & 8                      & 74                     \\ \hline
\multirow{3}{*}{L2 regularization}         & \textbf{0}       & 6.2                    & 68.4                   \\ \cline{2-4} 
                                           & 0.5              & 8                      & 74                     \\ \cline{2-4} 
                                           & 1                & 8                      & 71.4                   \\ \hline
\multirow{3}{*}{Exploration}               & \textbf{0}       & 6.2                    & 68.4                   \\ \cline{2-4} 
                                           & 0.5              & 6.8                    & 68.8                   \\ \cline{2-4} 
                                           & 1                & 8                      & 74                     \\ \hline
\multirow{3}{*}{Likelihood ratio clipping} & \textbf{0.25}    & 6.2                    & 68.4                   \\ \cline{2-4} 
                                           & 0.5              & 7.8                    & 73.8                   \\ \cline{2-4} 
                                           & 1                & 7.2                    & 69.8                   \\ \hline
\end{tabular}
}
\end{table*}
The results on these tables show up to 2 more bugs detected when applying different values of the hyperparameters that the Tensorforce framework offers (i.e., variable noise, discount factor, entropy/l2 regularization, and exploration).

\end{itemize}
\begin{tcolorbox}
\textbf{Recommendation 3: In the context of testing through exploration, we recommend an DRL framework that offers effective exploration strategies such as Tensorforce.}
\end{tcolorbox}

The performance of A2Cs and PPOs algorithms from the selected DRL frameworks indicate that their faster convergence rate led to the detection of bugs quickly, as well as a wider state coverage capability.

\begin{tcolorbox}
    Recommendation 4: For small search space environments, we recommend A2Cs and PPOs algorithms for a faster convergence rate and a wider state coverage capability of DRL agents.
\end{tcolorbox}

The performance of the DRL algorithms when applying to the datasets of the test case prioritization problem indicates that the DRL agents are not able to learn an accurate policy when it comes to  simple datasets. 

\begin{tcolorbox}
    Recommendation 5: For the DRL agents to learn an accurate policy for complex software systems, we recommend enriched datasets.
\end{tcolorbox}
\section{Related work}
\label{Related work}

Incorporating DRL algorithms in software engineering tasks has long been an active area of research \cite{singh2013architecture,bahrpeyma2015adaptive,chen2020enhanced, vuong2018reinforcement}.
In the case of SE testing, wuji by Zheng et al. \cite{zheng2019wuji}, is a framework that applies EA, MOO, and DRL to facilitate automatic game testing. EA and MOO are designed to explore states and DRL ensures the completion of the mission of the game. Further, the authors use the Block Maze game and two commercial online games to evaluate wuji. This work is used as a baseline in this paper. We compare the DRL part of wuji to state-of-the-art DRL algorithms from DRL frameworks. Specifically, we implement DRL algorithms from DRL frameworks to detect bugs in the Block Maze game and assess their performance against the DRL part of wuji.

Bagherzadeh et al. \cite{bagherzadeh2021reinforcement} leveraged state-of-the-art DRL algorithms from the Stable-baselines framework in CI regression testing. They investigate pointwise, pairwise, and listwise ranking models as DRL problems to find the optimal prioritized test cases. The authors also conducted experiments on eight datasets and compared their solutions against a small subset of non-standard DRL implementations. Again, we use this work as a baseline and implement DRL algorithms from DRL frameworks to rank test cases in a CI environment. As the authors implement the Stable-baselines framework, we leverage 2 other DRL frameworks (Tensorforce and Keras-rl) and compare them to Stable-baselines.

Koroglu et al. \cite{koroglu2018qbe} proposed QBE, a Q-learning framework to automatically test mobile apps. QBE generates behavior models and uses them to train the transition prioritization matrix with two optimization goals: activity coverage and the number of crashes. Its goal is to improve the code coverage and the number of detected crashes for Android apps. Bottinger et al. \cite{bottinger2018deep} introduced a program fuzzer that uses DRL to learn reward seed mutations for testing software. This technique obtains new inputs that can drive a program execution towards a predefined goal, e.g., maximizing code coverage. Kim et al. \cite{kim2018generating}, leveraged DRL  to automatically generate test data from structural coverage. Particularly, a Double DQN agent is trained in a Search-based Software Testing (SBST) environment to find a qualifying solution following the feedback from the fitness function. Chen et al. \cite{chen2020enhanced} proposed RecBi, the first compiler bug approach via structural mutation that uses DRL. RecBi uses the A2C algorithm to mutate a given failing test program. Then, it uses that failed test program to identify compilers' bugs.
Adamo et al.\cite{adamo2018reinforcement} Reichstaller et al.\cite{reichstaller2018risk} Dai et al. \cite{dai2019learning} used DRL to generate test cases. Adamo et al. \cite{adamo2018reinforcement}, build a DQN-based testing tool that generates test cases for Android applications. The tool is guided by the code coverage to generate suitable test suites. Reichstaller et al. \cite{reichstaller2018risk}, proposed a framework to test a Self-Adaptive System (SAS) where the tester is modeled as a Markov Decision Process (MDP). The MDP is then solved by using both model-free and model-based DRL algorithms to generate test cases that will adapt to SAS as they have the ability to take decisions at runtime. Soualhia et al. \cite{Soualhia20} leveraged DRL algorithms to propose a dynamic and failure-aware framework that adjusts Hadoop's scheduling decisions based on events occurring in a cloud environment. Each of these previous approaches from the literature either implements a DRL algorithm from scratch or uses an implemented one from a DRL framework. None of them has evaluated the performance of DRL frameworks on software testing tasks. Moreover, it is not clear what motivates the choice of DRL frameworks in the literature, as there are several of them. In our work, we investigate various state-of-the-art DRL algorithms from popular DRL frameworks, to assess DRL configurations on a game and regression testing environments.

\section{Threats to validity}\label{threats}

\textbf{Conclusion validity.}
Conclusion limitations concern the degree to which the statistical conclusions about which algorithms/frameworks perform best are accurate. 
We use Welch's ANOVA and Games-Howell's post-hoc test as statistical tests. The significance level is set to 0.05 which is standard across the literature as shown by Welch et al.\cite{welch1947generalization}, Games et al. \cite{games1976pairwise}. The non-deterministic nature of DRL algorithms can threaten the conclusions made in this work. We address this by collecting results from 10 independent runs in the case of the game testing problem. Regarding the test case prioritization problem, the results are collected from 5 independent runs and on multiple cycles (the MATH dataset has 55 cycles which is the least number of cycles among all datasets). 

\textbf{Internal validity.} Regarding the game testing problem, the fact that we only consider the DRL part of the wuji framework for comparison with the DRL strategies we studied, might threaten the validity of this work. Although we only compare the algorithms on sub-optimal solutions, it is necessary to make a fair comparison among the DRL algorithms. A potential limitation is the number of frameworks used and the algorithms chosen among these frameworks. We have chosen to evaluate some of the available frameworks and have not evaluated all the algorithms they offer. However, the frameworks used are among the most popular on GitHub, as well as the algorithms (see Section \ref{sec:Reinforcement Learning}). This ensures good coverage in terms of the usage of DRL in SE. In the future, we plan to expand our study to cover more algorithms. 

\textbf{Construct validity.}
A potential threat to validity is related to our evaluation metrics, which are standard across the literature. We 
use these metrics to make a fair comparison amongst framework/algorithms under identical circumstances. We discussed some of their limitations and how they can be interpreted in Sections \ref{sec:REINFORCEMENT LEARNING FOR GAME TESTING} and \ref{sec:Reinforcement Learning for Test case prioritization}. 


\textbf{External validity.} Since our goal is to compare DRL frameworks and their implemented algorithms for SE testing tasks, a potential limitation is the choice of the testing tasks for comparing frameworks. We address this threat by choosing game testing and test case prioritization problems that are totally different in SE testing to achieve enough diversity. While test case prioritization focuses on optimizing the order of test cases, one needs to find bugs in the game testing as early as possible. The results we found on the two studied problems might mitigate this threat as we consistently found some algorithms performing similarly among the frameworks. For example, the A2C algorithm had good performance whether applied to the game testing problem or test case prioritization problem.

\textbf{Reliability validity.} To allow other researchers to replicate or build on our research, we provide a detailed replication package \cite{replication-package} including the code and obtained results.

\section{Conclusion and Discussions} \label{Conclusion}
In this paper, we study the application of state-of-the-art implemented DRL algorithms from well-known frameworks on two important software testing tasks: test case prioritization and game testing. We rely on two baseline studies to apply and evaluate the performance of DRL algorithms from several frameworks (i) in terms of detecting bugs in a game, and (ii) in the context of a CI environment to rank test cases. Our results show that the same algorithm from different DRL frameworks can have different performances. Each framework provides hyperparameters unique to its implementation, therefore depending on the underlying SE tasks, the framework that has the most suitable hyperparameters will lead to better performance. We formulate recommendations to help SE practitioners to make an informed decision when leveraging DRL frameworks for the  development of SE tasks. In the future, we plan to expand our study to investigate more DRL algorithms/frameworks, and more SE activities.

Regarding the game testing problem, the DQN algorithm among all the studied frameworks has poor performance when detecting bugs, implying poor exploration capability of DQN. For the test case prioritization problem, Table \ref{tab:Welch ANOVA and Games-Howell post-hoc tests regarding the ranking model} shows that for the pairwise configuration and some enriched datasets DQN's performance is close to the A2C. It shows that DQN has a good ranking capability. The DQN algorithm computes the Q-values of each state-action pair in order to predict the next action to take. So, it is suitable for the pairwise ranking model as its action space is discrete (0 or 1), half the size of the action space of the game testing problem (0, 1, 2 or 4). This makes us wonder whether the discrete nature of the action space could be a factor in the obtained results. In the future, we plan to investigate it in more detail.

Regardless of the studied frameworks, PPO and A2C algorithms have shown good performance when applying on the game testing problem. The PPO has performed slightly better as it has detected 1 to 2 more bugs than the A2C when used to detect bugs in the Block Maze game. The PPO from Stable-baselines framework has detected 1 to 2 more bugs than the PPO algorithm from Tensorforce. Nevertheless, the hyperparameters tuning that was reported in Section \ref{futurework} by changing the variable noise parameter in PPO-TF has increased its detection capability to be better than PPO-SB. Therefore, in the future we plan to investigate PPO-SB and PPO-TF implementations in more detail.

\section{Conflict of interest}
The authors declared that they have no conflict of interest. 
\bibliographystyle{spbasic}
\bibliography{references.bib}
\end{document}